\RequirePackage{lineno}
\documentclass[twoside,12pt]{article}
\usepackage{epsfig}
\usepackage{graphicx}
\usepackage{subfigure}
\usepackage{multirow}
\usepackage[dvips,usenames]{color}
\usepackage{hyperref}
\usepackage{breakurl}
\usepackage{float}

\def\Journal#1#2#3#4{{#1} {#2} (#4) #3 }

\def\NPA{{\em Nucl. Phys.} A}

\def\PLB{{\em Phys. Lett.} B}

\def\PRL{\em Phys. Rev. Lett.}

\def\PREP{\em Phys. Rep.}

\def\PRD{{\em Phys. Rev.} D}
\def\PRC{{\em Phys. Rev.} C}

\newcommand{\be}{\begin{equation}}
\newcommand{\ee}{\end{equation}}
\newcommand{\bea}{\begin{eqnarray}}
\newcommand{\eea}{\end{eqnarray}}

\begin{document}


\title{ \vspace{1cm} Flavor Structure of the Nucleon Sea}
\author{Wen-Chen Chang$^{1}$ and Jen-Chieh Peng$^{2}$\\
\\
$^1$Institute of Physics, Academia Sinica, Taipei 11529, Taiwan\\
$^2$Department of Physics, University of Illinois at Urbana-Champaign\\
Urbana, Illinois 61801, USA
}
\maketitle

\begin{abstract} 
We review the current status and future prospects on the subject of flavor
structure of the nucleon sea. The flavor structure of the nucleon sea
provides unique information on the non-perturbative aspects of strong
interactions allowing stringent tests of various models on the
partonic structures of the nucleons as well as lattice QCD
calculations. The scope of this review covers the unpolarized,
polarized, and the transverse-momentum dependent sea-quark
distributions of the nucleons. While the main focus of this review 
is on the physics motivation and recent progress on the subject of
the nucleon sea, we also discuss future prospects of
addressing some outstanding issues on the flavor structure of the nucleon
sea. 
\end{abstract}
\eject
\tableofcontents
\eject


\section{Introduction}
\label{sec:intro}

The observation of the Bjorken-$x$ scaling behavior in deep inelastic 
scattering (DIS) revealed quarks as the point-like
constituents of the nucleons~\cite{friedman}. The existence of sea
quarks was further suggested from the DIS data, showing a prominent
rise of the partonic density at small $x$. The discovery of the
Drell-Yan process in hadron-hadron collision, which involves the
annihilation of a quark-antiquark pair, provided a further experimental
evidence for the existence of sea quarks in the nucleons.

Unlike the situation in atomic systems, where the effects of
particle-antiparticle pairs are relatively minor, the quark-antiquark
pairs have important roles in describing the internal structures of
nucleons and other hadrons. The large magnitude of the coupling
constant $\alpha_s$ in QCD implies that quark-antiquark pairs are readily
produced. The sea quarks, like the valence quarks and gluons, form an
integral part of the nucleon's structure. After several decades of
extensive experimental and theoretical studies, the valence quark
distributions of the nucleons are quite well known. In contrast, some
important aspects of the sea quarks, such as their flavor and spin
dependence, are just beginning to be explored.

In this article, we review the current status and future prospects for
our understanding of the flavor and spin structures of the sea quark
contents in the nucleons.  While the valence quarks in the nucleons are
restricted to the up and down flavors, the flavor structure of the sea
quarks is potentially richer, extending beyond the light quarks.  For
the light quark flavor dependence, a major surprise was found when DIS
and Drell-Yan experiments showed that the $\bar u$ and $\bar d$ have
very different $x$ distributions. As discussed later, this unexpected
finding has generated much interest and has provided new insights on
the origins of the nucleons sea. At the Large Hadron Collider (LHC) or
the future Electron Ion Collider (EIC), the role of sea quarks becomes
more important, since the densities of the sea quarks, especially for
the heavy flavors, are expected to rise at the large $Q^2$ and small
$x$ regions explored in these high energy colliders. This offers the
opportunity to study the flavor structures of the light quark
sea ($\bar u, \bar d$)
at new kinematic regions, as well as the heavy quark sea 
($s, c, b$) of which much less is known.

Another intriguing aspect of sea quarks is their spin structure.
Following the discovery of the ``spin puzzle" in the 1980s, the
decomposition of the proton's spin in terms of the spin and orbital
motion of the constituent quarks, antiquarks, and gluons continued to
be a major unresolved issue in high energy spin physics. The flavor
and $x$ dependence of the antiquark helicity distributions could offer
important insights for solving this puzzle. Moreover, many theoretical
models capable of explaining the difference between the unpolarized
$\bar u$ and $\bar d$ momentum distributions also have distinct predictions for
the flavor dependence of the polarized sea. New information on the
flavor structure of the sea quark polarization would provide a
stringent test for these models and would further advance our
understanding on the origin of the $\bar d/ \bar u$ flavor
asymmetry.

During the recent decades, significant progress has been made in
delineating the roles of the transverse momentum and transverse spin
of the quarks in nucleon structure. The properties of some novel
parton distributions involving these transverse degrees of freedom
have been investigated theoretically and first data are becoming
available from the semi-inclusive DIS measurements. The possibility to
measure the transverse spin and transverse momentum dependent
sea-quark distributions will offer another interesting view on the
nature of the nucleon sea.

In this review article, we discuss the recent progress and future
prospects for understanding the flavor and spin structure of the 
nucleon sea. This article
is organized as follows. In Section 2, we present a brief overview of the
relevant experimental tools in exploring the sea quarks. In
Section 3, we discuss several intriguing aspects of the
flavor structure of unpolarized nucleon sea. This is followed by discussions
on the helicity distributions of the nucleon sea in Section 4. We then
briefly review the topics of transverse spin or 
transverse-momentum dependent sea quark structures, which is just 
beginning to be explored. In Section 5, we highlight future 
prospects for advancing our
current knowledge on various outstanding issues in the subject of
nucleon sea, followed by summary and conclusion in Section 6.


\section{Probing the Nucleon Sea} 
\label{sec:overview}

Our knowledge about the partonic structure of the
nucleon~\cite{PDF_Roeck2011,PDF_Perez2013,PDF_Forte2013,PDF_Delgado2013,polPDF_Lampe1998,polPDF_deFlorian2011,polPDF_Aidala2012}
mainly comes from the deep inelastic scattering (DIS) using either
charged lepton or neutrino beams~\cite{DIS_Blumlein2012}. The DIS is
in general not effective in separating the sea quarks from the valence
quarks.  However, by detecting other particles (hadrons or charged
leptons) in coincidence with the scattered leptons, the semi-inclusive
DIS (SIDIS) process is sensitive to the flavor of the struck quark or
antiquark. Complementary information could be obtained from the
Drell-Yan process~\cite{DrellYan} where the quark and antiquark from
two interacting hadrons annihilate into a virtual photon, $Z$, or $W$
bosons~\cite{Chang:2013opa,DY_Peng2013}. This process yields
particularly important information on the sea quark content of the
nucleons.

Below we briefly introduce the kinematics of DIS, Drell-Yan and SIDIS
processes, following the formulation of
Refs.~\cite{StructFunc,CTEQ-Handbook,nuDIScross} and provide the leading
order expressions of the production cross sections.
We end this Section with a table listing the various
reactions and the associated subprocesses which are relevant for
probing sea quarks in the nucleon.

\subsection{Deep Inelastic Scattering with Charged Lepton Beams}

\begin{figure}[htb]
\centering
\subfigure[]
{\includegraphics[width=0.40\textwidth]{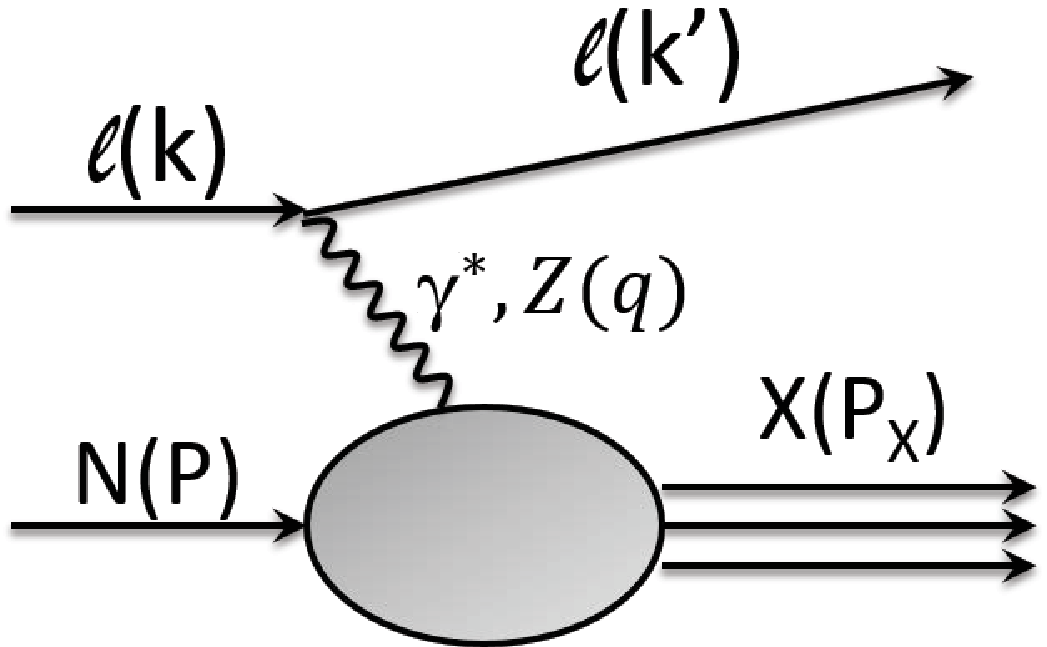}}
\subfigure[]
{\includegraphics[width=0.40\textwidth]{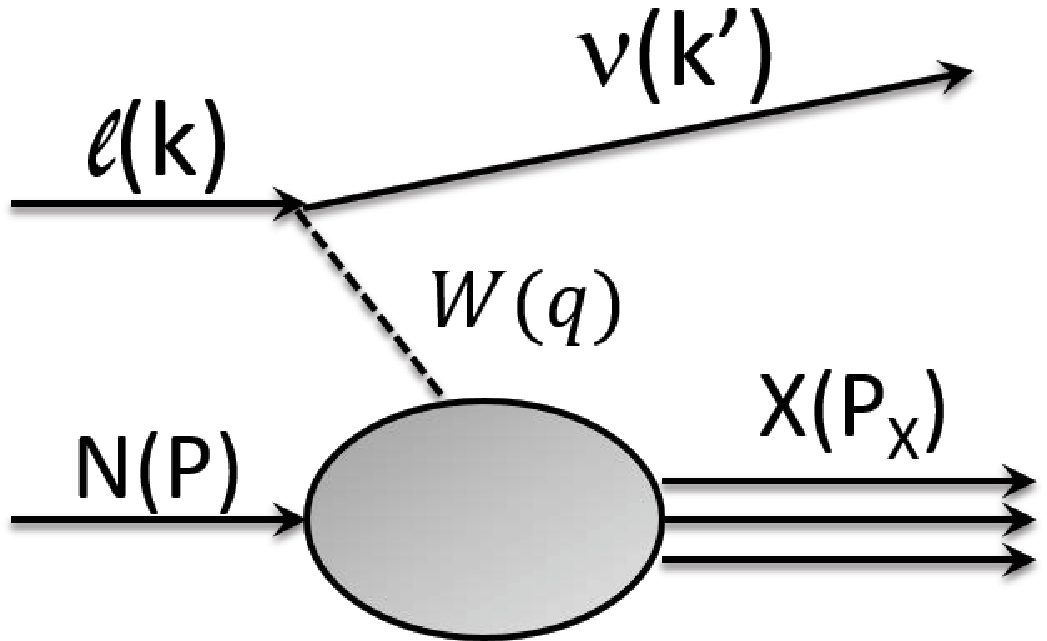}}
\caption{Inclusive lepton-hadron scattering via (a) $\gamma^*, Z$ (b)
  $W$ exchange.}
\label{fig:DIS}
\end{figure}

Figure~\ref{fig:DIS} illustrates the deep inelastic scattering process
of charged lepton beam off a nuclear $l N \rightarrow l X$ or $l N
\rightarrow \nu X$ where $l$ denotes the charged leptons, $\nu$ the
neutrino, $N$ the nucleon target and $X$ the undetected hadronic final
state. At leading order, the lepton couples to the nucleon either
through a virtual photon $\gamma^*$, $Z$ or $W$ boson. In the past
these processes were studied in the fixed-target experiments except
the collider experiments at HERA.

\noindent

The four-momentum vectors of the incoming and outgoing leptons are
denoted as $k$ and $k'$ respectively, and $P$ is that of the nucleon.
The momentum of the exchanged boson is $q = k - k'$, and the momentum
of the hadronic final state system $X$ is $P_X = P + q$. The
virtuality of the exchanged boson in DIS is spacelike, i.e. $q^2 <
0$.

Below we list several Lorentz invariants and kinematic variables
useful for the description of the kinematics of the process:

\begin{itemize}

\item $s = (k + P)^2$: the center of mass energy squared for the
  $lN$ system.

\item $Q^2 = -q^2 \equiv -(k - k')^2$: the magnitude of the invariant mass squared of the
  exchanged boson.

\item $\nu = (P \cdot q) / M_N = E_{k}-E_{k'}$: the energy transferred from the
  incident lepton to the target nucleon in the target rest frame.
 
\item $x = Q^2/(2 P \cdot q) = Q^2/(2 M_N \nu)$: the dimensionless
  Bjorken scaling variable, which represents the nucleon momentum
  fraction of the struck quark in the quark-parton.

\item $W^2 = (P + q)^2$: the invariant mass squared of the hadronic
  system $X$ in the final state.

\item $y = (P \cdot q) / (P \cdot k) = \nu / E_{k}$: the ratio of the
  transferred energy over the total lepton energy in the target
  nucleon rest frame.

\end{itemize}
The $x$, $Q^2$ and $y$ are related to $s$ by the relation of $Q^2 =
sxy$ in the massless limit and thus they are not completely
independent variables. Usually the differential cross sections are
given in two of them: $x, Q^2$ or $x, y$.

For charged lepton-nucleon scattering, mediated by the neutral current
(NC), i.e. the exchange of $\gamma^*$ or $Z$, the differential
cross section may be written in terms of a leptonic and a hadronic
tensor and their coupling via the exchanged boson. The hadronic tensor
is not known from first principle and represented by three structure
functions, $F_2$, $F_L$ and $xF_3$. The differential cross section is
expressed as~\cite{StructFunc}
\begin{equation}
\frac {d^2\sigma^{NC} (l^{\pm}N) } {dxdQ^2} =  \frac {2\pi\alpha^2} {x Q^4}  
\left[Y_+\,F_2^{NC}(x,Q^2) - y^2 \,F_L^{NC}(x,Q^2)
\mp Y_-\, xF_3^{NC}(x,Q^2) \right],
\label{eq:NCxsec}
\end{equation}
where $\alpha$ is the fine structure constant. The factor $Y_\pm
\equiv 1\pm(1-y)^2$ is due to the helicity dependence of the
electroweak interaction.

Provided that $Q^2 \gg 4 M^2_N x^2$ and $Q^2$ is well below that of
the $Z$ mass squared ($M_Z^2$), the parity violating structure
function $xF_3^{NC}$ is negligible. Furthermore there is no
longitudinal absorption cross section for $\gamma^*$ scattering on
quarks, i.e. $F_L$=0, because of the spin-1/2 nature of
quarks. Therefore the differential cross section of charged
lepton-nucleon scattering comes only from the first term in
Eq.~\ref{eq:NCxsec},
\begin{equation} 
\frac {d^2\sigma^{NC} (l^{\pm}N) } {dxdQ^2} = \frac {2\pi\alpha^2}
      {x Q^4} Y_+\,F_2^{NC}(x,Q^2).
\end{equation}

In the parton model, only charged partons of the hadron, the quarks
and antiquarks, couple to the electroweak currents at leading
order. The structure function $F_2^{NC}(x,Q^2)$ could be expressed in terms
of the parton density of the quark $q_i$ and antiquark $\bar q_i$ in
the nucleon and the quark charge squared $e_i^2$ as follows:
\begin{equation}
    F_2^{NC}(x,Q^2) = \sum_{i=1}^{n_f} e_i^2 \left[ xq_i(x,Q^2) + 
    x \bar q_i (x,Q^2) \right],
\end{equation}
where index $i$ refers to the quark flavor. Therefore the
lepton-nucleon scattering process has been the main experimental tool
for measuring the quark distribution functions of nucleons and
investigating their $Q^2$ dependence.

When $Q^2$ becomes comparable to $M_Z^2$, the $Z$ exchange and the
effect of $\gamma^*-Z$ interference have to be taken into account and
the parity violating structure function $xF_3$ can no longer be
neglected. The complete expression is referred to, for example,
Ref.~\cite{StructFunc}.

As for the charged-current (CC) where $W^{\pm}$ is exchanged and the
final state lepton is a (anti)neutrino, the charged lepton-nucleon
differential cross sections are expressed as~\cite{StructFunc}
\begin{equation} \frac {d^2\sigma^{CC}(l^\pm N) }
    {dxdQ^2} = \frac {G_F^2} {4\pi x} \left[ \frac { M^2_W} { (Q^2 + M_W^2)} \right]^2
    \left[ Y_+\,F_2^{CC}(x,Q^2) - y^2 \, F_L^{CC}(x,Q^2) \mp Y_-\, xF_3^{CC}(x,Q^2)
      \right],
\label{eq:CCxsec}
\end{equation}
and the Fermi coupling constant $G_F$ is
\begin{equation}
        G_F = \frac{\pi \alpha} {\sqrt 2 \sin^2\theta_W M^2_W},
\end{equation}
where $\theta_W$ is the Weinberg angle and $M_W$ the mass of $W$
boson.

In the parton model, $F_L$ vanishes in LO pQCD. Neglecting the
contribution from top($t$) and bottom($b$) quarks, the structure
functions $F_2^{CC}$ and $xF_3^{CC}$ are expressed by the sums and
differences of quarks and antiquarks densities as
\begin{eqnarray}
    F_2^{CC}(l^- N) & = & 2x( u + c + \bar d + \bar s ) \nonumber \\
    xF_3^{CC}(l^- N) & = & 2x( u + c - \bar d - \bar s)
\end{eqnarray}
for the lepton beam of left-handed polarization, and
\begin{eqnarray}
    F_2^{CC}(l^+ N) & = & 2x( \bar u + \bar c + d + s ) \nonumber\\
    xF_3^{CC}(l^+ N) & = & 2x( d + s - \bar u - \bar c)
\end{eqnarray}
for that of right-handed polarization. Hence the differential
cross section of lepton scattering is given by~\cite{StructFunc}
\begin{equation}
\frac {d^2\sigma^{CC}(l^-N) } {dxdQ^2} = (1 - P)\frac {G_F^2} {2\pi x} \left[ \frac { M^2_W} { (Q^2 + M_W^2)} \right]^2   \sum_{i=1}^{n_f} \left[ 
 xq_i(x,Q^2) + (1-y)^2 x\bar q_i (x,Q^2) \right]
\label{eq:CCxsec1}
\end{equation}
whereas that of antilepton scattering,
\begin{equation}
\frac {d^2\sigma^{CC}(l^+N) } {dxdQ^2} = (1 + P)\frac {G_F^2} {2\pi x } \left[ \frac { M^2_W} { (Q^2 + M_W^2)} \right]^2  \sum_{i=1}^{n_f} \left[ 
(1-y)^2 xq_i(x,Q^2) +  x\bar q_i (x,Q^2) \right]
\label{eq:CCxsec2}
\end{equation}
where the summation goes over only the quarks or antiquarks relevant for
the charge of the exchanged current and $P$ is the polarization degree
of the lepton beam, $P = (N_R - N_L)/(N_R + N_L)$. 

\subsection{Deep Inelastic Scattering with Neutrino or Antineutrino Beams}
\label{sec:nudis}

\begin{figure}[htb]
\centering \includegraphics[width=0.40\textwidth]{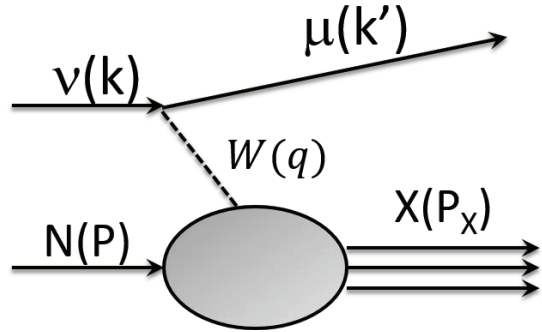}
\caption{Inclusive neutrino scattering via $W$ exchange.}
\label{fig:nuDIS}
\end{figure}
\noindent

The charged-current interactions were most extensively studied in the
neutrino and antineutrino inelastic scattering processes, where
\begin{equation}
               \nu(\bar \nu) N \rightarrow \mu^-(\mu^+) X
\end{equation}
at $Q^2 \ll M_W^2$, as illustrated in Fig.~\ref{fig:nuDIS}, and the
cross sections are expressed as
\begin{equation} \frac {d^2\sigma^{CC}(\nu,\bar \nu ) }
    {dx dQ^2} = \frac{(1 \pm P)}{2} \frac {G_F^2} {4\pi x} \left[ Y_+\,F_2^{CC}(x,Q^2) - y^2
      \,F_L^{CC}(x,Q^2) \pm Y_-\,xF_3^{CC}(x,Q^2) \right].
\end{equation}
Since the polarizations of the (anti)neutrino probes are intrinsically
(right)left-handed, both polarization factors of the beam $(1 \pm P)$
become 2 now.

In LO pQCD the differential cross sections of neutrino scattering
expressed by the nucleon parton densities are
\begin{equation}
\frac {d^2\sigma(\nu) } {dx dQ^2} = \frac {G_F^2} {\pi x} \sum_{i=1}^{n_f} \left[ 
 xq_i(x,Q^2) + (1-y)^2 x\bar q_i (x,Q^2) \right]
\end{equation}
and those for antineutrino scattering,
\begin{equation}
\frac {d^2\sigma(\bar \nu) } {dx dQ^2} = \frac {G_F^2} {\pi x}
\sum_{i=1}^{n_f} \left[ (1-y)^2 xq_i(x,Q^2) + x\bar q_i (x,Q^2)
  \right]
\end{equation}
where the sums go through only the appropriate quarks or antiquarks
for the charge of the current. Because of the different coupling
strength of the exchanged $W$ boson with the quarks and antiquarks,
the combination of the data from neutrino and antineutrino DIS
experiments are used to extract the individual quark and antiquark
densities. The other combination of data from proton and a heavy
isoscalar target could lead to flavor separation of parton densities.

\subsection{Drell-Yan Process}

\begin{figure}[htb]
\centering
\subfigure[]
{\includegraphics[width=0.40\textwidth]{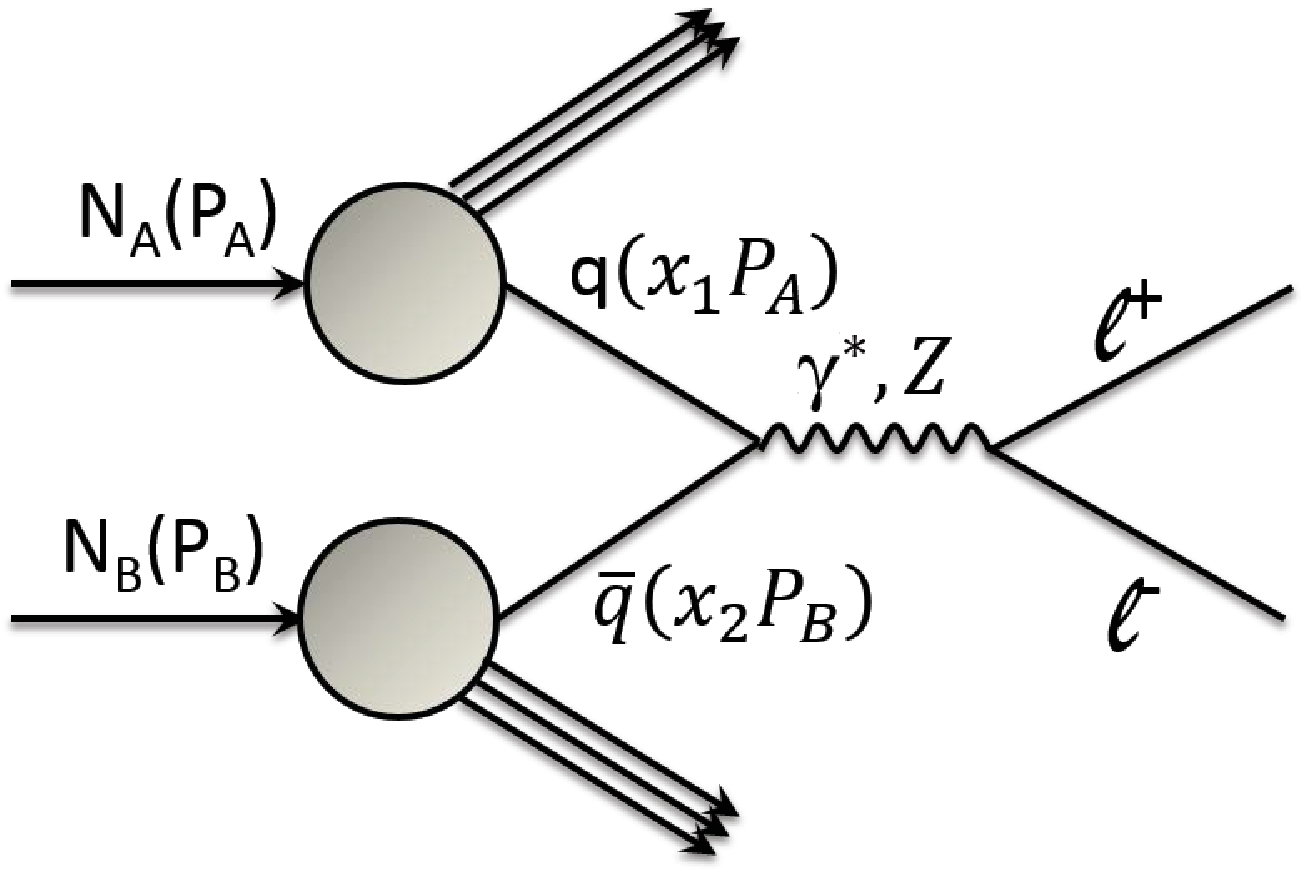}}
\subfigure[]
{\includegraphics[width=0.40\textwidth]{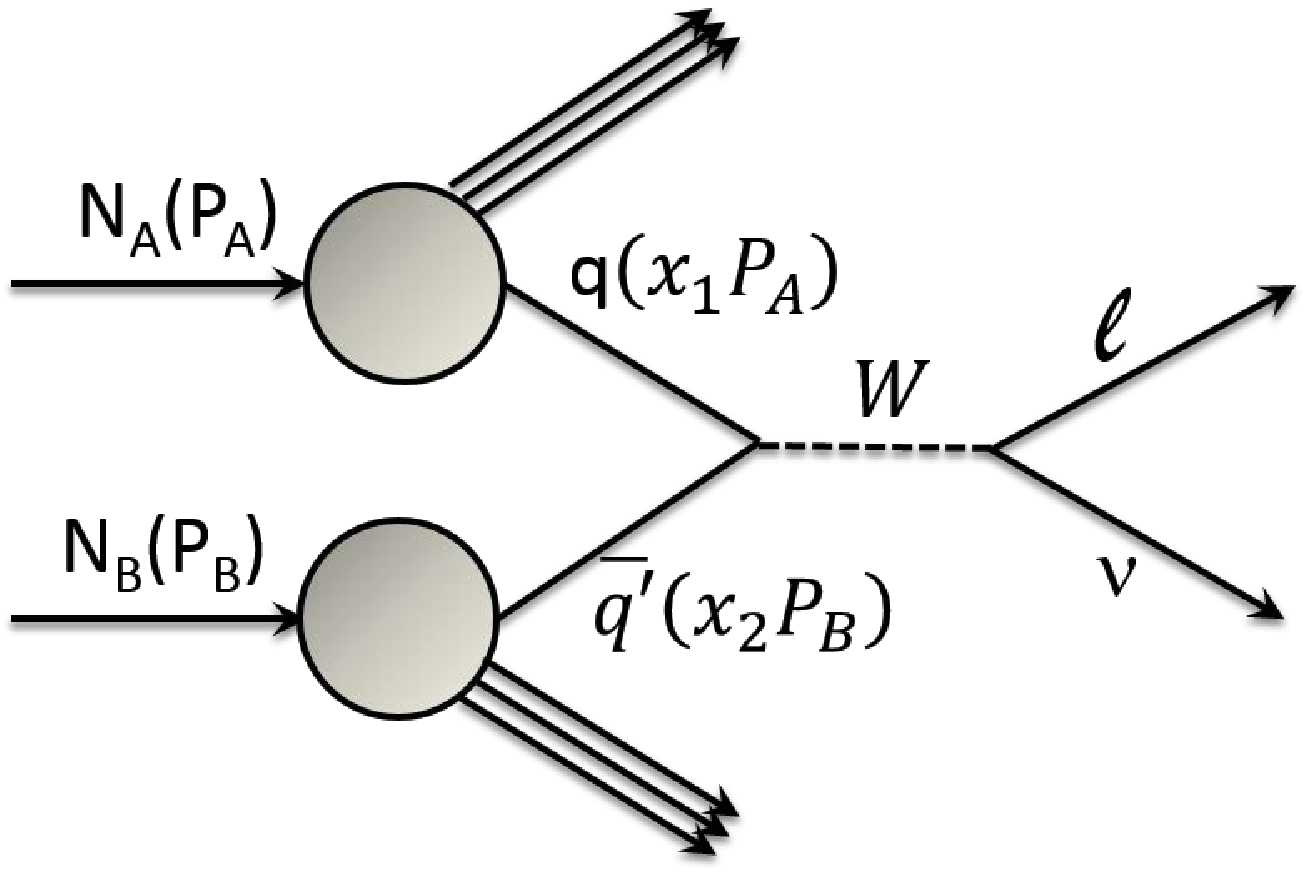}}
\caption{Drell-Yan process for (a) $l^+ l^-$ and (b) $l \nu$ production
in the collisions of two nucleons.}
\label{fig:DY}
\end{figure}

As illustrated in Fig.~\ref{fig:DY}, the Drell-Yan (DY) process is the
production of a lepton pair with large invariant mass ($l^+l^-$ and $l
\nu$) in the collisions of two hadrons.
Different from DIS, the virtual
boson formed in DY is timelike, i.e. $q^2 = Q^2 >0$. The square of 
the center of
mass energy for two colliding hadrons is $s \equiv (P_A + P_B)^2$. In
the parton model, the intermediate bosons ($\gamma^*, Z, W$)
are produced by the
annihilation of quark-antiquark pair, which then couple to the
lepton pair through electromagnetic or weak interactions. In general
the DY process yields information complementary to what is revealed in
DIS and is particularly useful in probing the distributions of antiquarks.

\noindent

The inclusive DY cross section of collisions of hadron $A$ and $B$
is expressed as~\cite{CTEQ-Handbook}
\begin{eqnarray}
\frac{d \sigma^V}{d Q^2} & = & \sigma_0^V W_{AB}^V(\tau) \\
W_{AB}^V(\tau) & = & \int_0^1 dx_1 \int_0^1 dx_2 \delta(\tau - x_1 x_2)D_{AB}^V
\end{eqnarray}
with $V$ = $\gamma^*$, $Z$ or $W$, and $\tau=Q^2/s$ is the DY scaling
variable. The factor $\sigma_0^V$ contains the full dimensions of $d
\sigma^V / d Q^2$ and the dimensionless function $W_{AB}^V$ is the
convoluted integral over the product of the parton distributions in
the projectile and target hadrons denoted as $D_{AB}^V$. The $x_1$ and
$x_2$ are the momentum fractions of interacting partons in the hadron
$A$ and $B$ respectively.

For the production of virtual photon $\gamma^*$, we have
\begin{eqnarray}
\sigma_0^{\gamma^*} & = & \frac{4 \pi \alpha^2}{3 N_c Q^2 s} \\ 
D_{AB}^{\gamma^*} (x_1, x_2) & = & \sum_{i=1}^{n_f} e_i^2 \{
q_i^A(x_1)\bar{q_i}^B(x_2)+(A \leftrightarrow B) \}
\end{eqnarray}
where $N_c$ is the number of color charge.

In the case of $Z$ boson production, the $\sigma_0^V$ and $D_{AB}^V$ become
\begin{eqnarray}
\sigma_0^{Z} & = & \frac{\pi \alpha^2}{192 N_c \sin^4\theta_W  
\cos^4\theta_W} \frac{q^2}{s} \frac{1+(1-4 \sin^4\theta_W)^2 }{(Q^2-M_Z^2)^2 + M_Z^2 \Gamma_Z^2}\\
D_{AB}^{Z} (x_1, x_2) & = & \sum_{i=1}^{n_f} [1+(1-4|e_i|\sin^2\theta_W)^2]\{ q_i^A(x_1)\bar{q_i}^B(x_2)+(A \leftrightarrow B) \}
\end{eqnarray}
where $\theta_W$ is the Weinberg angle and $M_z$ and $\Gamma_z$ are
the mass and total width of the $Z$ boson.

Similarly, the corresponding results for the case of $V=W^-$ are
\begin{eqnarray}
\sigma_0^{W^-} & = & \frac{\pi \alpha^2}{12 N_c \sin^4\theta_W} \frac{q^2}{s} \frac{1}{(Q^2-M_W^2)^2 + M_W^2 \Gamma_W^2}\\
D_{AB}^{W^-} (x_1, x_2) & = & \cos^2 \theta_C [ d^A(x_1)\bar{u}^B(x_2) + s^A(x_1)\bar{c}^B(x_2)] \nonumber \\
& + & \sin^2 \theta_C [ s^A(x_1)\bar{u}^B(x_2) + u^A(x_1)\bar{c}^B(x_2)] \nonumber \\
& + & \{ A \leftrightarrow B \}
\end{eqnarray}
where $\theta_C$ is the Cabibbo mixing angle and $M_W$ and $\Gamma_W$
are the mass and total width of the $W$ boson. The results for $V=W^+$
could be obtained by charge conjugation operation on the
parton densities.

\subsection{Semi-Inclusive Deep Inelastic Scattering}

\begin{figure}[htb]
\centering
\subfigure[]
{\includegraphics[width=0.40\textwidth]{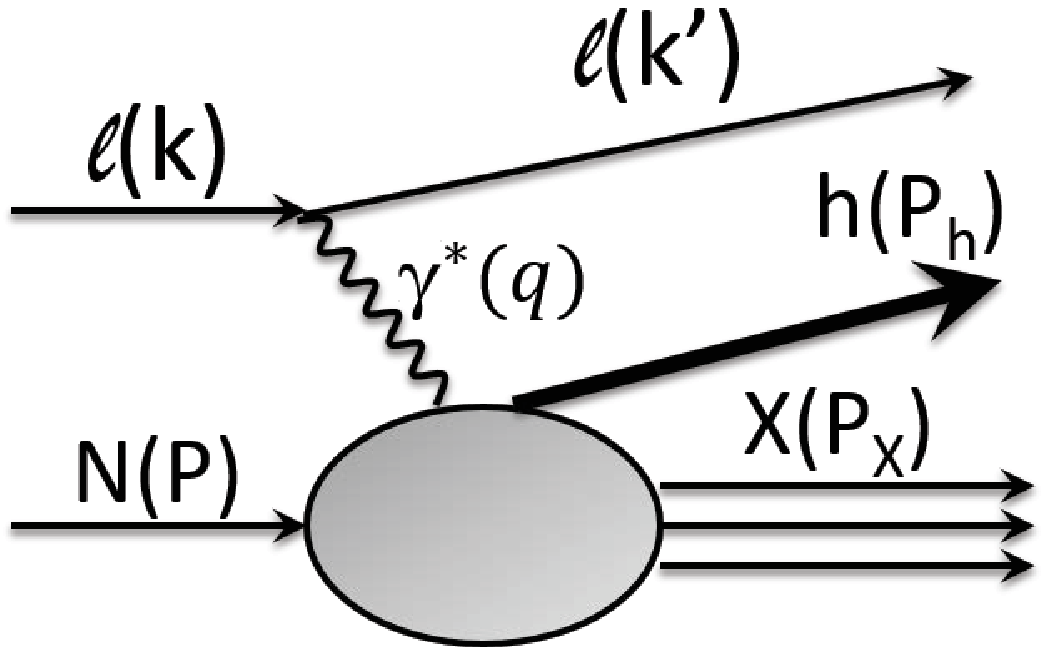}}
\subfigure[]
{\includegraphics[width=0.40\textwidth]{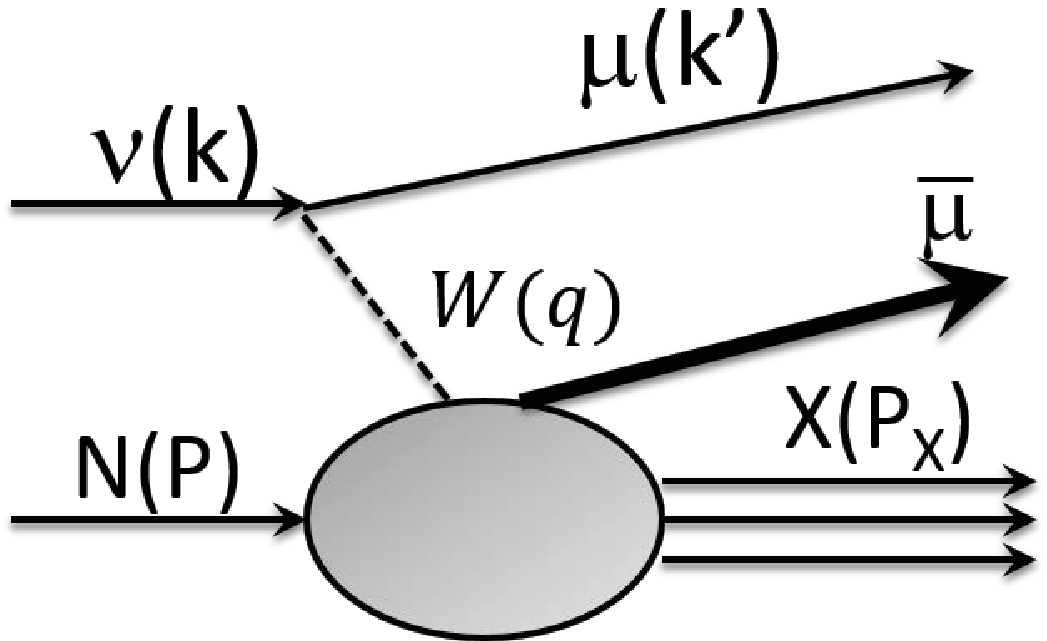}}
\caption{Semi-inclusive deep inelastic scattering via (a) $\gamma^*$
  (b) $W$ exchange.}
\label{fig:SIDIS}
\end{figure}
\noindent

When certain hadrons or leptons produced in the DIS are detected
together with the scattered lepton, the process is then called
``semi-inclusive DIS'' (SIDIS). As illustrated in
Fig.~\ref{fig:SIDIS}(a) for the case of SIDIS production of hadron
$h$, the leading order cross section is proportional to the
combination of nucleon parton density $q(x,Q^2)$ and the fragmentation
function $D_q^h(z,Q^2)$ expressed as:
\begin{equation}
\frac{d\sigma^h}{dxdydz}(x,y,z) = \frac{2 \pi \alpha^2}{y Q^2}Y_+
\sum_{q, \bar q} e_q^2 q(x,Q^2) D_q^h(z,Q^2),
\end{equation}
where $Y_+=1+(1-y)^2$ and $D_q^h(z,Q^2)$ is the probability of a quark
$q$ fragmenting into a hadron $h$ of four-momentum $P_h$ with an
energy fraction $z$ ($= P_h \cdot P / q \cdot P$). The measured hadron
multiplicities of hadron $h$ after being normalized to the number of
DIS events could be expressed as:
\begin{equation}
\frac{dN^h(x,z)}{dN^{DIS}} = \frac{\sum_{q, \bar q} e_q^2 q(x,Q^2)
  D_q^h(z,Q^2)}{\sum_{q, \bar q} e_q^2 q(x,Q^2)}.
\end{equation}

The neutrino and antineutrino scattering with $W$ exchange
could lead to the production of charmed quarks and such process
can be viewed as an example of SIDIS, illustrated in
Fig.~\ref{fig:SIDIS}(b).  It has been used as an experimental approach
to differentiate the nucleon strange quark distributions, $s(x)$ and
$\bar s(x)$. The signature for the production of charmed quarks in
neutrino- and antineutrino-nucleon scattering is the presence of 
a $\mu^+ \mu^-$ pair in the final state.

In the case of neutrino scattering, the underlying process is a
neutrino interacting with an $s$ or $d$ quark, producing a charm quark
which fragments into a charmed hadron, i.e. $\nu_\mu {\rm N}
\rightarrow \mu^{-} c {\rm X}; c \rightarrow D^+ {\rm X}$, and 
followed by the semi-leptonic decay of charmed hadron which produces
a second muon of opposite sign, $ D^+ \rightarrow \mu^+
\nu_{\mu} X$. Likewise the process with an incident antineutrino
involves interaction with an $\bar s$ or $\bar d$
antiquark, again leading to oppositely-signed muons in the final
state.

The differential cross section for dimuon production is expressed
generally as~\cite{nuDIScross}
\begin{equation}
{d^3\sigma (\nu_\mu N\rightarrow \mu^-\mu^+ X) \over dx\: dy\: dz } =
{d^2\sigma (\nu_\mu N\rightarrow \mu^- c X) \over dx\: dy} 
\: D_c^h(z) \: B_c(c\rightarrow \mu^+ X), 
\label{eq:dimuon}
\end{equation}
where $x$ is the momentum fraction of the struck quark, the function
$D_c^h(z)$ the hadronization of charmed quarks and $B_c$ the weighted
average of the semi-leptonic branching ratios of the charmed hadrons
produced in neutrino interactions.

The leading order differential cross section for an isoscalar target,
neglecting target mass effects, is given by:
\begin{eqnarray}
{d^2\sigma (\nu_\mu N\rightarrow \mu^- c X) \over dx\: dy}
& = & 
\frac{G_F^2 M_N E_\nu }{\ \pi (1+Q^2/M_W^2)^2 } \; \{ \; [x u(x,Q^2)+  
x d(x,Q^2 )]\: |V_{cd}|^2  \nonumber \\  
 &  & + \; 2x s(x,Q^2) \:
|V_{cs}|^2 \; \} \left( 1-\frac{m_c^2}{2 M_N E_\nu x }\right), 
\label{eq:nuDIS}
\end{eqnarray}
where $x u(x,Q^2 )$, $x d(x,Q^2)$ and $x s(x,Q^2)$ represent the
momentum distributions of the $u$, $d$ and $s$ quarks within the
proton (the corresponding $\bar{\nu }_\mu $ process has the quarks
replaced by their antiquark partners) respectively and $|V_{cd}|$ and $|V_{cs}|$
are the CKM matrix elements. The difference between neutrino and antineutrino induced dimuon differential cross section is expressed as:
\begin{eqnarray}
{d^2\sigma (\nu_\mu N\rightarrow \mu^- c X) \over dx\: dy} - {d^2\sigma ({\bar \nu}_\mu N\rightarrow \mu^+ \bar c X) \over dx\: dy}
& = & 
\frac{G_F^2 M_N E_\nu }{\ \pi (1+Q^2/M_W^2)^2 } \; \{ \; [x u_v(x,Q^2)+  
x d_v(x,Q^2 )]\: |V_{cd}|^2  \nonumber \\  
 &  & + \; 2x (s(x,Q^2) - \bar s(x,Q^2))\:
|V_{cs}|^2 \; \} \left( 1-\frac{m_c^2}{2 M_N E_\nu x }\right), 
\label{eq:nuDIS2}
\end{eqnarray}
The contribution of valence quarks in Eq.~\ref{eq:nuDIS2} is
suppressed relative to the strange contribution because of the large
difference in the coefficients: $ |V_{cd}|^2 \sim 0.05$ and
$|V_{cs}|^2 \sim 0.9$. Therefore the difference of cross sections in
two reactions is sensitive to the strange distribution asymmetry $x
(s(x,Q^2) - \bar s(x,Q^2))$.

We conclude this section by summarizing the experimental reactions and
the associated subprocesses which are sensitive to the nucleon sea
quarks in Table~\ref{tab:process}.

\begin{table}[htb]
  \begin{center}
    \begin{tabular}{llll}
      \hline
      \hline
      \noalign{\smallskip} 
      Process & Subprocess & Partons Probed \\ 
      \noalign{\smallskip} 
      \hline
      $lN \to l X$ & $\{\gamma^*,Z\} {q, \bar q} \to {q, \bar q}$ & $q,\bar{q}$ \\
      $l^-p \to \nu X$ & $W^- \{u,c,\bar d, \bar s \}\to \{d,s, \bar u, \bar c \}$ & $u,c, \bar{d}, \bar{s}$  \\
      $l^+p \to \bar{\nu} X$ & $W^+ \{d,s, \bar u, \bar c \}\to \{u,c, \bar d, \bar s \}$ & $d,s,\bar{u},\bar{c}$  \\
      $lN \to l \{\pi,K\} X$ & $\{\gamma^*,Z\} {q, \bar q} \to \{\pi,K\}$ & $u, d ,s,\bar{u}, \bar{d}, \bar{s}$  \\
      \hline
      $\nu p \to \mu^-\,X$ & $W^- \{u,c,\bar d, \bar s \}\to \{d,s, \bar u, \bar c\}$ & $u,c, \bar{d}, \bar{s}$  \\
      $\bar{\nu} p \to \mu^+\,X$ & $W^+ \{d,s, \bar u, \bar c \}\to \{u,c, \bar d, \bar s \}$ & $d,s,\bar{u},\bar{c}$  \\
      $\nu\,N \to \mu^-\mu^+\,X$ & $W^+ s\to c$ & $s$ & \\
      $\bar{\nu}\,N \to \mu^+\mu^-\,X$ & $W^- \bar{s}\to\bar{c}$ & $\bar{s}$ \\
      \hline
      $NN \to \mu^+\mu^-\,X$ & $q\bar{q} \to \{\gamma^*,Z\}$ & $q, \bar{q}$ \\
      $NN \to W^- X$ & $q\bar{q'} \to W^-$ & $d,s,\bar{u},\bar{c}$ \\
      $NN \to W^+ X$ & $q\bar{q'} \to W^+$ & $u,c,\bar{d},\bar{s}$ \\
      \hline
      \hline
    \end{tabular}
  \end{center}
  \caption{The experimental reactions and the subprocesses sensitive
    to the parton density of nucleon sea quarks.}
  \label{tab:process}
\end{table}
%


\section{Unpolarized Distributions of Sea Quarks}
\label{sec:unpolsea}

A simple picture of the quark sea from the perturbative
quark-antiquark pair production by gluons would lead to the following
expectations:

\begin{itemize}

\item The sea is composed of equal amount of $\bar u$ and $\bar d$
  because gluon is flavor-blind and $u$ and $d$ quarks roughly have
  equal mass.

\item Assuming SU(3) symmetry, the $x$ distribution of $\bar u$, $\bar
  d$, $\bar s$ (or $s$) will be identical.

\item The sea quarks would be distributed mainly at the small-$x$
  region where gluons are abundant.

\end{itemize}

\begin{figure}[H]
\begin{minipage}{0.33\textwidth}
\includegraphics[width=\textwidth]{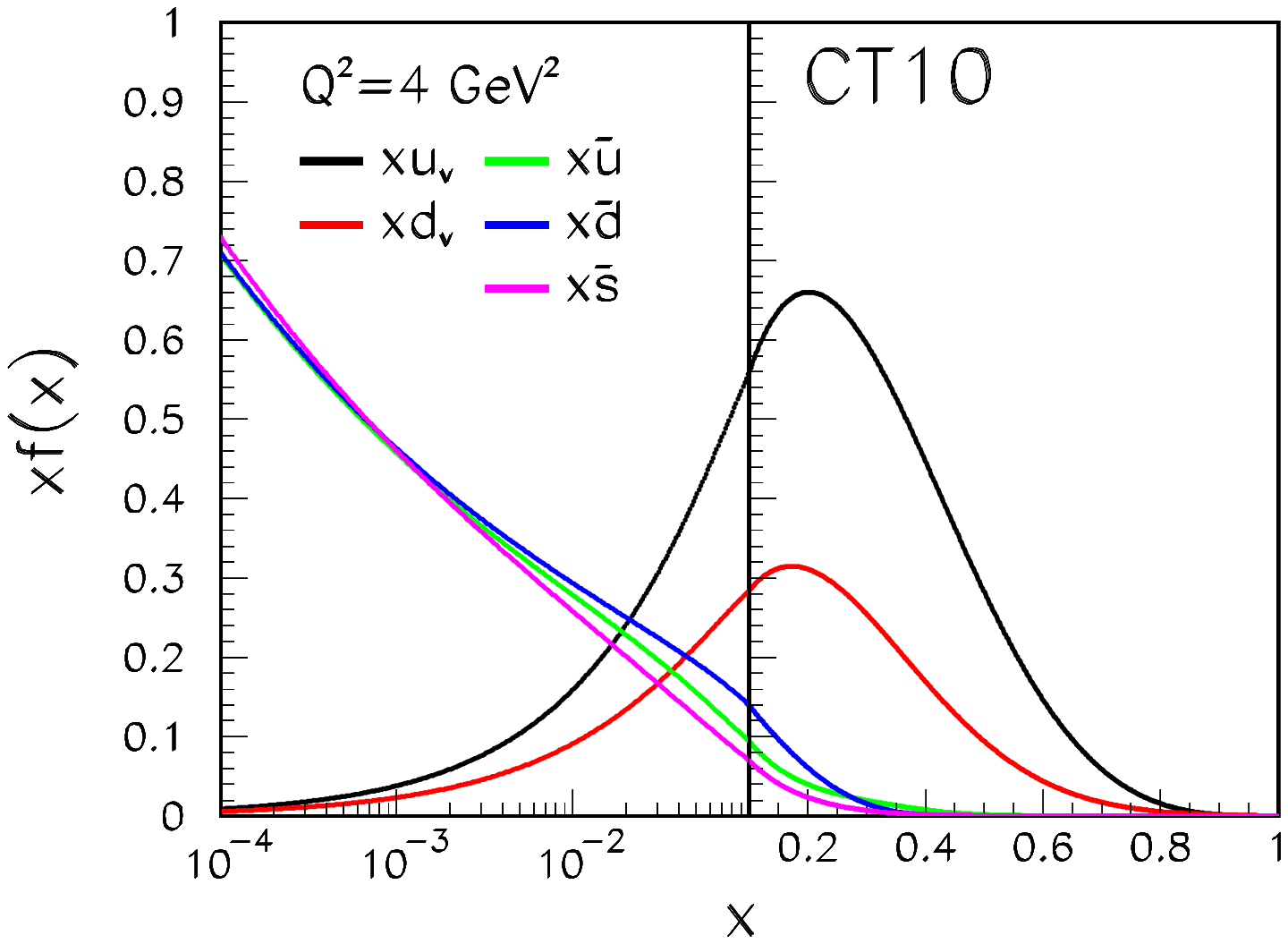}
\end{minipage}
\begin{minipage}{0.33\textwidth}
\includegraphics[width=\textwidth]{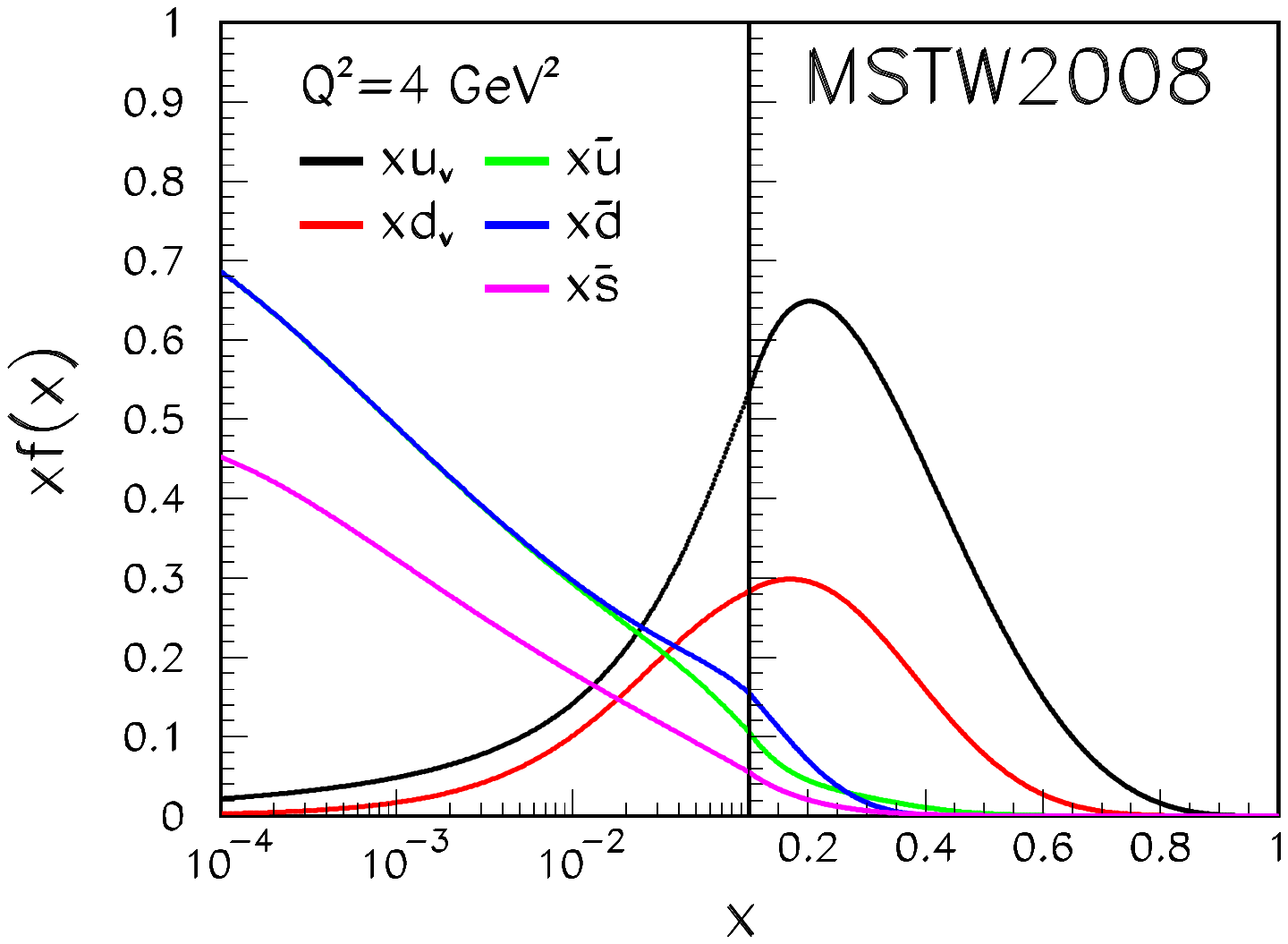}
\end{minipage}
\begin{minipage}{0.33\textwidth}
\includegraphics[width=\textwidth]{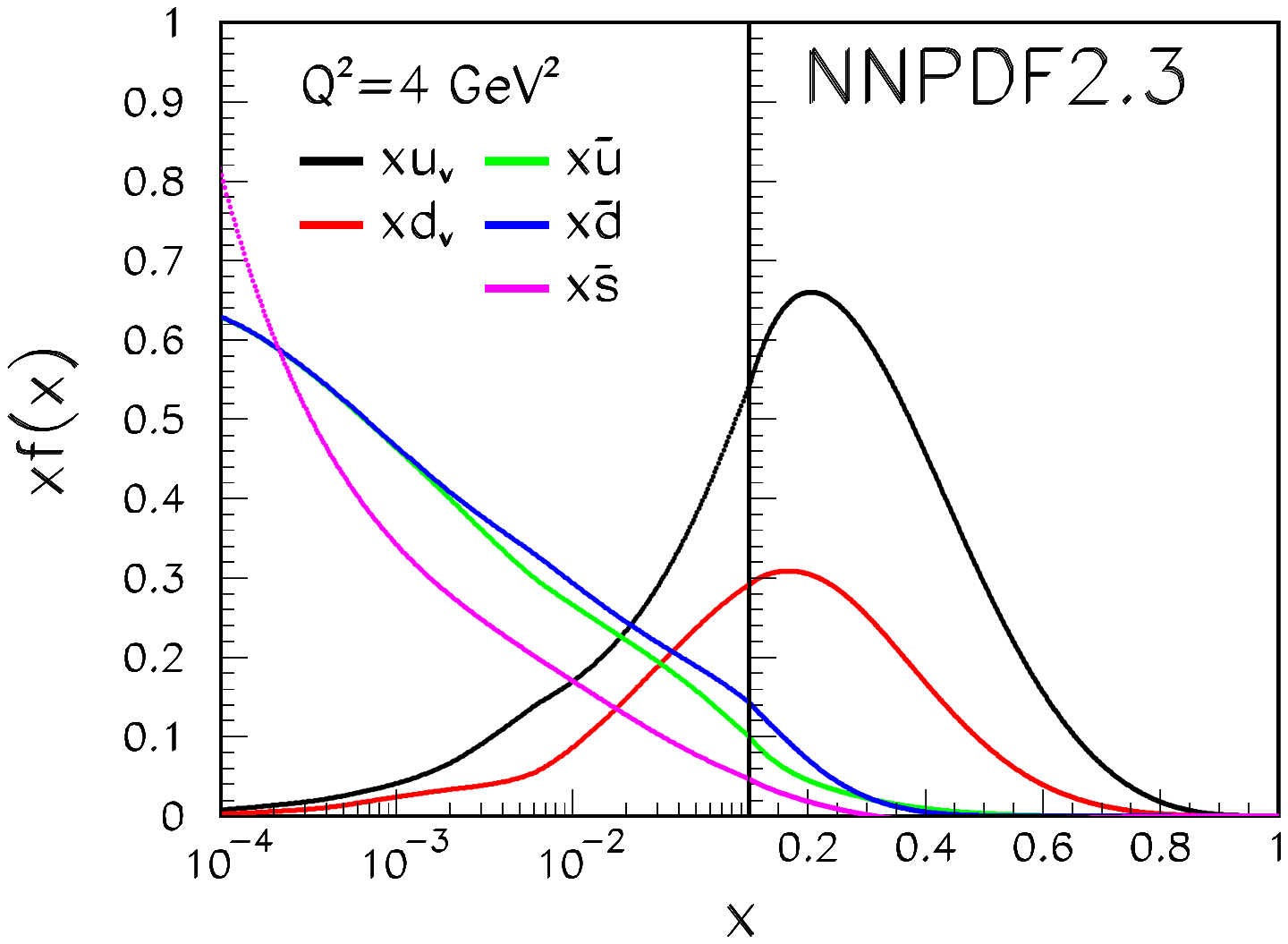}
\end{minipage}
\caption{Unpolarized parton distribution $xf(x,Q^2)$ of valence quarks
  $u_v, d_v$ and sea quarks $\bar u, \bar d, \bar s$ from
  CT10~\cite{CT10}, MSTW2008~\cite{MSTW08} and
  NNPDF2.3~\cite{NNPDF2.3} PDFs at $Q^2 = 4$ GeV$^2$.}
\label{fig:PDF}
\end{figure}

Figure~\ref{fig:PDF} shows the $x$ distributions of unpolarized parton
density for valence quarks $u_v, d_v$ and sea quarks $\bar u, \bar d,
\bar s$ from next-to-next-to-leading order (NNLO) global analysis,
CT10~\cite{CT10}, MSTW2008~\cite{MSTW08} and NNPDF2.3~\cite{NNPDF2.3},
at $Q^2$ = 4 GeV$^2$. Obviously the naive expectations of SU(2) and
SU(3) flavor symmetry of sea quarks are not observed in the
experimental data and there exists interesting flavor structure. In
this Section, we review the recent experimental and theoretical
progress in understanding this flavor structure of the unpolarized nucleon
sea of $u, d, s$ and the valence-like charm sea.

\subsection{Breaking of SU(2) Flavor Symmetry of Light Quarks Sea}

The first hint of flavor symmetry breaking of light quark sea came
from the measurements of the Gottfried sum ($S_G$) 
in the DIS experiments. The Gottfried sum is defined as
\begin{eqnarray}
S_G & = & \int_0^1 \left[F^p_2 (x) - F^n_2 (x)\right]{dx \over x} \nonumber \\
& = & \int_0^1 \sum_i e_i^2 \left[q_i^p(x)+\bar{q}_i^p(x)-q_i^n(x)-\bar{q}_i^n(x)\right]dx \nonumber \\
& = &{1\over 3}\int_0^1 \left[u_v^p(x) - d_v^p(x)\right]dx +{2\over 3}\int_0^1 \left[\bar u^p(x)-\bar d^p(x)\right]dx \nonumber \\
& = &{1\over 3}+{2\over 3}\int_0^1 \left[\bar u^p(x)-\bar d^p(x)\right]dx,
\label{eq_GS}
\end{eqnarray}
where $x$ is the Bjorken variable, $F^p_2$ and $F^n_2$ are the proton
and neutron structure functions. Eq.~(\ref{eq_GS}) is derived assuming
charge symmetry at the partonic level, namely, $u^p(x)=d^n(x),~d^p(x)=
u^n(x),~ \bar u^p(x) = \bar d^n(x),$ and $ \bar d^p(x) = \bar
u^n(x)$. If the nucleon sea is $\bar u, \bar d$ flavor symmetric, the
Gottfried Sum Rule (GSR), $S_G = 1/3$, is obtained.

The New Muon Collaboration (NMC)~\cite{nmc91, nmc94} determined the
Gottfried sum to be $ 0.235\pm 0.026$ at $Q^2 = 4$ GeV$^2$, which is
significantly below 1/3. This cast doubt on the validity of assuming
an symmetric $\bar u$, $\bar d$ sea of the proton. Even though the
violation of the GSR could also be caused by unusual behavior of the
parton distributions at unmeasured small-$x$ region, as well as by the
violation of the charge symmetry at the partonic level, this
surprising result is usually interpreted as an evidence of flavor
asymmetry of the nucleon light sea quarks~\cite{Preparata1991}.

\begin{table}[H]
\begin{center}
\begin{tabular}{lllll}\hline
\noalign{\smallskip} 
 & CT10 & MSTW2008 & NNPDF2.3 & Experiment  \\
\noalign{\smallskip}
\hline
\hline
\noalign{\smallskip}

$S_G$ (4 GeV$^2$) & $0.2443(7)$ & $0.2807(4)$ & $0.2394(24)$ &
$0.235(26)$ (NMC~\cite{nmc94}) \\

%




\hline
\hline
\end{tabular}
\end{center}
\caption{Values of the Gottfried sum, $S_G$, for CT10~\cite{CT10},
  MSTW2008~\cite{MSTW08} and NNPDF2.3~\cite{NNPDF2.3} PDFs at $Q^2 =
  4$ GeV$^2$ and from the NMC experiment.}
\label{tab:GSR}       
\end{table}

\begin{figure}[H]
\centering
\subfigure[]
{{\includegraphics[width=0.45\textwidth]{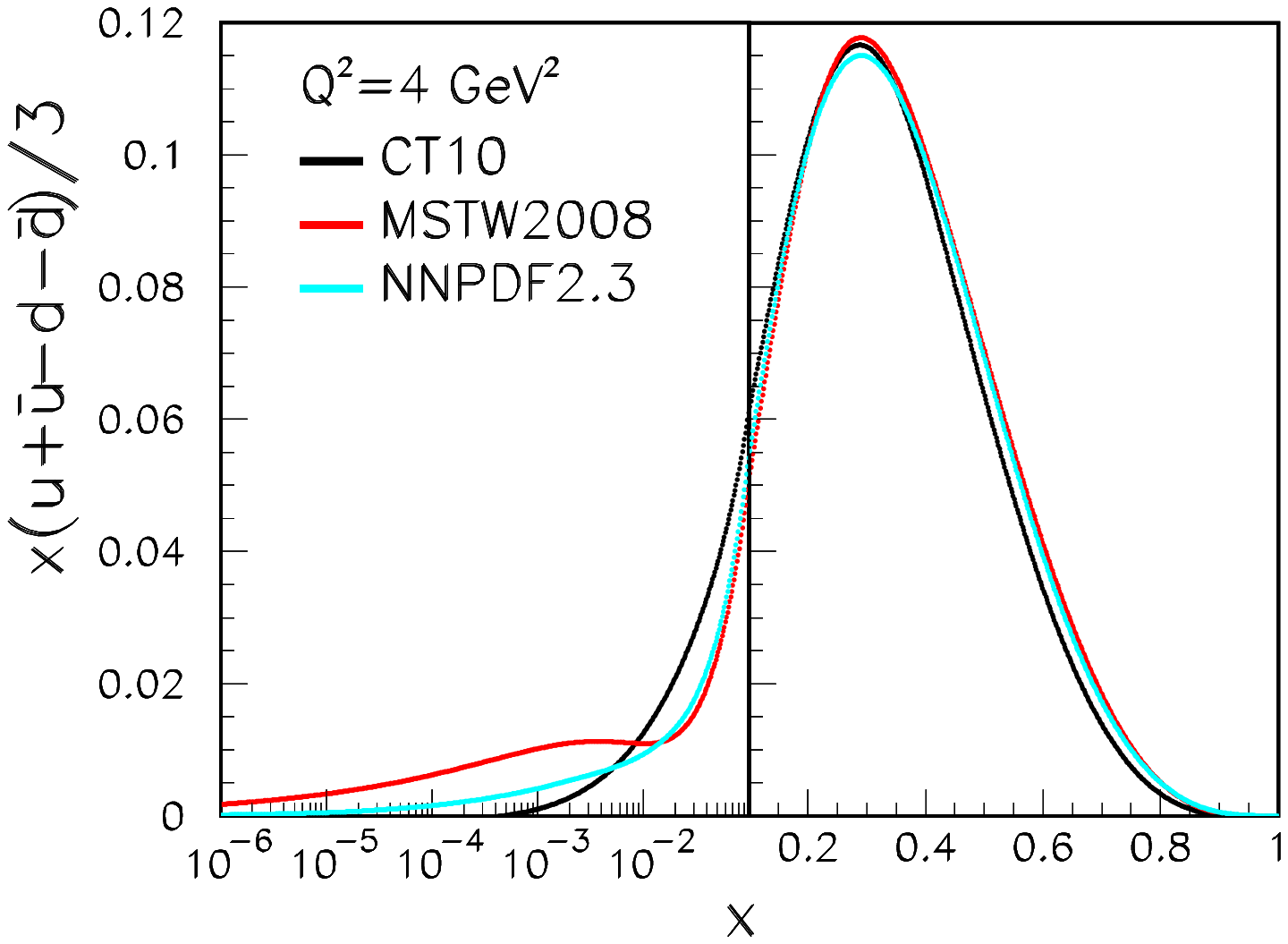}}
\label{fig:GSR1}}
\subfigure[]
{{\includegraphics[width=0.45\textwidth]{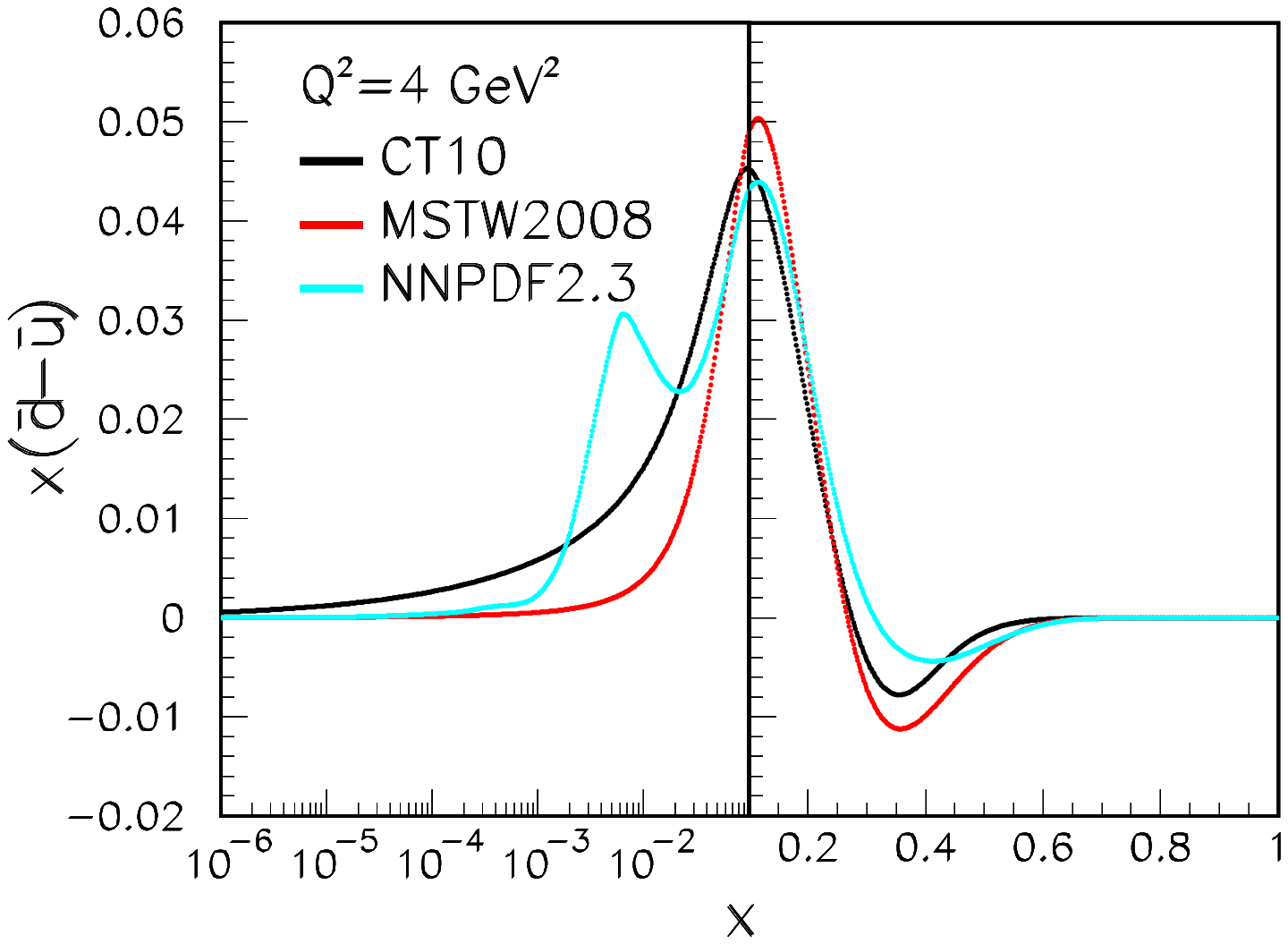}}
\label{fig:GSR2}}
\caption{The $x$ distributions of (a) $x(u + \bar u - d - \bar d)/3$ ,
  and (b) $x(\bar d - \bar u)$ from CT10~\cite{CT10},
  MSTW2008~\cite{MSTW08} and NNPDF2.3~\cite{NNPDF2.3} PDFs at $Q^2 =
  4$ GeV$^2$.}
\label{fig:GSR}
\end{figure}

In Table~\ref{tab:GSR} we evaluate the Gottfried sum, $S_G$, estimated
by the parton distribution functions (PDFs) from some recent global
analysis, CT10~\cite{CT10}, MSTW2008~\cite{MSTW08} and
NNPDF2.3~\cite{NNPDF2.3}, at $Q^2 = 4$ GeV$^2$. There are visible
differences in $S_G$ among them which could be understood from their
$x$ distributions of $x(u + \bar u - d - \bar d)/3$ and $x(\bar d -
\bar u)$ of these PDFs, shown in Figs.~\ref{fig:GSR1}
and~\ref{fig:GSR2}, respectively. Clearly the discrepancy mostly
originates from the considerable uncertainties of sea quark densities
in both small- and large-$x$ regime, and this calls for the need of
exploring the structure of sea quarks by the on-going Fermilab
E906/Seaquest Drell-Yan experiment and the proposed electron ion
colliders.

\begin{figure}[H]
\centering
\subfigure[]
{{\includegraphics[width=0.45\textwidth]{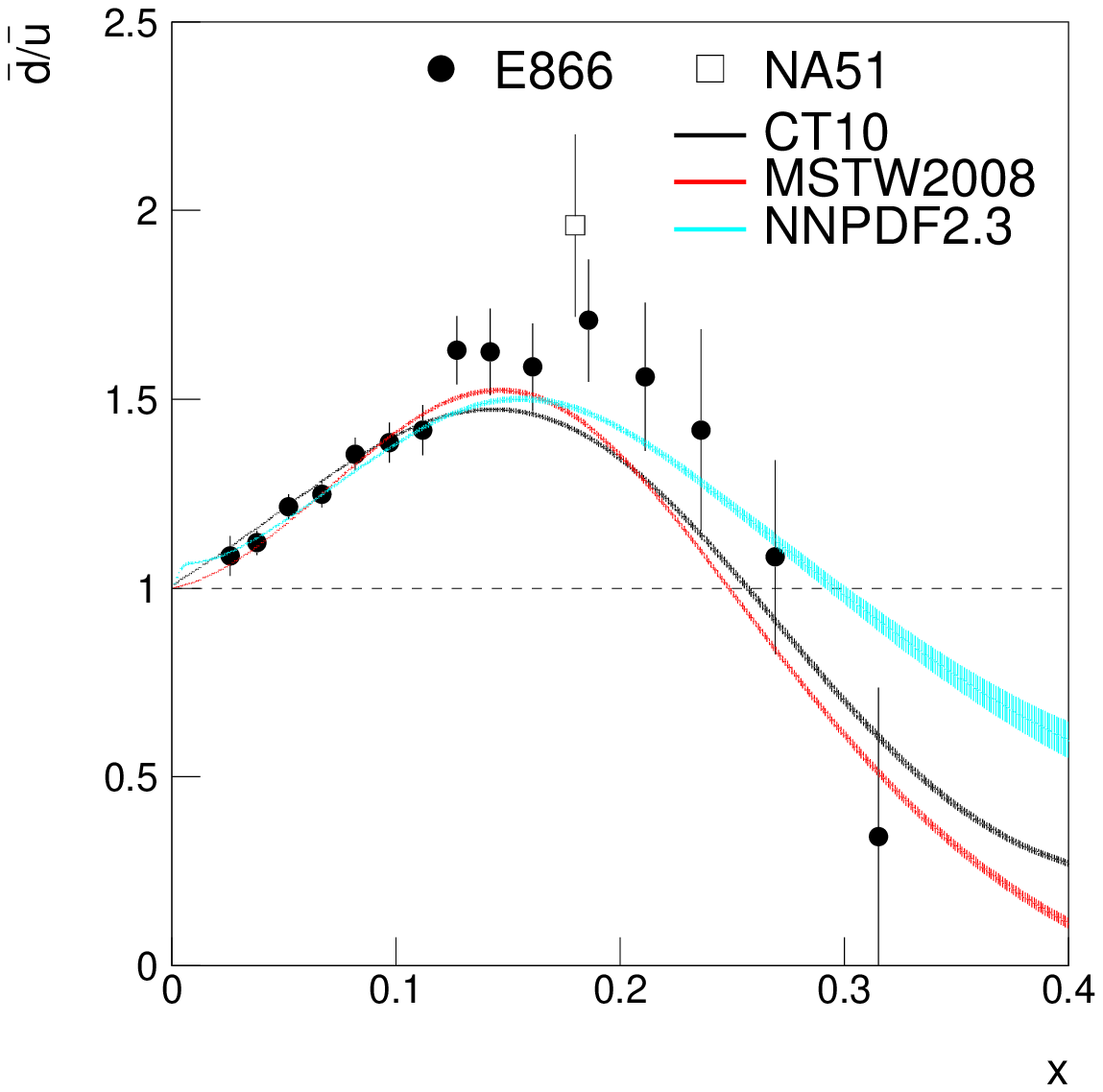}}
\label{fig:Rdbarubar}}
\subfigure[]
{{\includegraphics[width=0.45\textwidth]{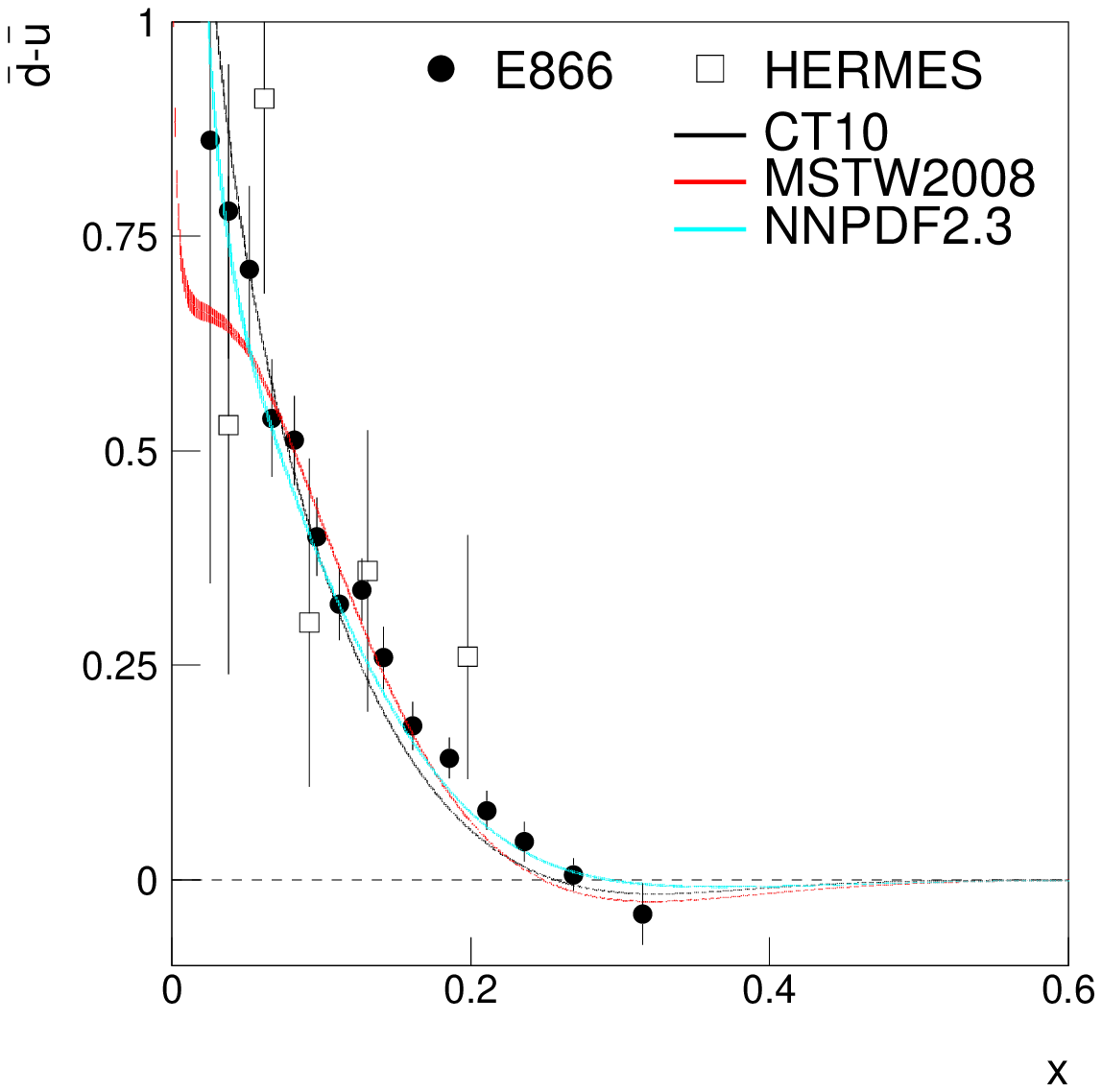}}
\label{fig:dbarubar}}
\caption{Data of (a) $\bar{d}(x)/\bar{u}(x)$ and (b) $\bar{d}(x) -
  \bar{u}(x)$ from NA51~\cite{NA51}, E866/NuSea~\cite{E866} and
  HERMES~\cite{hermes98}, compared to the results of CT10~\cite{CT10},
  MSTW2008~\cite{MSTW08} and NNPDF2.3~\cite{NNPDF2.3} PDFs at $Q^2 = 54$
  GeV$^2$.}
\label{fig:lightsea}
\end{figure}

After the DIS result from NMC, an independent and elegant test of the
flavor asymmetry of sea quark was done by the proton-induced Drell-Yan
(DY) experiments. The DY cross section at forward rapidity measured by
fixed-target experiments is dominated by the annihilation of the
$u(x_1)$ quark in the proton beam with the $\bar u(x_2)$ antiquark in
the target nucleon. For $x_1 \gg x_2$ and $d(x_1) \ll 4u(x_1)$, the
ratio of DY cross sections using deuterium and hydrogen target could
be simplified as,
\begin{equation}
{\sigma^{pd} \over 2\sigma^{pp}} \approx {1 \over 2}{[1+{1 \over
      4}{d(x_1) \over u(x_1)}] \over [1+{1 \over 4}{d(x_1) \over
      u(x_1)}{\bar d(x_2) \over \bar u(x_2)}]}[1+{\bar d(x_2) \over
    \bar u(x_2)}] \approx {1 \over 2}[1+{\bar d(x_2) \over \bar
    u(x_2)}].
\end{equation}
Therefore, the $x$ dependence of $\bar d/\bar u$ could be determined
by the DY process. The NA51 experiment at CERN gave the first result
of $\bar d/\bar u$ = $0.51 \pm 0.04 \pm 0.05$ at $\langle x
\rangle$=0.18 and $\langle M_{\mu \mu} \rangle$=5.22 GeV using 450 GeV
proton beam~\cite{NA51}. The Fermilab E866/NuSea experiment later
performed the measurement over a broad range of $x$ using 800 GeV
proton beam~\cite{E866}. The extracted $\bar d/\bar u(x)$ ratios are
shown in Fig.~\ref{fig:Rdbarubar} and they increase linearly from 1
at $x=0$ up to $x \sim 0.15$, reaching a maximum of 1.75 and then
drops off at higher $x$. The $\bar d/\bar u$ ratio falls below unity
at the largest $x$. Figure~\ref{fig:lightsea} compares $\bar d(x)/\bar
u(x)$ and $\bar d(x) - \bar u(x)$ with the results of recent PDFs:
CT10~\cite{CT10}, MSTW2008~\cite{MSTW08} and NNPDF2.3~\cite{NNPDF2.3}.

The other results on the $\bar d / \bar u$ flavor asymmetry came from HERMES
collaboration~\cite{hermes98} where charged pions produced from SIDIS
process off hydrogen and deuterium targets in the kinematic regions of
$0.02 < x < 0.3$ and $1 < Q^2 < 10$ GeV$^2$/c$^2$ were measured. The value of
$\bar d(x) -\bar u(x)$ was deduced from the measured yields as
follows:
\begin{equation}
{\bar d(x) - \bar u(x) \over u(x) - d(x)}={J(z) [1-r(x,z)]-[1+r(x,z)]
  \over J(z) [1-r(x,z)]+[1+r(x,z)]}
\end{equation}
where
\begin{equation}
r(x,z)={d\sigma^{\pi^-}_p/dz - d\sigma^{\pi^-}_n/dz \over
  d\sigma^{\pi^+}_p/dz - d\sigma^{\pi^+}_n/dz}, \mbox{ and } J(z)={3
  \over 5} \left( {1+D^{\pi^-}_u (z)/D^{\pi^+}_u (z)\over
  1-D^{\pi^-}_u (z)/D^{\pi^+}_u (z)} \right),
\end{equation}
and $D^{\pi^\pm}_u(z)$ are the corresponding favored and disfavored
pion fragmentation functions. The ratio $r(x,z)$ was determined from
the hydrogen and deuterium data. HERMES's results of $\bar d(x) - \bar
u(x)$ are consistent with E866's ones at larger $Q^2$ as shown in
Fig.~\ref{fig:dbarubar}.

Table~\ref{tab:dbarubar} summarizes three experimental determinations
of the integral $\int_0^1 (\bar d(x)-\bar u(x))dx$ at different energy
scales and the corresponding evaluation using various PDFs. This
integral calculated with three different PDFs shows very small $Q^2$
dependence. The fact that the MSTW2008 integral is much smaller than
those of the other two PDFs could be understood from the structure of
$\bar d(x)-\bar u(x)$ displayed in Fig.~\ref{fig:GSR2}. The E866
integral is smaller than those from NMC and HERMES, but consistent
with them within the quoted errors.

\begin{table}[H]
\begin{center}
\begin{tabular}{lllll}\hline
\noalign{\smallskip} 
$\langle Q^2 \rangle$ & CT10 & MSTW2008 & NNPDF2.3 & Experiment  \\
\noalign{\smallskip}
\hline
\hline
\noalign{\smallskip}

2.5 GeV$^2$ & $0.129(1)$ & $0.079(1)$ & $0.137(3)$ & $0.16(3)$
(HERMES~\cite{hermes98})\\\

4 GeV$^2$ & $0.129(1)$ & $0.079(1)$ & $0.138(3)$ & $0.148(39)$
(NMC~\cite{nmc94}) \\

54 GeV$^2$ & $0.130(1)$ & $0.080(1)$ & $0.139(3)$ & $0.118(12)$
(E866~\cite{E866}) \\

\hline
\hline
\end{tabular}
\end{center}
\caption{Energy scale ($\langle Q^2 \rangle$) and the corresponding
  estimation of $\int_0^1 (\bar d(x)-\bar u(x))dx$ using
  CT10~\cite{CT10}, MSTW2008~\cite{MSTW08} and
  NNPDF2.3~\cite{NNPDF2.3} PDFs together with the results determined
  by three experiments.}
\label{tab:dbarubar}       
\end{table}
%


Many theoretical attempts have been made to understand the origin of
the nucleon sea and the flavor asymmetry. In 1977 Field and
Feynman~\cite{Field:1976} pointed out that the $\bar d = \bar u$ would
not strictly hold in the perturbative QCD because the additional
valence $u$ quark in proton could lead to a larger suppression of $g
\rightarrow u \bar u$ via Pauli blocking. This provided a simple and
qualitative interpretation of the observed flavor asymmetry of $\bar
d$ and $\bar u$ in the nucleons. However later NLO perturbative QCD
calculations confirmed~\cite{ross79} that the Pauli blocking effect in
the gluon splitting is too small to account for the sizable violation
of GSR. Hence it is generally believed that the observed large
difference between $\bar d(x)$ and $\bar u(x)$ suggests a
non-perturbative origin.

The observed $\bar d/\bar u$ asymmetry could be reasonably explained
by meson-cloud model, chiral quark model, chiral quark soliton model,
and instanton model. Details of these models have been reviewed in
several articles~\cite{speth98,kumano98,Vogt00,garvey02}. For example,
the meson-cloud model~\cite{thomas,tony} treats the proton as a linear
combination of a bare proton plus pion-nucleon and pion-delta states:

\begin{eqnarray}
|p\rangle & = & \sqrt{Z}~|p_0\rangle + a_{N\pi/p}~\Large{[}-\sqrt{\frac{1}{3}}
|p_0\pi^0\rangle + \sqrt{\frac{2}{3}} |n_0 \pi^+\rangle\Large{]} \nonumber \\
& + & a_{\Delta\pi/p} ~\Large{[}\sqrt{\frac{1}{2}}|\Delta^{++}_0
\pi^-\rangle - \sqrt{\frac{1}{3}}
|\Delta^+_0\pi^0\rangle + \sqrt{\frac{1}{6}}|\Delta^0_0\pi^+\rangle\Large{]} \nonumber \\
& + & a_{\Lambda K/p}~|\Lambda_0 K^+\rangle + a_{\Sigma K/p}~\Large{[}-\sqrt{\frac{1}{2}}
|\Sigma^{+}_0 K^0\rangle + \sqrt{\frac{1}{2}} |\Sigma^{0}_0 K^+\rangle\Large{]} + ...
\label{eq:meson}
\end{eqnarray}

\noindent where the subscript zeros denote bare baryons with flavor
symmetric seas and $Z$ is the normalization constant and the parameter
$a_{BM/p}$ reflects the relative strength of proton splitting into the
virtual baryon-meson (BM) state. The $x$ distributions of sea quarks
$\bar u$, $\bar d$, $s$ and $\bar s$ could be obtained by convoluting
their distributions in either meson or baryon with the splitting
functions. The $\bar u$ and $\bar d$ seas receive contributions from
the valence antiquarks of the pion cloud. The excess of $\bar d$ over
$\bar u$ arises because of the dominance of the $n_0 \pi^+$
configuration over the less probable $\Delta_0^{++}\pi^-$
configuration. This leads to an overall excess of $\bar d$ over $\bar
u$. Similarly the different distributions of $\bar s(x)$ in the
fluctuating kaon, and $s(x)$ in the $\Lambda$ and $\Sigma$ hyperons
could introduce possible $s(x) / \bar s(x)$ asymmetries in the proton.

In the chiral quark model~\cite{szczurek96} the relevant degrees of
freedom are constituent quarks and Goldstone bosons ($\pi, K,
\eta$). The Goldstone bosons directly couple to the constituent quarks
of the proton as a consequence of the spontaneously broken chiral
symmetry at low energies. The Fock decomposition of the constituent
quarks $|U\rangle$ and $|D\rangle$ in the proton could be represented
as follows:

\begin{eqnarray}
|U\rangle & = & \sqrt{Z} |u \rangle + \sqrt{\frac{1}{3}}a_{\pi/U} | u \pi^0 \rangle +  \sqrt{\frac{2}{3}}a_{\pi/U}| d \pi^+ \rangle + a_{K/U}| s K^+ \rangle + ... \nonumber \\
|D\rangle & = & \sqrt{Z} |d \rangle + \sqrt{\frac{1}{3}}a_{\pi/D} | d \pi^0 \rangle +  \sqrt{\frac{2}{3}}a_{\pi/D}| u \pi^- \rangle + a_{K/D}| s K^0 \rangle + ... ,
\label{eq:chiral_quark}
\end{eqnarray}
where small letters denote bare constituent quarks. The $Z$
and $a_{M/q}$ are the normalization constant and parameter of
Goldstone bosons ($M$) coupling with the bare constituent quark
states. The descriptions of the asymmetries of $\bar d(x)/\bar u(x)$
and $s(x)/\bar s(x)$ could be derived in similar manners as what is
done in the meson-cloud model.

In the past decade, there are updated results from meson-cloud
model~\cite{alwall05}, chiral quark
model~\cite{ding05PLB,ding05PRD,song11}, and chiral quark soliton
model~\cite{wakamatsu03,wakamatsu05,wakamatsu09}. In
Ref.~\cite{alwall05}, Gaussian forms for the momentum distributions
and hadronic fluctuations of baryons into baryon-meson pairs were used
in the framework of meson-cloud model and proper QCD evolution is
taken into account. The asymmetry of $\bar u(x)$ and $\bar d(x)$ found
in the Drell-Yan ratio of $pp$ and $pd$ scattering can be properly
described. The chiral soliton model has been generalized into the
flavor SU(3) version~\cite{wakamatsu03} and the SU(3) breaking effect
due to the mass difference $\delta m_s$ between the $s$ quark and the
$u,d$ quarks is estimated using first order perturbation theory in the
parameter $\delta m_s$. The magnitude of the asymmetry $\bar d(x)-\bar
u(x)$ remains the same after this SU(3) generalization.

There are two new attempts to understand the sea flavor asymmetry
putting emphasis on the statistical properties of partons bound inside
the nucleon: statistical model and balance model. Bourrely et
al.~\cite{bourrely01,bourrely03,bourrely05,bourrely07,zhang09}
developed a new form of statistical parametrization, allowing an
$x$-dependent chemical potential. By incorporating QCD evolution they
can describe a variety of data. Zhang et
al.~\cite{zhang01,zou02,zhang02,alberg08,zhang10} constructed a model
using the principle of detailed balance without any free parameter,
and they obtained Gottfried sum in surprisingly good agreement with
the experimental values.

\begin{figure}[H]
  \subfigure[Meson cloud model. Figure from ~\cite{alwall05}]
  {\includegraphics[width=0.48\textwidth]{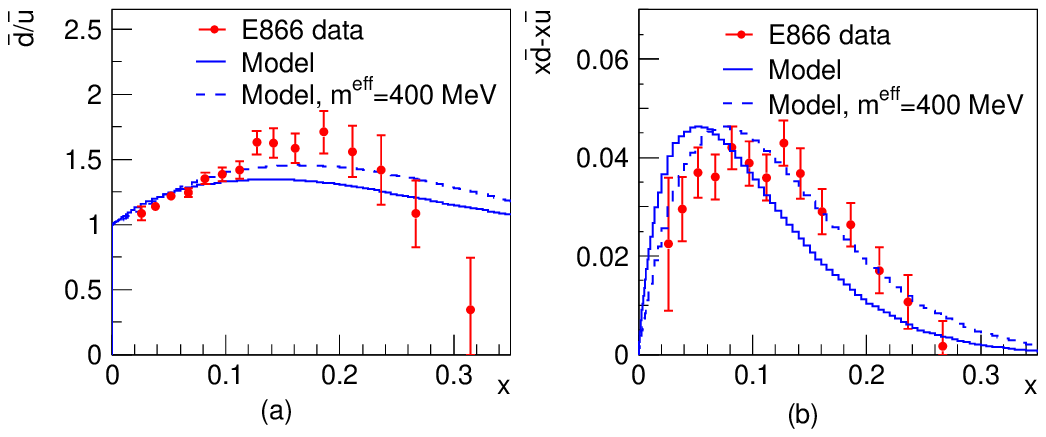}
    \label{fig:dbarubar_MBM}}
  \hspace{-0.5cm}
  \subfigure[Chiral quark model. Figure from ~\cite{song11}]
  {\includegraphics[width=0.32\textwidth]{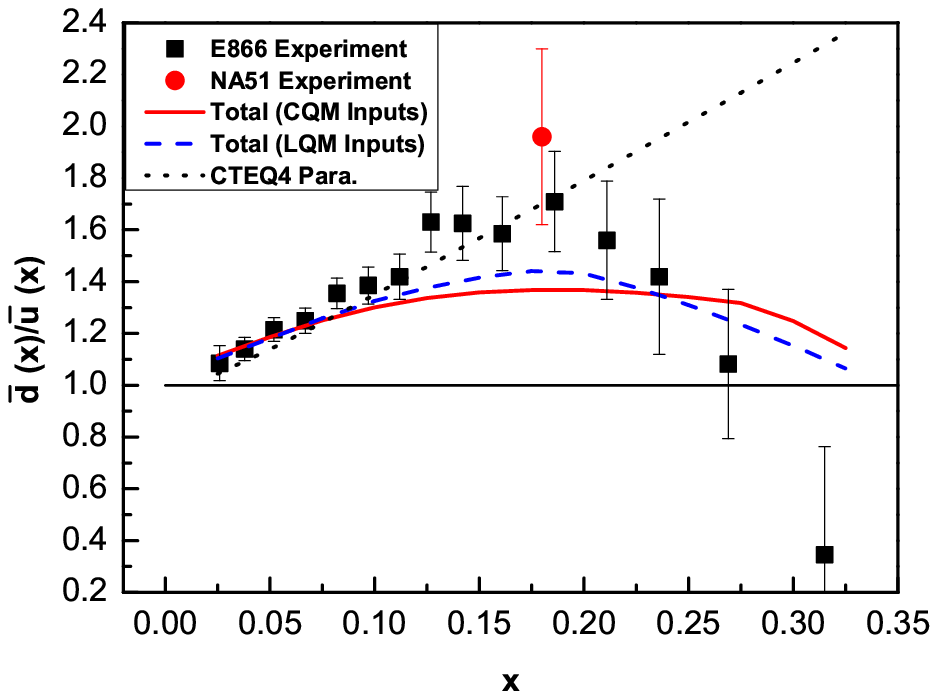}
  \hspace{-1.0cm}
   \includegraphics[width=0.32\textwidth]{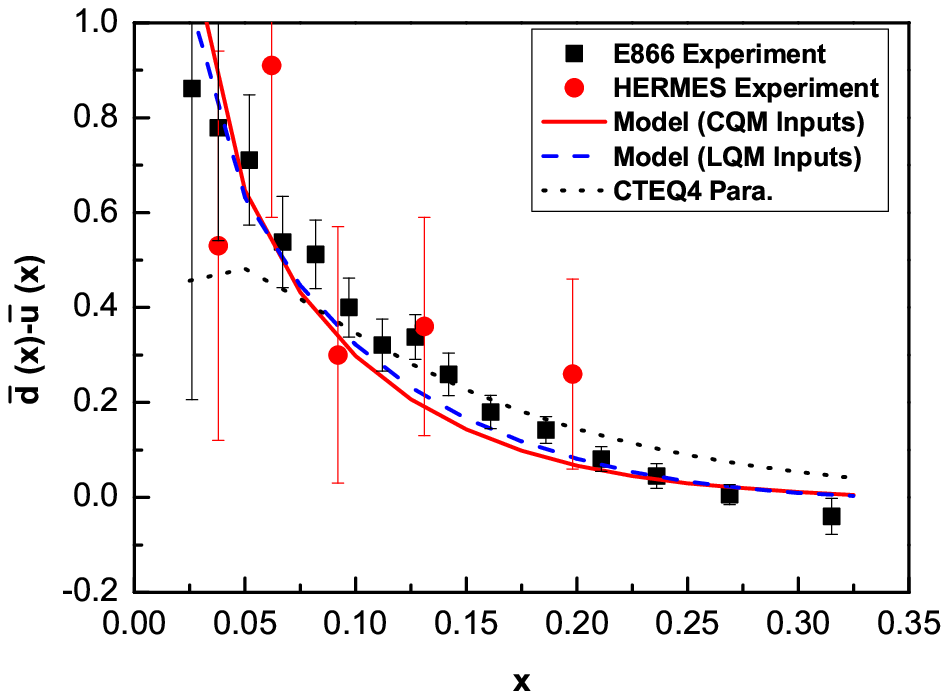}
    \label{fig:dbarubar_CQM}}
  \subfigure[Chiral quark soliton model. Figure from ~\cite{wakamatsu09}]
  {\includegraphics[width=0.50\textwidth]{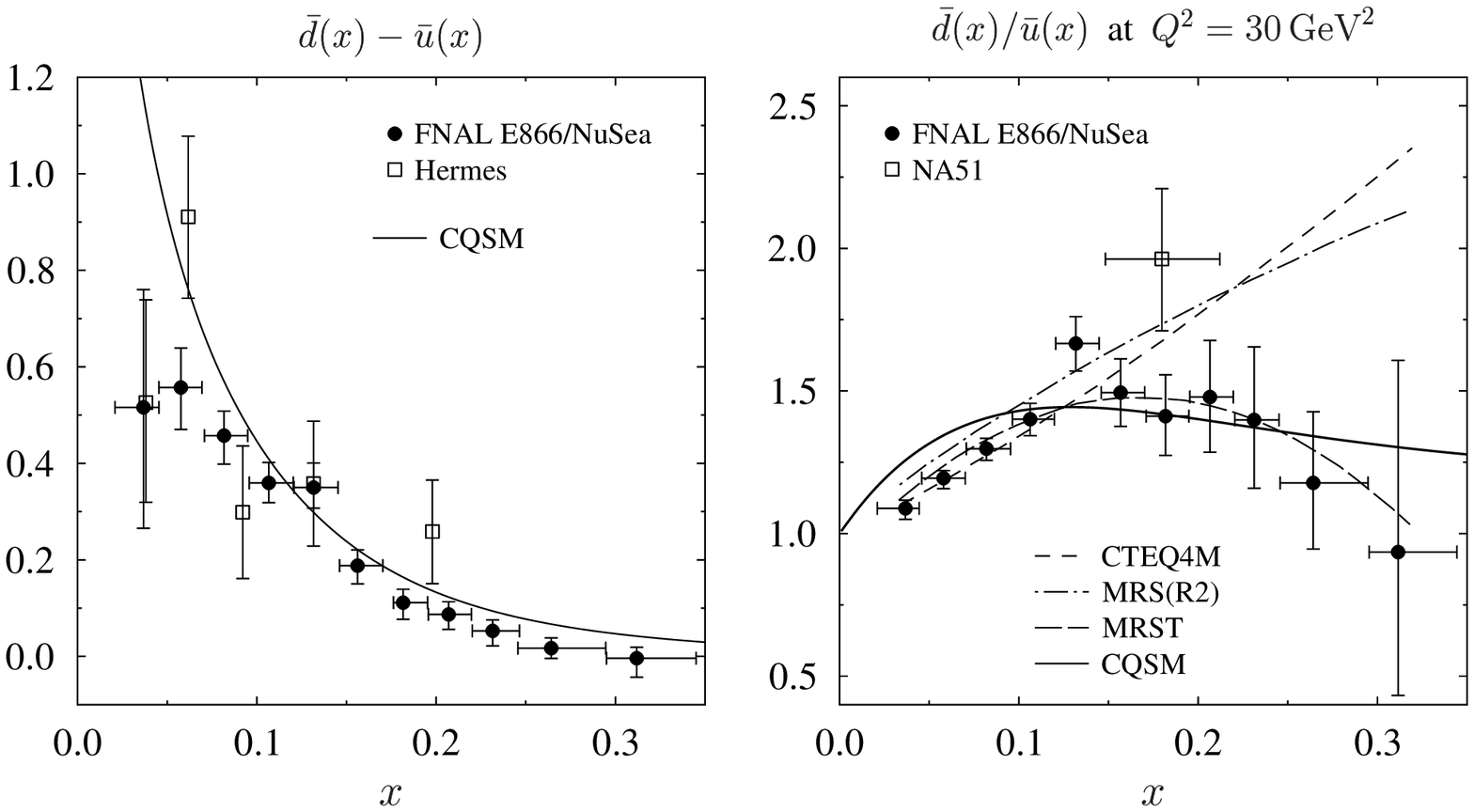}
    \label{fig:dbarubar_CQSM}}
  \hspace{-0.7cm}
  \subfigure[Statistic model. \newline
    Figure from ~\cite{bourrely01}]
  {\includegraphics[width=0.22\textwidth]{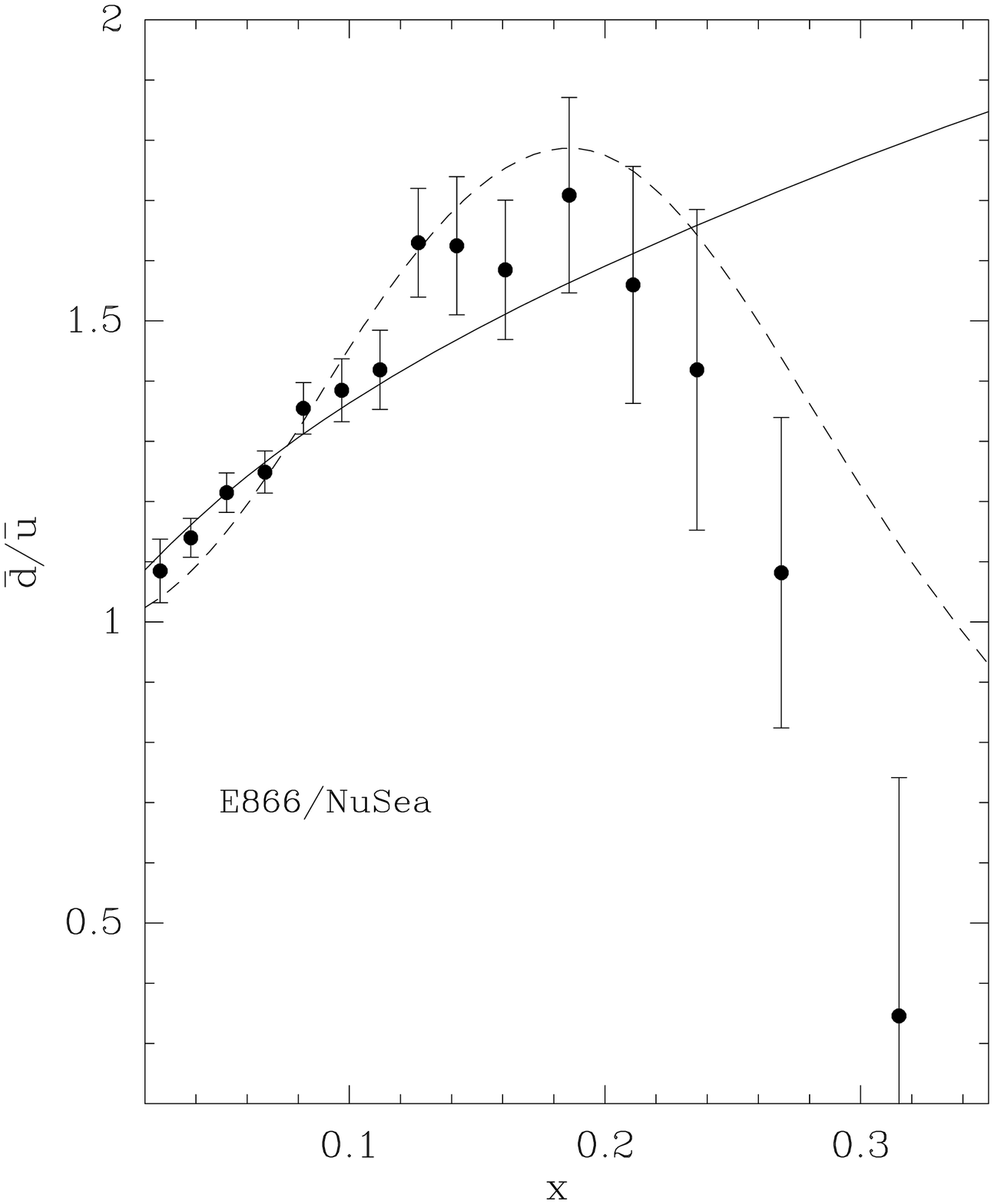}
    \label{fig:dbarubar_stat}}
  \subfigure[Balance model. \newline
    Figure from ~\cite{zou02}]
  {\includegraphics[width=0.28\textwidth]{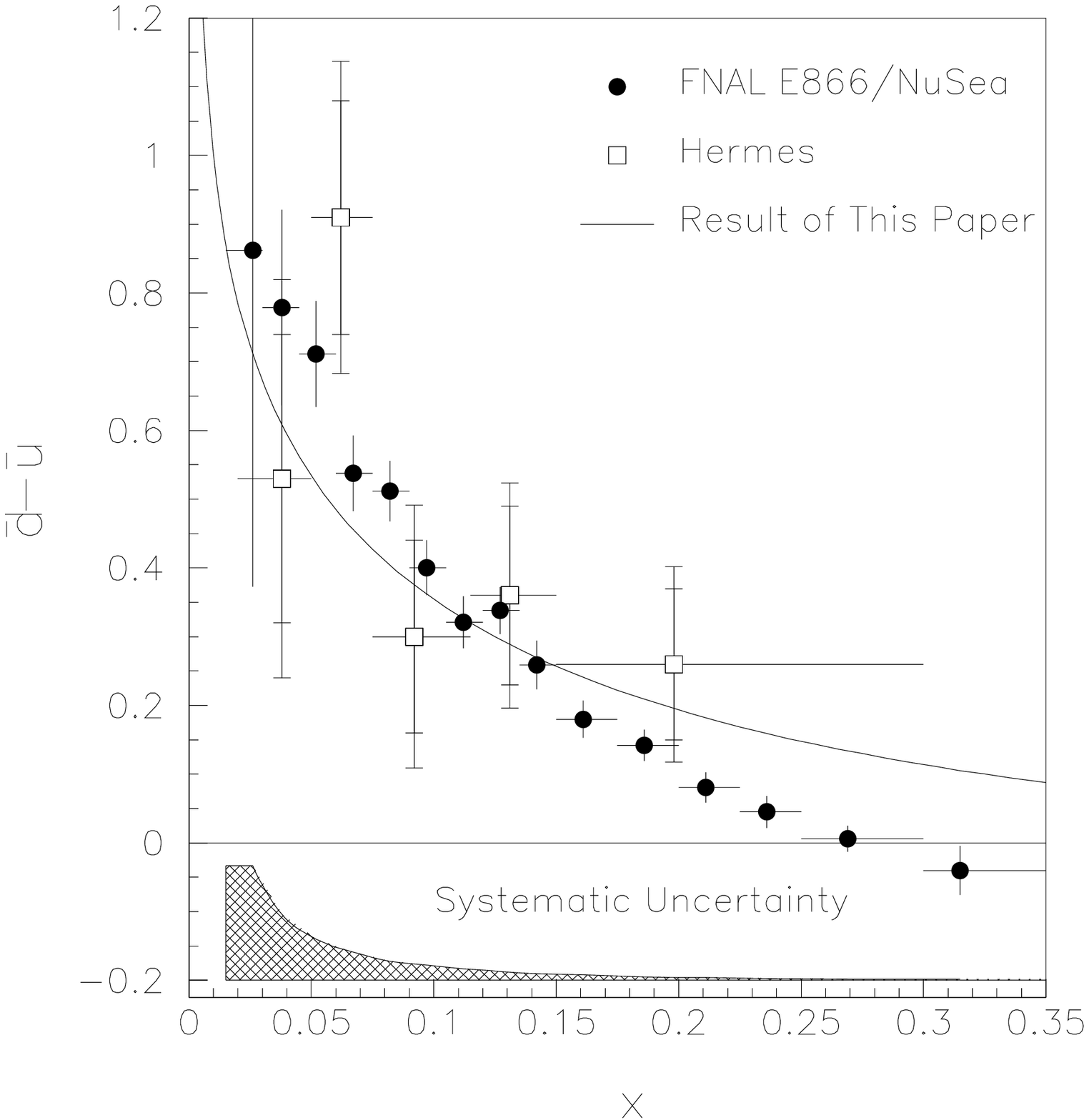}
    \label{fig:dbarubar_balance}}
\caption{Comparison of the data of $\bar{d}(x)/\bar{u}(x)$ and
  $\bar{d}(x) - \bar{u}(x)$ from NA51~\cite{NA51},
  E866/NuSea~\cite{E866} and HERMES~\cite{hermes98} with the
  predictions of theoretical models.}
\label{fig:dbarubar_allmodel}
\end{figure}

Figure~\ref{fig:dbarubar_allmodel} shows various theoretical attempts in
describing the data of $\bar d(x)/\bar u(x)$ and $\bar d(x) - \bar
u(x)$. In common, all the models could describe the general feature of
$\bar d > \bar u$ reasonably well but none of them could accommodate
the tentative behavior of $\bar d(x)/\bar u(x) < 1$ at large $x$.

\begin{figure}[H]
\centering
\includegraphics[width=0.6\textwidth]{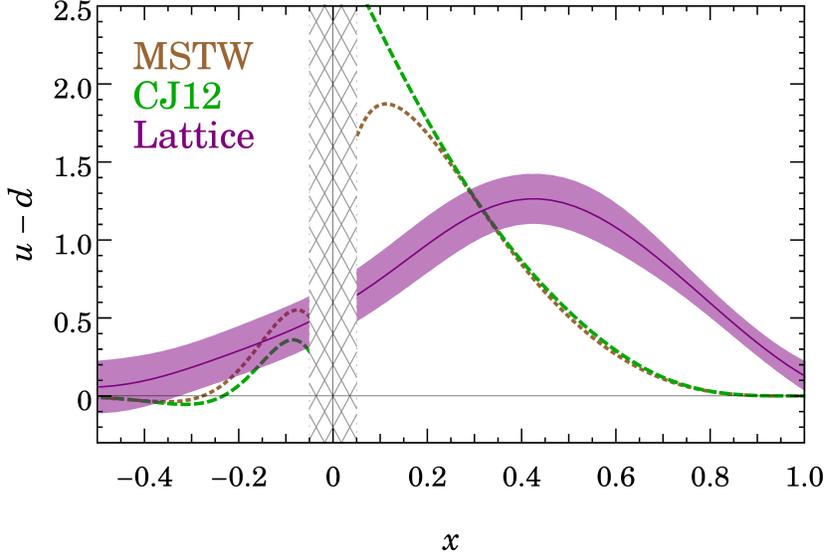}
\caption{Unpolarized isovector quark distributions computed by lattice
  QCD~\cite{lin14}(purple band), compared with the global analysis of
  MSTW2008~\cite{MSTW08}(brown dotted line) and CJ12~\cite{CJ12}(green
  dashed line) PDFs. The sea quark distribution is represented in the
  negative $x$ region as $\bar q(x) = -q(-x)$. Figure
  from~\cite{lin14}.}
\label{lattice_sea}
\end{figure}

Thomas et al.~\cite{thomas00} have derived a leading non-analytic
chiral behavior of $\bar{d} - \bar{u}$ which is a unique
characteristics of Goldstone boson loops in chiral theories. It
provided a possible theoretical connection between the flavor
asymmetry in the nucleon sea and the fundamental chiral symmetry
breaking in QCD. There exists difficulty in exploring the sea
structure from lattice QCD: only lower moments of the parton
distributions could be computed and the separation of quark and
antiquark distributions is not available. Very recently, the
qualitative feature of the nucleon sea flavor structure of $ \bar d(x)
> \bar u(x)$ (and also $\Delta \bar d(x) < \Delta \bar u(x)$) was
demonstrated directly from QCD by the lattice approach with a large
pion mass of 310 MeV~\cite{lin14}, as shown in
Fig.~\ref{lattice_sea}. By doing calculations with a large-momentum
nucleon, the light-cone quantities are connected to lattice-QCD
non-local time-independent matrix elements~\cite{ji13}. These results
are encouraging and promising for a realistic comparison with the
measured $x$ dependence of sea quark structure in the future.

\begin{figure}[H]
\begin{center}
\begin{minipage}{0.6\textwidth}
\includegraphics[width=\textwidth]{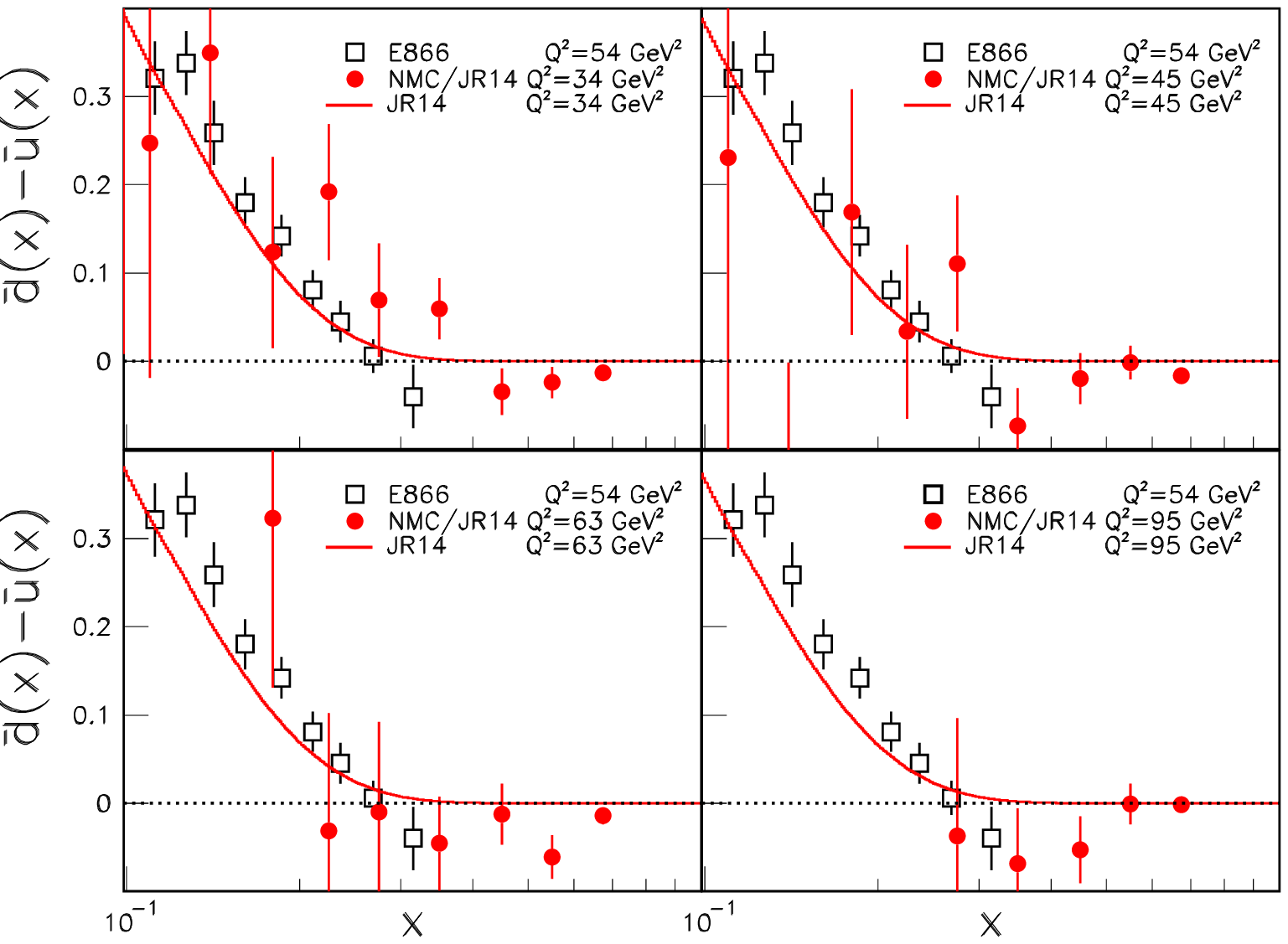}
\end{minipage}
\begin{minipage}{0.3\textwidth}
\includegraphics[width=\textwidth]{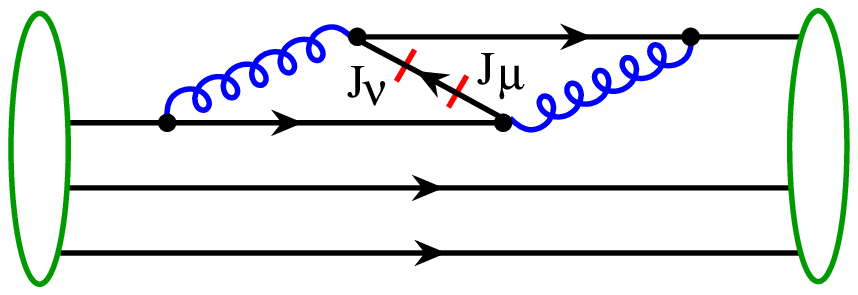}
\vskip 0.05in
time \ $\longrightarrow$
\vskip 0.05in
\includegraphics[width=\textwidth]{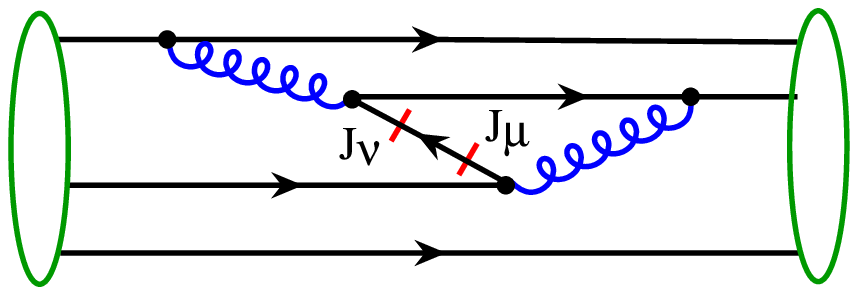}
\end{minipage}
\end{center}
\caption{Left (a): Values of $\bar d(x) - \bar u(x)$ evaluated using
  Eq.~(\ref{eq:f2pn2}) and the NMC data~\cite{nmc95,nmc97} of
  $F^p_2(x)-F^n_2(x)$ at $Q^2 = 34, 45, 63,$ and 95 GeV$^2$. The JR14
  parametrization for $u_v(x)-d_v(x)$ is used. The values of $\bar
  d(x) - \bar u(x)$ from E866 measurement at $Q^2 = 54$ GeV$^2$ are
  also shown. The solid curves are $\bar d(x) - \bar u(x)$ from
  JR14. Right (b): QCD quantum fluctuation capable of generating
  connected $\bar u(x)$ or $\bar d(x)$, involving one (top) or two
  (bottom) valence quarks, which could lead to more $\bar u(x)$ than
  $\bar d(x)$. Figure from~\cite{peng14}.}
\label{dbmud_new}
\end{figure}

Recently an independent evidence was reported for the $\bar d(x)-\bar
u(x)$ sign-change at $x \sim 0.3$ based on an analysis of the NMC DIS
data~\cite{peng14}. At the leading order in $\alpha_s$ the $\bar d(x)
- \bar u(x)$ can be extracted from the NMC measurement of
$F^p_2(x)-F^n_2(x)$ and the parametrization of $u_v(x) - d_v(x)$ from
the PDF as follows:
\begin{equation}
\bar d(x) - \bar u(x) = \frac{1}{2} [u_v(x) - d_v(x)] -
\frac{3}{2x}[F^p_2(x)-F^n_2(x)].
\label{eq:f2pn2}
\end{equation}
Figure~\ref{dbmud_new}(a) shows $\bar d(x) - \bar u(x)$ at
four values of $Q^2$ using Eq.~\ref{eq:f2pn2} with the NNLO
JR14~\cite{JR14} parametrization of the valence quark distributions
and the NMC data~\cite{nmc95,nmc97} for $F_2^p(x) - F_2^n(x)$. The
JR14 is a recent PDF where the nuclear corrections from the CJ
group~\cite{CJ11} is implemented and $\bar d(x) - \bar u(x) \ > 0$ is
assumed at all $x$ in the global analysis. Very similar results are
obtained if more recent PDFs are used for the valence quark
distributions. Figure~\ref{dbmud_new}(a) shows that both the NMC and the
E866 data show evidence that $\bar d(x) - \bar u(x)$ changes sign at
$x\sim 0.3$.

Specific examples of diagrams possibly responsible for generating more
$\bar u$ than $\bar d$ at high $x$ involving one or two valence quarks
are shown in Fig.~\ref{dbmud_new} (b). The antiquark
mode of the QCD quantum fluctuation of a quark is probed by the
currents $J_\mu$ and $J_\nu$.  In time sequence the valence quark
first radiates a highly virtual gluon which splits into a
quark-antiquark pair. It is followed by the annihilation or
recombination of the valence quark and the newly produced antiquark
into a highly virtual gluon, which is finally absorbed by the
quark. These diagrams could generate roughly a factor of 2 more $\bar
u(x)$ than $\bar d(x)$ due to $u_v/d_v$ the 2-to-1 ratio of valence
quarks in the proton.

The significance of the sign-change of $\bar d(x) - \bar u(x)$ for $x
> 0.3$, if confirmed by future experiments, e.g. E906/SeaQuest
Drell-Yan experiment, is that it would severely challenge existing
theoretical models which can successfully explain $\bar d(x) - \bar
u(x)$ at $x<0.25$, but predict no sign-change at higher $x$.

\subsection{Strange Quark Sea}

\subsubsection{Breaking of SU(3) flavor symmetry of quark sea}

Most of the information about the strange sea is obtained from the
neutrino deep inelastic scattering~\cite{nuDIS1,nuDIS2} via the
process of $W$ boson exchange as introduced in
Sec.~\ref{sec:nudis}. The measurements of charm production with dimuon
events in the final state have been performed by several experiments:
CDHS~\cite{CDHS1982}, CCFR~\cite{CCFR93,CCFR95},
CHARMII~\cite{CHARM1999}, NOMAD~\cite{NOMAD2000,NOMAD2013},
NuTeV~\cite{NuTeV2001, NuTeV2002, NuTeV2002PRD, NuTeV2007} and
CHORUS~\cite{CHORUS2008,CHROUS2011}. The integrated strange over
non-strange nucleon sea $\kappa_s$, defined as
\begin{equation}
\kappa_s = {\int_0^1 x [s(x,\mu^2)+ \bar s(x,\mu^2)] dx \over \int_0^1
  x [\bar u(x,\mu^2)+ \bar d(x,\mu^2)] dx},
\end{equation}
and the strange sea over non-strange quark, $\eta_s$, defined as
\begin{equation}
\eta_s = {\int_0^1 x [s(x,\mu^2) + \bar s(x,\mu^2)] dx \over \int_0^1
  x [u(x,\mu^2)+d(x,\mu^2)] dx},
\end{equation}
are determined by these experiments at the energy scale $\mu^2$ and
the results are summarized in Table~\ref{tab:strange}. It is noted
that the extracted values of $\kappa_s$ and $\eta_s$ depend on the
order of perturbative QCD correction for charm quark production and
the charm fragmentation function. Since the determination was done
with several parameters simultaneously, $\kappa_s$ or $\eta_s$ also
correlates with the other parameters e.g. the mass of charm $m_c$ and
the branching ratio $B_{c}$.

The most recent result of $\kappa_s$ is $0.591 \pm 0.019$ at $\mu^2 =
20$ GeV$^2$ determined at NNLO by NOMAD
collaboration~\cite{NOMAD2013}. Overall the results of $\kappa_s$
indicate that the momentum fraction carried by strange sea is about
30-50\% of that carried by non-strange sea quarks and the SU(3) flavor
symmetry is broken. The momentum fraction ratio of the strange quark
to the total non-strange quark content, $\eta_s$, is about 5-10\%.

\begin{table}[H]
\begin{center}
\begin{tabular}{llllll}\hline
\noalign{\smallskip} 
Experiment (year) & QCD order & $\kappa_s$ & $\eta_s$ & $\mu^2$ (GeV$^2$) & Ref.  \\
\noalign{\smallskip}
\hline
\hline
\noalign{\smallskip}

CDHS (1982) & LO & $0.52 \pm 0.09$ & $0.052 \pm 0.004 $  & 20 & \cite{CDHS1982} \\

CCFR (1993) & LO & $0.373^{+0.048} _{-0.041}$$\pm 0.018$ & $0.064^{+0.008} _{-0.007}$$ \pm 0.002$ & 22.2 & \cite{CCFR93} \\

CCFR (1995) & NLO & $0.477^{+0.051} _{-0.050}$$^{-0.017} _{+0.036}$ & $0.099 \pm 0.008 \pm 0.004$ & 22.2 & \cite{CCFR95} \\

CHARMII (1999) & LO & $0.388 ^{+0.074} _{-0.061}$$\pm 0.067$ & $0.068
\pm 0.014$ & 20 & \cite{CHARM1999} \\

NOMAD (2000) & LO & $0.48^{+0.09} _{-0.07}$$^{+0.17} _{-0.12}$ & $0.071^{+0.011} _{-0.009}$$^{+0.020} _{-0.015}$ &  & \cite{NOMAD2000} \\

NuTeV (2001) & LO & $0.38 \pm 0.08 \pm 0.043$ &  &  &\cite{NuTeV2001} \\

NuTeV (2007) & NLO & & $0.061 \pm 0.001 \pm 0.006$  &  &\cite{NuTeV2007} \\

CHORUS (2008) & NLO & $0.33 \pm 0.05 \pm 0.05$ & & 20 & \cite{CHORUS2008} \\

NOMAD (2013) & NNLO & $0.591 \pm 0.019$ & & 20 & \cite{NOMAD2013} \\

\hline
\hline
\end{tabular}
\end{center}
\caption{List of neutrino-induced dimuon experiments, the order of QCD
  analysis for charm quark production, the strange quark content
  parameters $\kappa_s$ and $\eta_s$ and the energy scale ($\mu^2$),
  in the order of the year of published results. Errors are
  statistical and systematic.}
\label{tab:strange}       
\end{table}
%


\begin{figure}[H]
\centering 
\includegraphics[width=0.6 \textwidth]{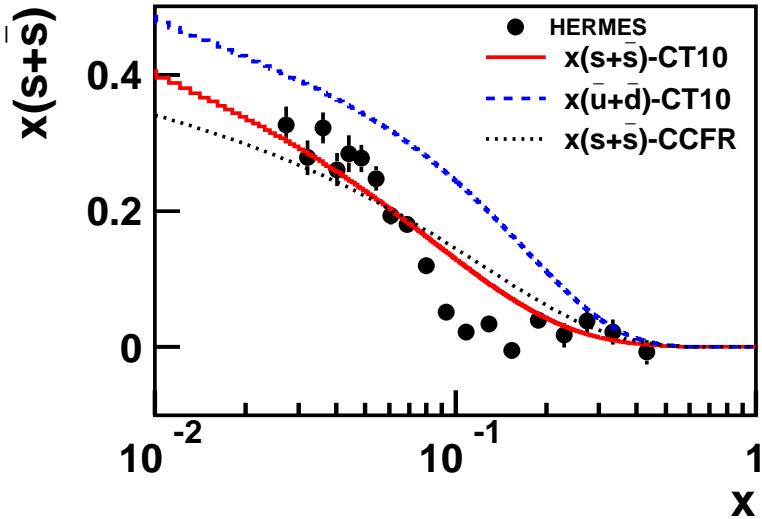}
\caption{Data of $x(s(x) + \bar s(x))$ from HERMES~\cite{hermes08} at
  $Q^2 = 2.5$ GeV$^2$ in comparison with the predictions of $x(s(x) +
  \bar s(x))$ and $x(\bar u(x) + \bar d(x))$ from CT10~\cite{CT10} at
  $Q^2 = 2.5$ GeV$^2$ and $x(s(x) + \bar s(x))$ from
  CCFR~\cite{CCFR95} at $Q^2 = 1$ GeV$^2$.}
\label{fig:hermes2008}
\end{figure}

In SIDIS reaction, the HERMES collaboration reported the determination
of $x(s(x) + \bar s(x))$ over the range of $0.02 < x < 0.5$ at $Q^2 =
2.5$ GeV$^2$ from their measurement of charged kaon production on a
deuteron target~\cite{hermes08}. The extraction of strange quark
densities is done with leading-order expression. The HERMES data,
shown in Fig.~\ref{fig:hermes2008}, exhibits an intriguing feature
that $x(s(x) + \bar s(x))$ for $x<0.1$ shows a strong rise towards
small $x$ and gradually deviates from the parametrization of CCFR at
$Q^2 = 1$ GeV$^2$~\cite{CCFR95} but agrees with the prediction of
CT10~\cite{CT10}, which does not require $ (s(x) + \bar s(x)) = (\bar
u(x) + \bar d(x))/ 2 $ for the parton densities at the initial
scale. The $x(s(x) + \bar s(x))$ for $x<0.05$ actually become close to
$x(\bar u(x) + \bar d(x))$ of CT10 PDF. This reflects a restoration of
SU(3) flavor symmetry at small $x$ even at low $Q^2$. Another
intriguing feature is that beyond $x \sim 0.1$ the data become
relatively independent of $x$ and HERMES results suggest the presence
of two different components of the strange sea, one of which dominates
at small $x$ $(x<0.1)$ and the other at larger $x$ $(x>0.1)$. Recently
HERMES results have been re-evaluated~\cite{hermes2013}, however the
results strongly depend on the kaon fragmentation function in the
extraction process. The strange quark sea has been interpreted as a
mixture of intrinsic and extrinsic sea components, to be discussed in
Sec~\ref{sec:IntrinsicModel}. The role of the strange sea in
separating the light-quark connected sea from disconnected sea is
discussed in Sec~\ref{sec:CSDS}.


At LHC, the measurement of $W^{\pm}$ boson production in proton-proton
collisions, either inclusive~\cite{kusina12} or associated with a
single charm quark~\cite{Baur:1993zd,stirling13} allows the
determination of the strange parton distributions in conjunction with
DIS data. First the ATLAS experiment determined the strange-to-down
antiquark ratio $r_s$ ($=(s + \bar s)/2 \bar d)$ to be
1.00$^{+0.25}_{-0.28}$ at $x=0.023$ and $Q^2 = 1.9$ GeV$^2$ from the
global analysis of inclusive $W$ and $Z$ production in $pp$ collisions
at 7 TeV and DIS inclusive data from HERA
(ATLAS-epWZ12)~\cite{Atlas_WZ}. The result supports the hypothesis of
an SU(3)-symmetric light-quark sea, i.e $\bar s = \bar d$ in the small
$x$ region at small $Q^2$, consistent with what was observed by HERMES
in SIDIS reaction.

\begin{figure}[H]
\centering 
\includegraphics[width=0.6 \textwidth]{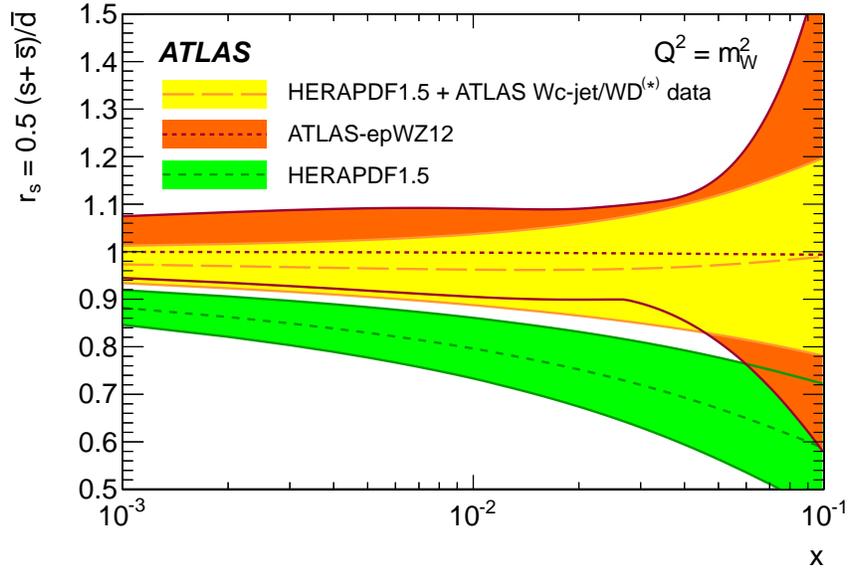}
\caption{Ratio of strange-to-down sea-quark distributions $r_s =
  0.5(s+\bar{s})/\bar{d}$ as a function of $x$ at $Q^2=M_W^2$ obtained
  from HERA data only (HERAPDF1.5), ATLAS $W,Z$ production only
  (ATLAS-epWZ12) and HERA plus ATLAS $W+c$ production
  (HERAPDF1.5+ATLAS-Wc). Figure from~\cite{Atlas_Wc}.}
\label{fig:ATLAS_s}
\end{figure}

Recently ATLAS determined the same ratio $r_s$ to be $0.96^{+0
  .26}_{-0.30}$ at $Q^2 = 1.9$ GeV$^2$ using the measurement of cross
sections of the associated $W+c$ production at LHC~\cite{Atlas_Wc} and
the DIS data from HERA (represented by HERAPDF1.5~\cite{HERAPDF}), in
good agreement with the values determined in the previous combined
analysis of $W,Z$ production. Figure~\ref{fig:ATLAS_s} shows the
$x$-dependence of $r_s$ at $Q^2=m_W^2$, obtained from three global
analyses: HERAPDF1.5, ATLAS-epWZ12 and HERAPDF1.5+ATLAS-Wc. At large
$Q^2=m_W^2$, the sea quarks are dominantly produced by QCD evolution
and SU(3) symmetry is clearly observed as seen in the $r_s(x) \sim 1 $
across the whole $x$ region.

The same measurement of associated $W+c$ production from
proton-proton collision at LHC is also done by the CMS
collaboration~\cite{CMS_Wc}. Recently CMS performed a
next-to-leading-order global analysis using this result together with
those of muon charge asymmetry in the inclusive $W$ production and the
inclusive DIS cross sections at HERA for the determination of valence
quark and the strange quark distributions~\cite{CMS_Wasym}. The
integrated strange fraction of the nucleon sea $\kappa_s$ is
determined at $Q^2 = 20$ GeV$^2$ to be
$0.52^{+0.12}_{-0.10}\,\rm{(exp.)}^{+0.05}_{-0.06}\,\rm{(model)}
^{+0.13}_{-0.10}\,\rm{(parametrization)}$, in agreement with the
value obtained by the neutrino-DIS NOMAD experiment~\cite{NOMAD2013}.

\begin{figure}[H]
\centering
\includegraphics[width=0.4 \textwidth]{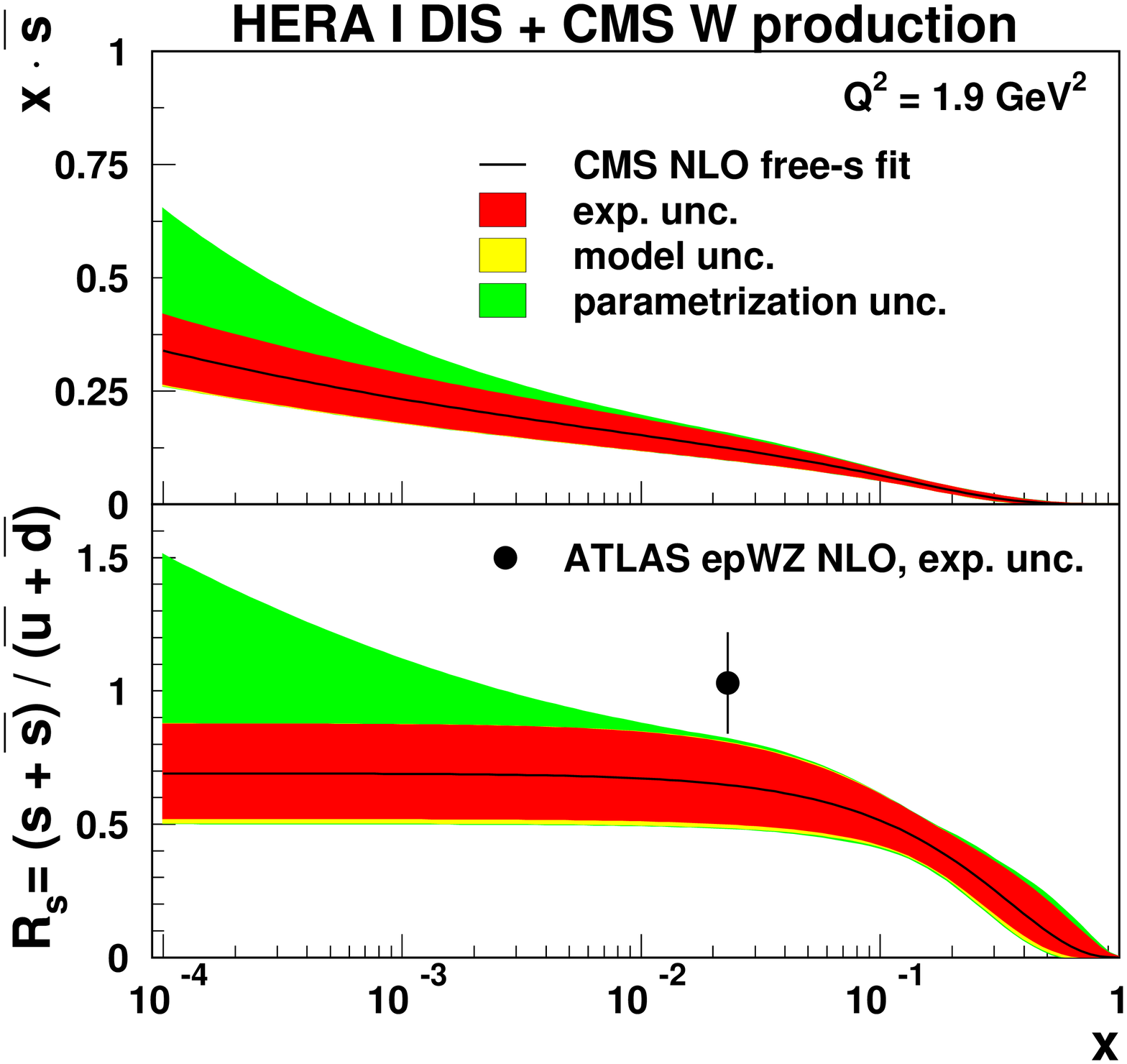}
\includegraphics[width=0.4 \textwidth]{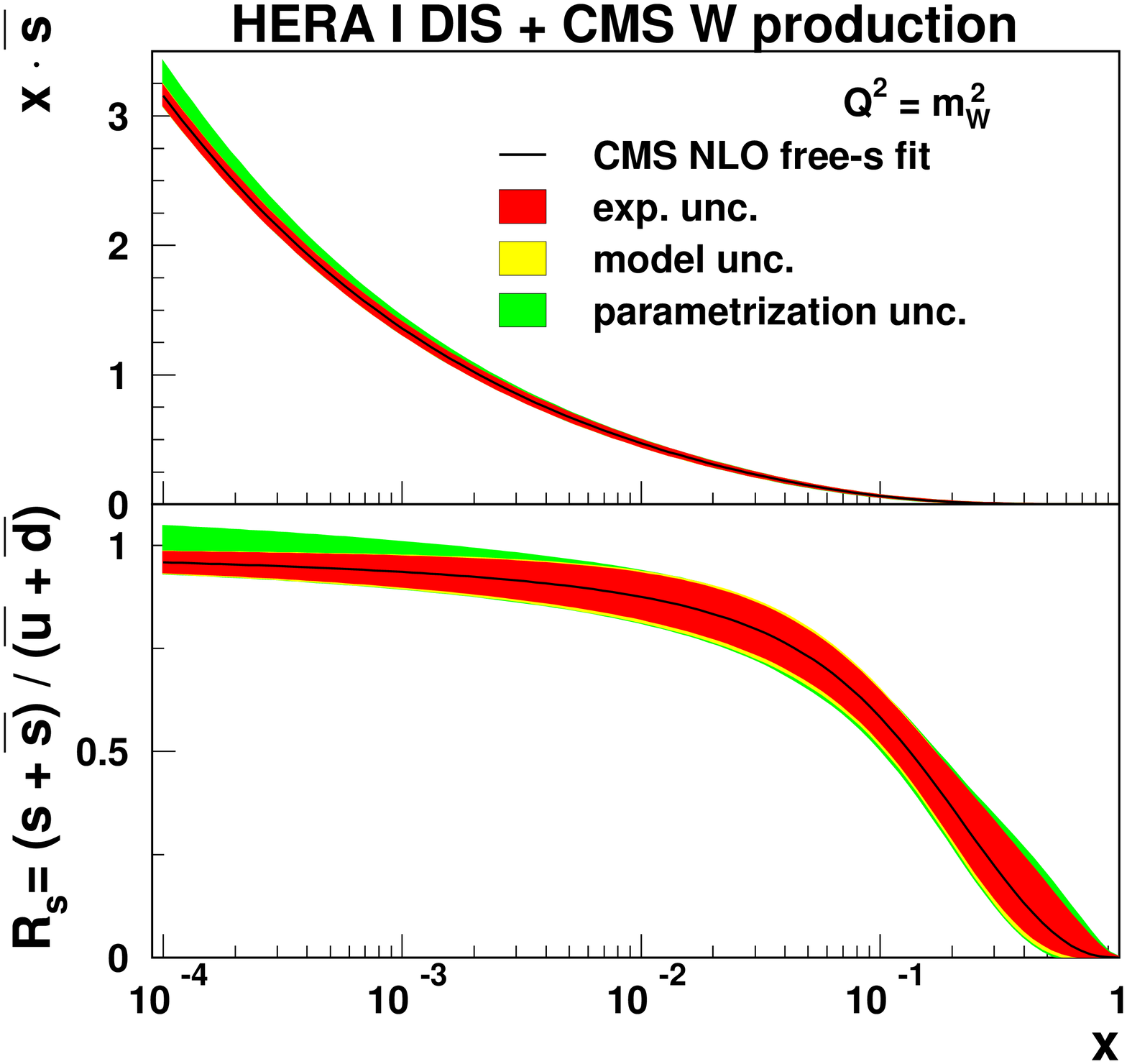}
\caption{Antistrange-quark distribution $x \bar s (x,Q)$ and the
  strangeness ratio $R_{s}(x,Q)$, obtained in the QCD analysis of HERA
  and CMS data, shown as functions of $x$ at $Q^2 = 1.9$ GeV$^2$
  (left) and $Q^2 = M^2_W$ (right). The NLO ATLAS result of $r_s = (s
  + \bar s)/2 \bar d$~\cite{Atlas_WZ} is presented by a closed symbol
  in the left-bottom figure for comparison. Figure
  from~\cite{CMS_Wasym}.}
\label{fig:CMS_s}
\end{figure}

The strange quark distribution $s(x,Q^2)$ and the strangeness ratio
$R_{s}(x,Q)=(s + \bar s)/(\bar u + \bar d) (x, Q^2)$ at $Q^2 = 1.9$
GeV$^2$ and $m_W ^2$ are shown in Fig.~\ref{fig:CMS_s}. The strange
quark fraction rises with the energy scale and the $s$ quark becomes
comparable to $\bar u$ and $\bar d$ at the regions of intermediate to
low $x$. In the left plot of Fig.~\ref{fig:CMS_s} the $r_s$ determined
by ATLAS is displayed together with the distributions of $R_{s}(x)$ by
CMS at the same scale $Q^2 = 1.9$ GeV$^2$. Since the difference of
$\bar u$ and $\bar d$ at this $x$ region is expected to be very minor,
this comparison shows a $2 \sigma$ difference in the strangeness
fraction determined by the LHC experiments. A discussion on the
possible cause for this discrepancy is given in
Sec.~\ref{sec:StrangeGlobal}.

\subsubsection{Asymmetry between $s (x)$ and $\bar s (x)$}

There is no fundamental reason why the $x$ distributions of strange
quark and antistrange quark in the nucleons are the same as long as
the integral of the asymmetry over all values of $x$ vanishes,
i.e. $\left< s - \bar s \right>$ = $\int_0^1 [s(x) - \bar s (x)]dx =
0$. The strangeness asymmetry could in principle affect the extraction
of the Weinberg angle from neutrino-nucleus deep inelastic scattering
and thus non-zero asymmetry could partially explain the anomaly seen
by the NuTeV experiment~\cite{NuTeV2002} in their measurement of
Weinberg
angle~\cite{alwall05,ding05PLB,wakamatsu05,davidson2002,cao03,kretzer04,alwall04}.

\begin{figure}[H]
\centering
\includegraphics[width=0.4\textwidth]{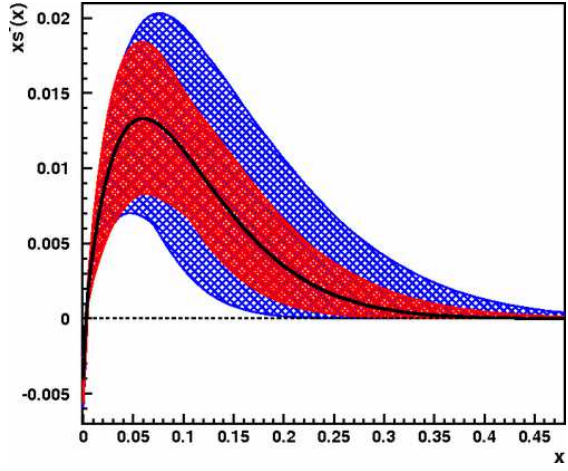}
\caption{The extracted distribution of $xs^- (x) = x(s(x) - \bar
  s(x))$ at $Q^2 = 16$ GeV$^2$ by NuTeV. Blue error band represents
  the total uncertainty while the red band is the uncertainty without
  the error of charm semileptonic branching ratio. Figure from
  ~\cite{NuTeV2007}.}

\label{fig:NuTeV}
\end{figure}

The CCFR NLO evaluation of the strange quark distributions did not
exclude the possible difference between $s(x)$ and $\bar s
(x)$~\cite{CCFR95}. The NuTeV collaboration determined $\left< x(s -
\bar s) \right>$ from LO analysis of dimuon events from neutrino
(antineutrino) DIS with the nucleon~\cite{NuTeV2002PRD} and found a
negative value of $-0.0027 \pm 0.0013$. Afterward it performed a
complete NLO analysis~\cite{NuTeV2007} and found a slightly positive
asymmetry of $0.00196 \pm 0.00046 \pm 0.00045$, as shown in
Fig.~\ref{fig:NuTeV}. Clearly there is still a large uncertainty in
the extraction of $s(x)$ and $\bar s(x)$ distributions from the
experimental data. In Ref.~\cite{gao2008} it was suggested that
heavy-quark recombination effect where a heavy $c$ quark combines with
a light antiquark forming a $D$ meson should be taken into account in
extracting the strange asymmetry from the dimuon events in neutrino
(antineutrino) DIS. Their study showed that the value of $\left< x(s -
\bar s) \right>$ could increase by $0.0023$ with the inclusion of this
effect.

\begin{figure}[H]
\begin{minipage}{0.45\textwidth}
\includegraphics[width=\textwidth]{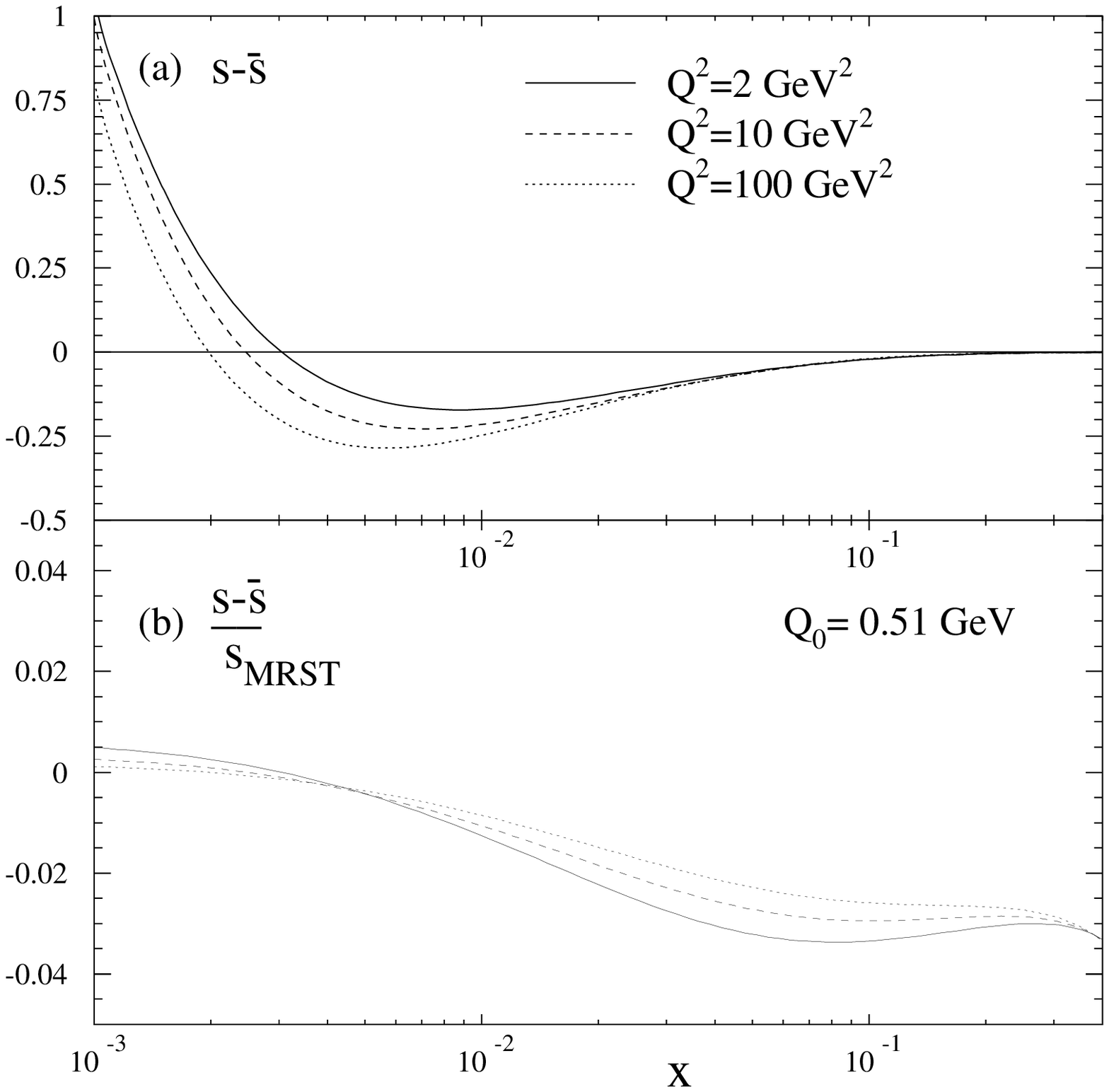}
\end{minipage}
\hspace{-0.5cm}
\begin{minipage}{0.55\textwidth}
\includegraphics[width=\textwidth]{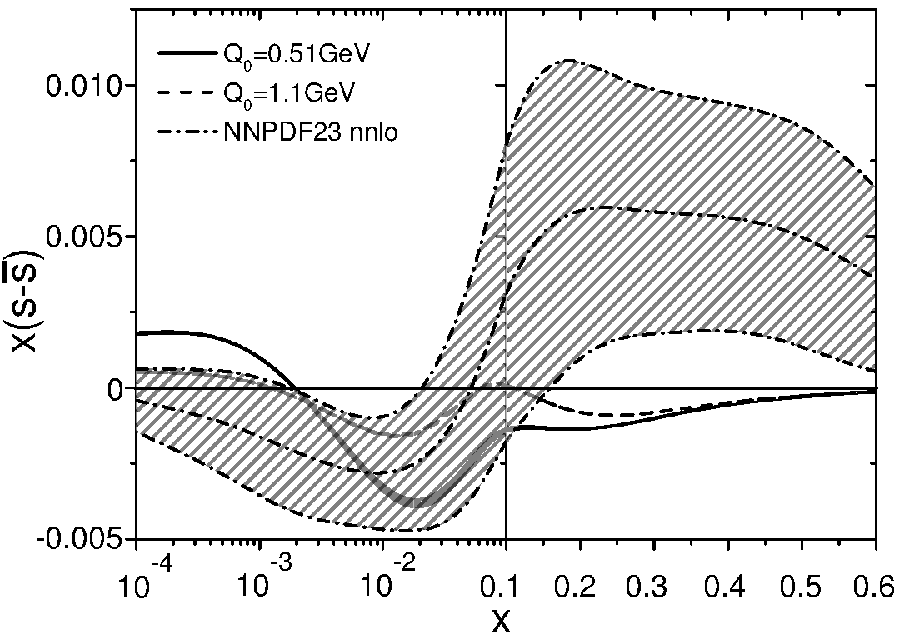}
\end{minipage}
\caption{Left: (a) The asymmetry $s(x) - \bar s(x)$ from perturbative
  NNLO QCD calculation (b) the ratio to the LO strange density. Figure
  from ~\cite{catani04}. Right: (c) The perturbative strangeness
  asymmetry $x(s(x) - \bar s(x))$ at $Q^2 = 16$ GeV$^2$ with the
  initial scale $Q_0 = 0.51$ GeV (solid line) and $1.1$ GeV (deashed
  line). The small bands represent the $1 \sigma$ range due to the
  uncertainties associated with the $d_v$ and $u_v$ of ABM11 PDF. The
  total asymmetry from NNLO NNPDF2.3 PDF and its $1 \sigma$
  uncertainty band is overlaid. Figure from ~\cite{feng12}.}
\label{fig:toKK}
\end{figure}

Using the 3-loop splitting functions~\cite{moch04}, a non-zero
$s(x)-\bar s(x)$ could be generated from NNLO QCD evolution starting
with the condition of $s(x)=\bar s(x)$ at the initial
scale~\cite{catani04}. The reason is that the probability of a $q
\rightarrow q'$ splitting is different from that of $q \rightarrow
\bar q'$ and the $u,d$ valence densities in the nucleon are
different. Figure~\ref{fig:toKK} (a) and (b) shows that the generated
asymmetry $s-\bar s$ is not negligible which is positive at small $x$
and negative at large $x$. The second moment $\left< x(s - \bar s)
\right>$ is evaluated to be about $-5 \times 10^{-4}$ at $Q^2 = 20$
GeV$^2$. The perturbative effect gives a $\langle x(s-\bar s) \rangle$
with a sign opposite to the latest NuTeV result.

The nonperturbative contributions of strangeness asymmetry could come
from nucleon fluctuating into virtual hyperon and kaon pairs such as
$\Lambda K$ and $\Sigma K$. This mechanism of kaon cloud is very
similar to that of non-strange meson cloud which leads to the $\bar d
/ \bar u$ asymmetry. Nevertheless theoretical predictions of
$s(x)-\bar s(x)$ are rather diverse. Signal and Thomas~\cite{signal87}
predicted that the $s$ and $\bar s$ have different distributions with
the bag model and the quantitative results depended on the bag
radius. Holtmann et al.~\cite{holtmann96} found that $s < \bar s$ in
small $x$ region and $s > \bar s$ in large $x$ region using the meson
cloud model with the fluctuation function calculated from time-ordered
perturbative theory in the infinite momentum frame. However Brodsky
and Ma~\cite{brodsky96} reached an opposite conclusion using a
light-cone two-body wave function model for the description of the
meson-baryon fluctuation. It was shown in Ref.~\cite{cao99} that the
difference between $s(x)$ and $\bar s(x)$ is sensitive to the
splitting functions of hyperon and kaon as well as the distributions
of strange partons inside the virtual hadrons.

In the chiral quark model~\cite{ding05PLB,ding05PRD} and the
chiral-quark soliton model~\cite{wakamatsu03,wakamatsu05}, the
coupling of meson cloud is via the constituent quark of the nucleon
and $s(x) > \bar s(x)$ in large $x$ region is predicted. In a recent
study of strangeness asymmetry~\cite{feng12}, both nonperturbative
meson cloud and perturbative NNLO contributions are taken into account
simultaneously. It is found that the nonperturbative contribution
dominates in the region of $x \geq 0.1$ while the perturbative ones
are more significant in the smaller $x$ region, as shown in
Fig.~\ref{fig:toKK} (c), compared with the results of NNPDF global
analysis~\cite{NNPDF2.3}. The best region to detect this asymmetry
experimentally is claimed to be $0.02<x<0.03$ and more than one node
of the $s(x) -\bar s(x)$ asymmetry is suggested.


\subsubsection{Strange quarks from the global analysis}
\label{sec:StrangeGlobal}

\begin{figure}[H]
\hspace{-0.7cm}
\subfigure[]
{{\includegraphics[width=0.35\textwidth]{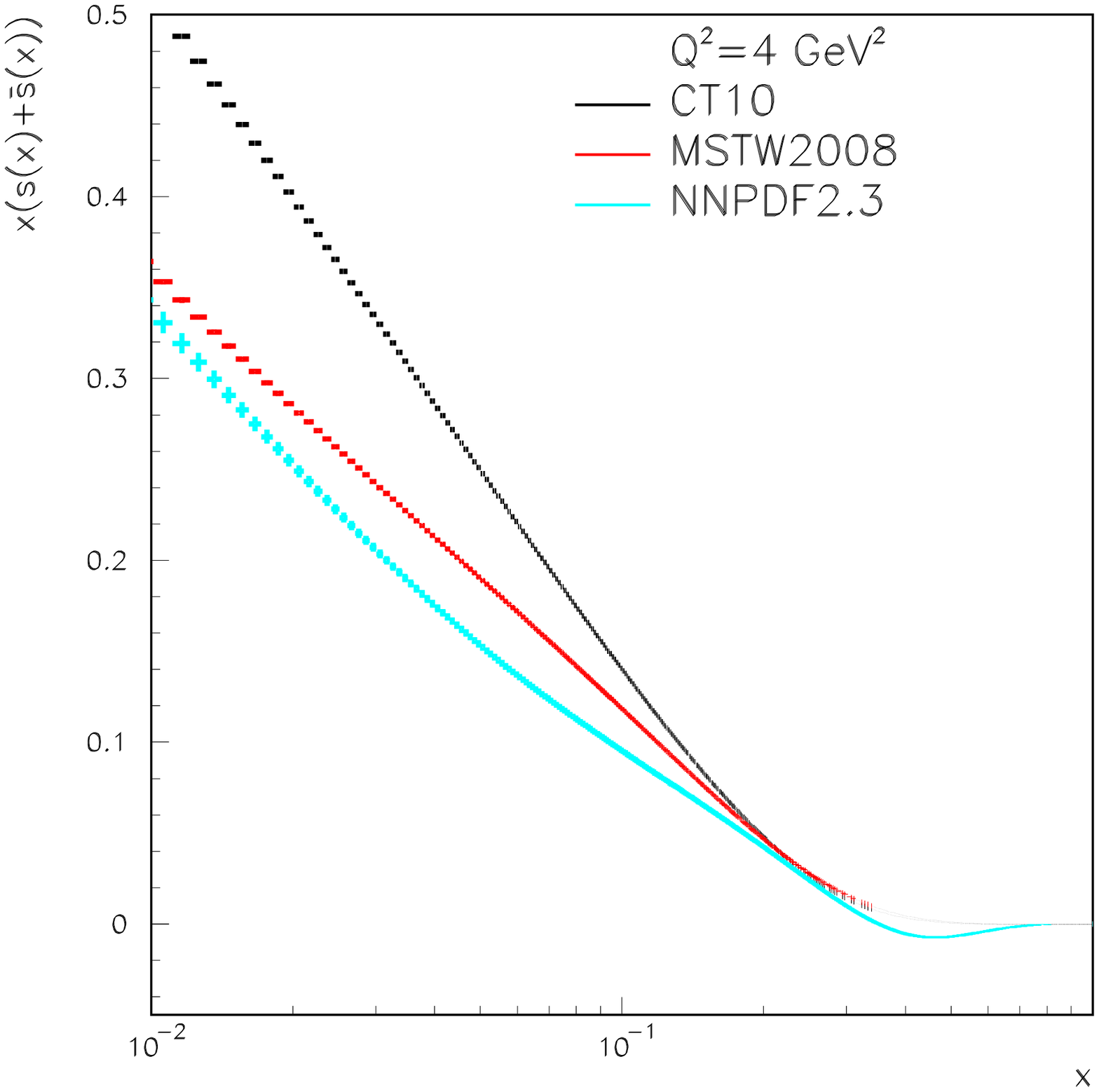}}
\label{fig:ssbarsum}}
\hspace{-0.7cm}
\subfigure[]
{{\includegraphics[width=0.35\textwidth]{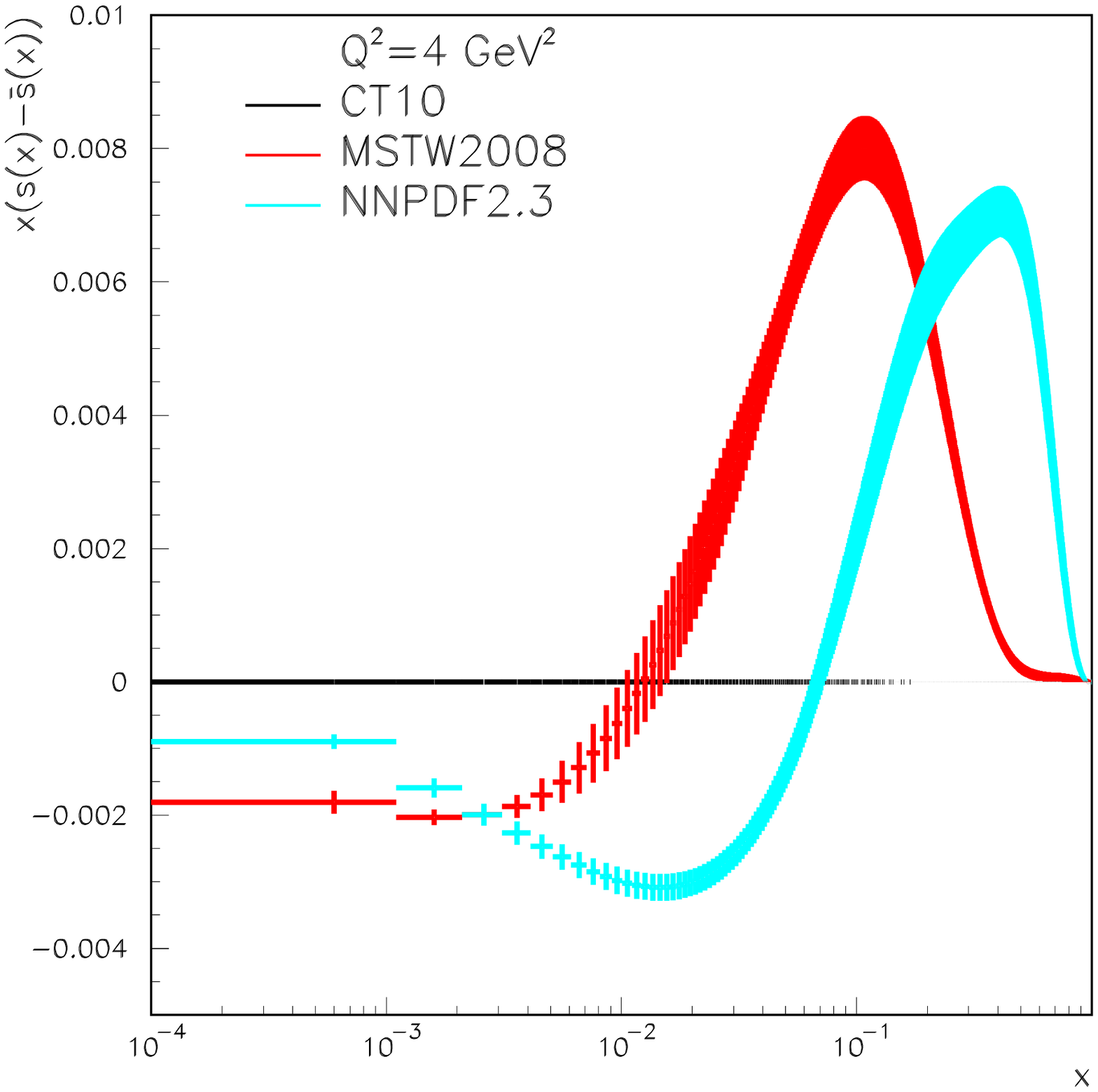}}
\label{fig:ssbarasym}}
\hspace{-0.7cm}
\subfigure[]
{{\includegraphics[width=0.35\textwidth]{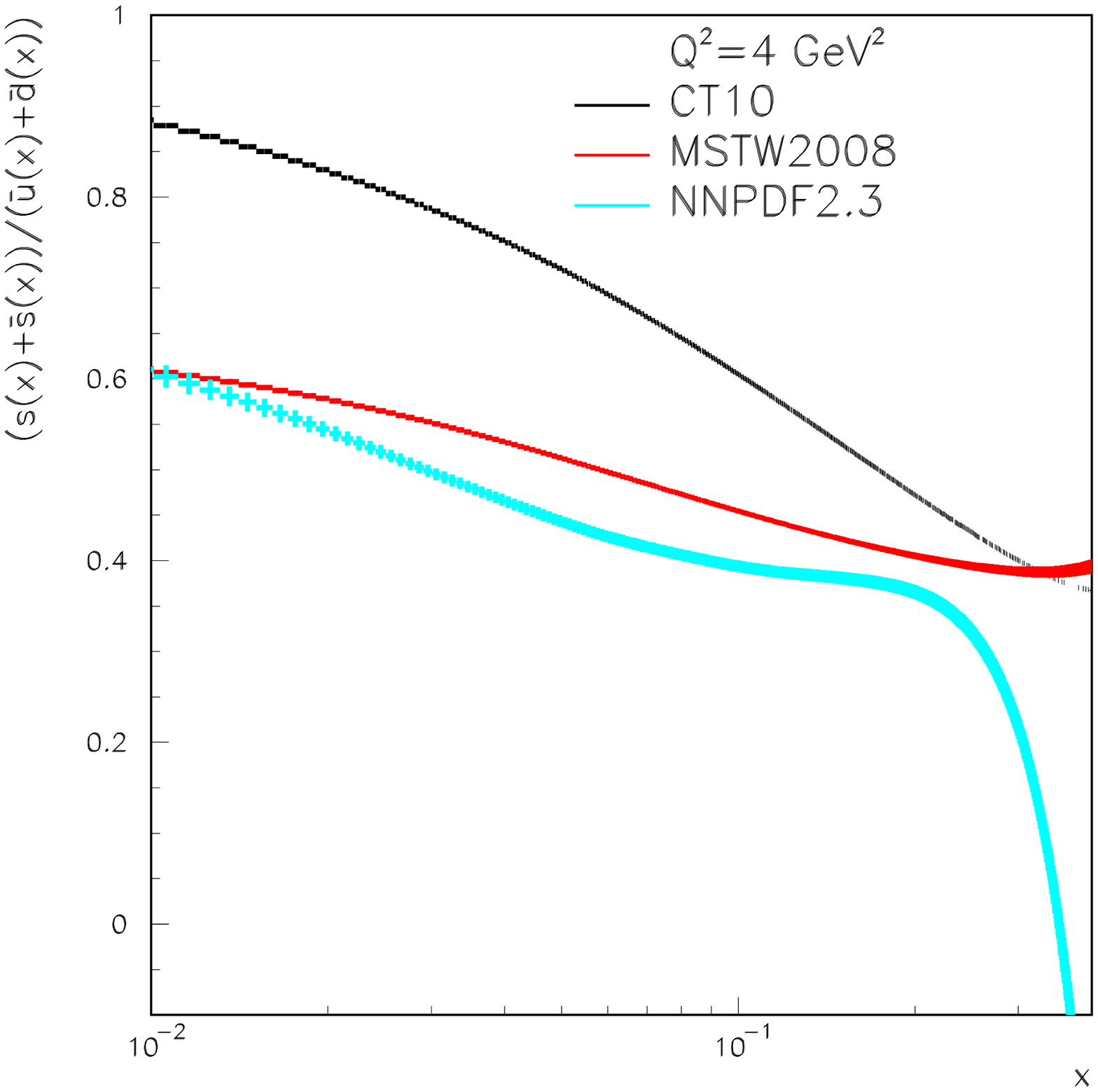}}
\label{fig:ssbar}}
\caption{The distributions of $x(s(x)+ \bar s (x))$, $x(s(x)- \bar s
  (x))$ and $(s(x)+ \bar s (x))/(\bar u(x) + \bar d(x))$ from
  CT10~\cite{CT10}, MSTW2008~\cite{MSTW08} and
  NNPDF2.3~\cite{NNPDF2.3} PDFs at $Q^2 = 4$ GeV$^2$.}
\label{fig:ssbarPDF}
\end{figure}

Figure~\ref{fig:ssbarPDF} shows the distributions of $x(s(x)+ \bar s
(x))$, $x(s(x)- \bar s (x))$ and $(s(x)+ \bar s (x))/(\bar u(x) + \bar
d(x))$ at $Q^2 = 4$ GeV$^2$ from CT10~\cite{CT10},
MSTW2008~\cite{MSTW08} and NNPDF2.3~\cite{NNPDF2.3}. There is an
assumption of $s$ and $\bar s$ symmetry in CT10 while $s - \bar s$
asymmetry is allowed in MSTW2008 and NNPDF2.3. For MSTW2008 and
NNPDF2.3, the strange quark sea $s + \bar s$ is suppressed relative to
the non-strange sea $\bar u + \bar d$ over the whole $x$ region.  The
strange quarks in CT10 are less suppressed than that in MSTW2008 and
NNPDF2.3.

\begin{figure}[H]
\centering
\subfigure[]
{\includegraphics[width=0.45\textwidth]{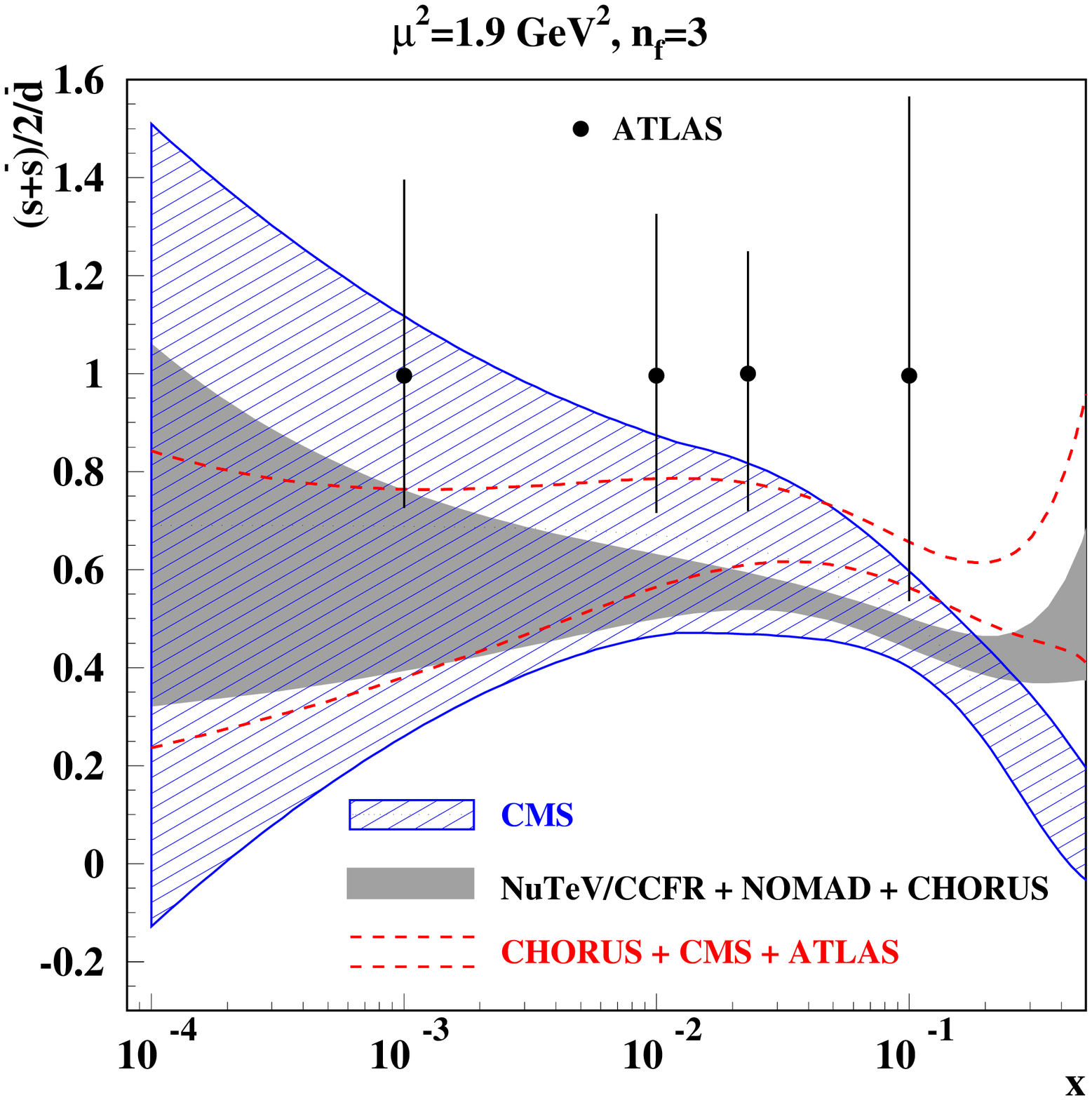}
\label{fig:ssup}}
\subfigure[]
{\includegraphics[width=0.45\textwidth]{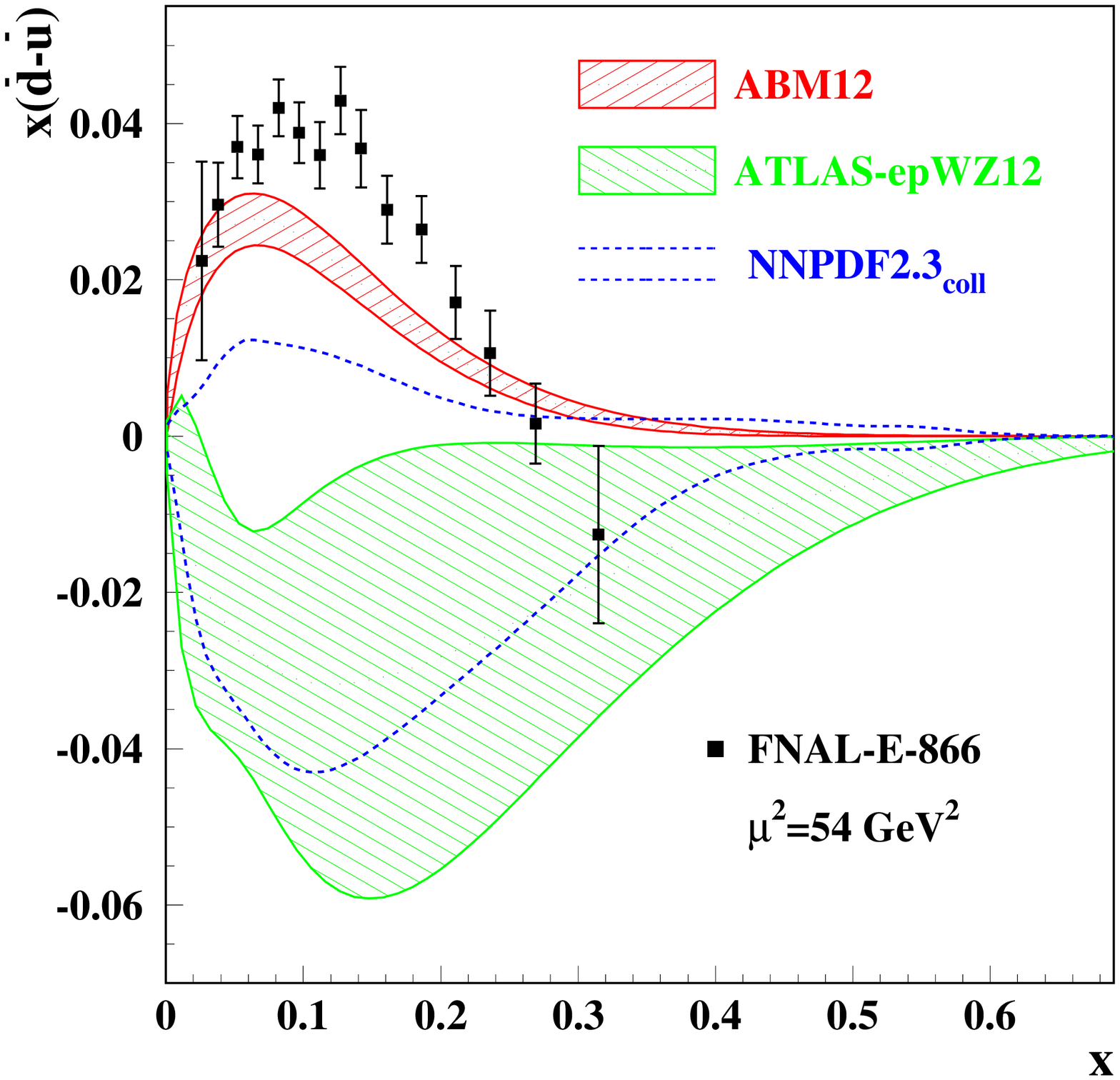}
\label{fig:udm}}
\caption{(a) The strange sea suppression factor
  $r_s=(s+\bar{s})/2\bar{d}$ and uncertainty band as a function of
  $x$ obtained in the analyses based on the combination of the data by
  NuTeV/CCFR~\cite{NuTeV2001} CHORUS~\cite{CHROUS2011}, and
  NOMAD~\cite{NOMAD2013} (shaded area) and CHORUS~\cite{CHROUS2011},
  CMS~\cite{CMS_Wc}, and ATLAS~\cite{Atlas_Wc} (dashed lines), in
  comparison with the results obtained by the CMS
  analysis~\cite{CMS_Wc} (hatched area) and by the ATLAS
  $epWZ$-fit~\cite{Atlas_WZ,Atlas_Wc} (full circles). The energy scale
  $\mu^2$ is 1.9 GeV$^2$. (b) The asymmetry of the non-strange sea
  $x(\bar{d}-\bar{u})$ at the scale of $\mu^2=54~{\rm GeV}^2$ as a
  function of $x$ obtained in the ABM12 fit~\cite{ABM2014}
  (right-tilted hatch), in comparison with the corresponding ones
  obtained by the ATLAS~\cite{Atlas_WZ} (left-tilted hatch) and the
  NNPDF~\cite{NNPDF2.3} (dashed lines) analyses using only the LHC and
  HERA collider data. Data of $x(\bar{d}-\bar{u})$ shown as full
  squares are from the FNAL E866 Drell-Yan
  experiment~\cite{E866}. Figures from ~\cite{ABM2014}.}
\label{fig:ABM12}
\end{figure}

It was noted that the dimuon data from neutrino scattering and
improved analyses are important to constrain the uncertainty of
strange parton distributions in the global
analyses~\cite{kusina12}. The inclusion of measurements of $W$ and $Z$
bosons and inclusive jet production at LHC should further improve the
constraints at the small $x$
region~\cite{NNPDF2.3,kusina12}. Very recently a new determination of
nucleon strange sea was obtained using data including neutrino DIS,
lepton DIS and $W$ production at LHC~\cite{ABM2014}. In this study it
is found that the $x$ dependence of the strange sea distribution is
similar to the non-strange one. Figure~\ref{fig:ssup} shows two $x$
distributions of $r_s$ at the energy scale $\mu^2$ = 1.9 GeV$^2$ which
are obtained by the combination of NuTeV/CCFR~\cite{NuTeV2001},
CHORUS~\cite{CHROUS2011} and NOMAD~\cite{NOMAD2013}, and the
combination of CHORUS~\cite{CHROUS2011}, CMS~\cite{CMS_Wc} and
ATLAS~\cite{Atlas_Wc}, respectively. This new determination is
consistent with the results obtained by CMS~\cite{CMS_Wc} from the
analysis of the $W+c$ production in combination with the HERA DIS
data~\cite{HERAPDF}. However, a much higher value of $r_s \sim 1$ was
obtained by ATLAS~\cite{Atlas_Wc} from the fit of the ATLAS data on
the $W,Z$ production~\cite{Atlas_WZ} together with the HERA DIS data.

This discrepancy is investigated and the speculation is that there is
a lack of sufficient constraints of non-strange sea from solely HERA
DIS data and ATLAS data. Figure~\ref{fig:udm} shows the $x$
distributions of non-strange sea asymmetry $x(\bar d(x) - \bar u(x))$
from several global analyses at $\mu^2$ = 54 GeV$^2$, compared to the
FNAL E866 Drell-Yan data~\cite{E866}. Obviously the value of $x(\bar
d- \bar u)$ obtained in ATLAS analysis~\cite{Atlas_Wc} is negative and
that from the analysis with only the LHC and HERA collider data in
NNPDF2.3 is also slightly negative, in disagreement with the E866
Drell-Yan data. This finding suggests that the strange sea enhancement
observed in the analysis of ATLAS~\cite{Atlas_WZ, Atlas_Wc} might be
achieved at the expense of a suppressed $\bar d$ distribution because
the HERA inclusive DIS data do not have enough sensitivity to the
flavor structure of antiquarks.

\subsection{Five-quark intrinsic sea model and lattice QCD}

Below we will review some recent progress in the theoretical
interpretations with regard to the observed breaking of SU(2) and
SU(3) flavor symmetry of light quark sea in terms of the five-quark
intrinsic sea and lattice QCD.

\subsubsection{Intrinsic and extrinsic sea}
\label{sec:IntrinsicModel}

\begin{figure}[H]
\centering
\subfigure[]
{{\includegraphics[width=0.30\textwidth]{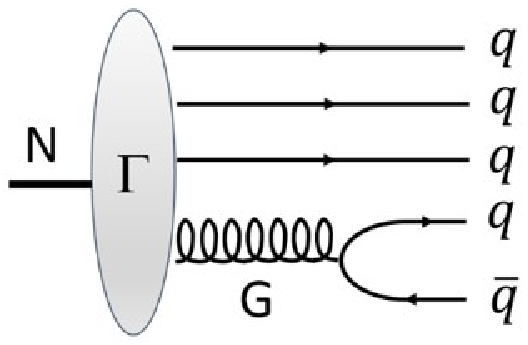}}}
\subfigure[]
{{\includegraphics[width=0.30\textwidth]{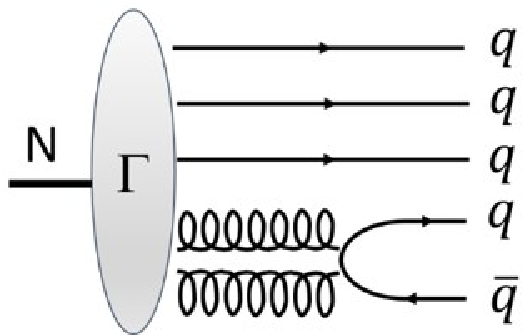}}}
\subfigure[]
{{\includegraphics[width=0.30\textwidth]{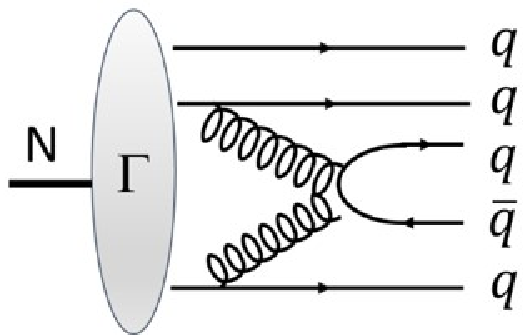}}}
\caption{(a) Gluon splitting contributing to the extrinsic sea quarks
  in the proton wave function~\cite{hoyer90}. (b) Gluon fusion and (c)
  light quark scattering contributing to the intrinsic sea quarks in
  the proton wave function~\cite{hoyer90}.}
\label{fig:BHPS}
\end{figure}

The possible existence of a significant $|u u d c \bar c \rangle$ five-quark
Fock component in the proton was proposed long time ago by Brodsky,
Hoyer, Peterson, and Sakai (BHPS)~\cite{brodsky80} to explain the
unexpectedly large production rates of charmed hadrons at large
forward $x_F$ region. The intrinsic charm originating from the
five-quark Fock state is to be distinguished from the ``extrinsic"
charm produced in the splitting of gluons into $c \bar c$
pairs. Typical diagrams of gluon splitting, gluon fusion and light
quark scattering characterizing the extrinsic and intrinsic
sea~\cite{hoyer90} are shown in Fig.~\ref{fig:BHPS}. The extrinsic
charm has a ``sea-like" characteristics with large magnitude only at
the small $x$ region. In contrast, the intrinsic charm is
``valence-like" with a distribution peaking at larger $x$.

Since the BHPS model predicts the probability for the $|u u d Q \bar Q \rangle$
configuration to be inversely proportional to $m_Q^2$, where $m_Q$ is
the mass of the quark $Q$~\cite{brodsky80}, the light quark sector
could in principle provide more clear evidences for the roles of the
five-quark Fock states, allowing the specific predictions of the BHPS,
such as the shape of the quark momentum distributions originating from
the five-quark configuration, to be tested. To search for evidence for
the intrinsic five-quark Fock states, it is essential to separate the
contributions of the intrinsic sea from those of the extrinsic one.

In Ref.~\cite{Chang:2011vx,Chang:2011du,Peng:2012rn}, the BHPS model
was extended to the light quark sector and the predictions were
compared with the experimental data of $\bar d - \bar u$, $s + \bar
s$, and $\bar u + \bar d - s -\bar s$ from E866~\cite{E866},
HERMES~\cite{hermes08} experiments and the PDF set
CTEQ6.6~\cite{CTEQ6.6}. The $\bar d - \bar u$ and $\bar u + \bar d - s
-\bar s$ are the SU(2) and SU(3) flavor-nonsinglet quantities which
are free from the contributions of the extrinsic sea
quarks. Non-perturbative effects could lead to different probabilities
for the $|u u d d \bar {d} \rangle$, $|u u d u \bar u\rangle$ and $|u
u d s \bar {s} \rangle$ configurations. The $x$ distributions of the
$\bar q$ sea in $|uudq\bar{q} \rangle$ were first constructed in the
BHPS model at a certain initial scale $\mu$ and then evolved to the
$Q^2$ of the experiments to compare with the data. As for the HERMES
$s + \bar s$ data, the sharp rise towards small $x$ is consistent with
the extrinsic sea, while the data at $x>0.1$ is interpreted as the
intrinsic component of strange sea, as described by the BHPS model.
 
\begin{figure}[H]
\centering
\subfigure[]
{{\includegraphics[height=0.3\textwidth]{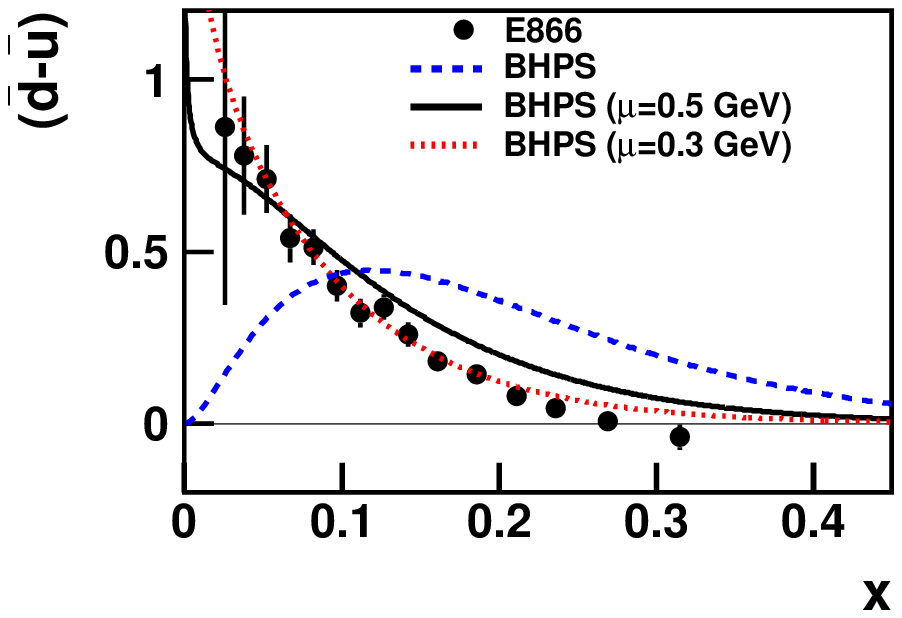}}
\label{fig:dbarubar2}}
\hspace{-1cm}
\subfigure[]
{{\includegraphics[width=0.3\textwidth]{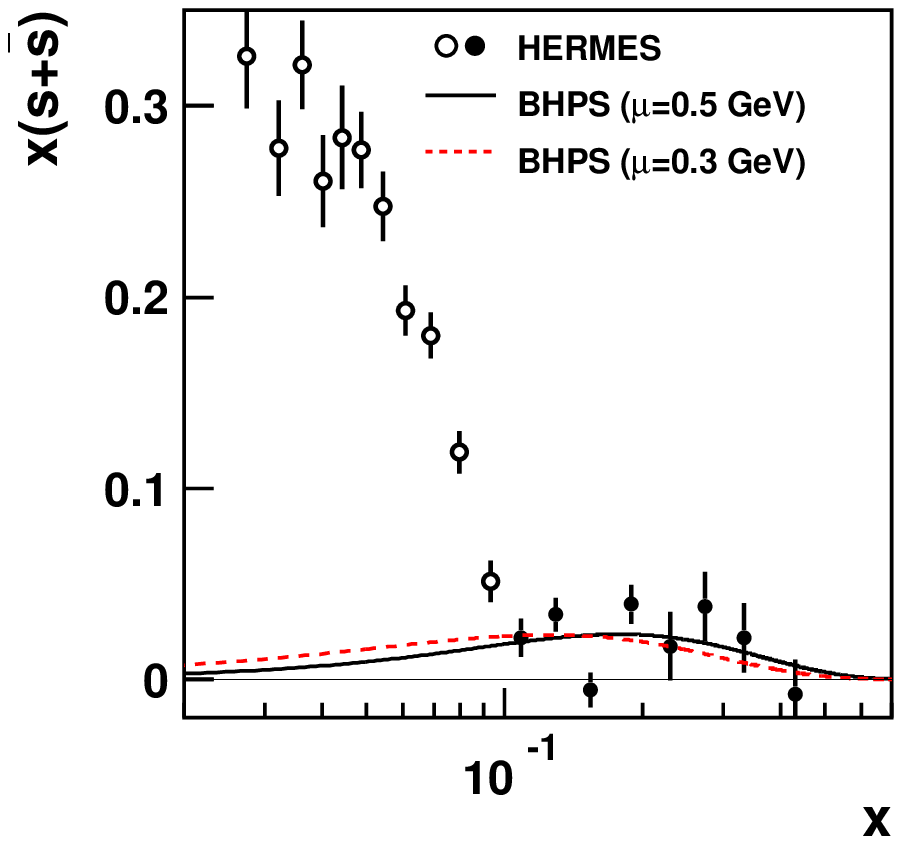}}
\label{fig:ssbar2}}
\hspace{-0.8cm}
\subfigure[]
{{\includegraphics[width=0.3\textwidth]{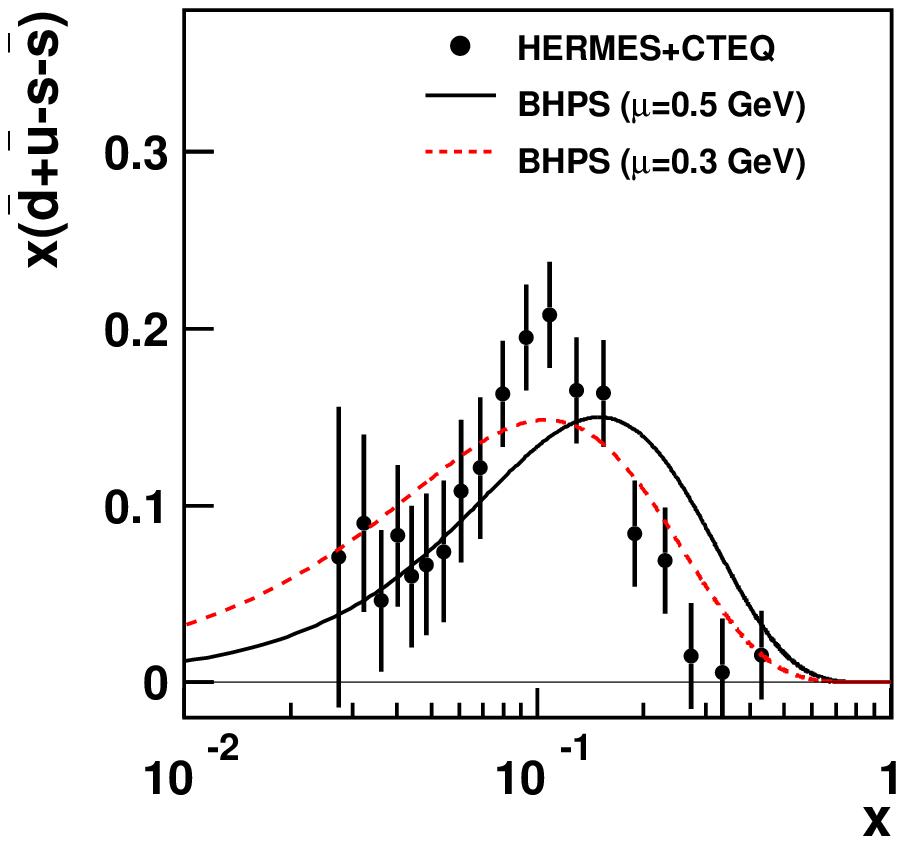}}
\label{fig:deltas}}
\caption{(a) Comparison of the $\bar d(x) - \bar u(x)$ data with the
  calculations based on the BHPS model. The blue dashed curve
  corresponds to the BHPS result at initial scale, and the solid and
  dotted curves are obtained by evolving the BHPS result to $Q^2 =
  54.0$ GeV$^2$ using $\mu = 0.5$ GeV and $\mu = 0.3$ GeV,
  respectively. (b) Comparison of the HERMES $x(s(x) + \bar s(x))$
  data with the calculations based on the BHPS model. The solid and
  dashed curves are obtained by evolving the BHPS result to $Q^2 =
  2.5$ GeV$^2$ using $\mu = 0.5$ GeV and $\mu = 0.3$ GeV,
  respectively. The normalizations of the calculations are adjusted to
  fit the data at $x > 0.1$. (c) Comparison of the $x(\bar d(x) + \bar
  u(x) - s(x) - \bar s(x))$ data with the calculations based on the
  BHPS model. The values of $x(s(x) + \bar s(x))$ are from the HERMES
  experiment~\cite{hermes08}, and those of $x(\bar d(x) + \bar u(x))$
  are obtained from the PDF set CTEQ6.6~\cite{CTEQ6.6}. The solid and
  dashed curves are obtained by evolving the BHPS result to $Q^2 =
  2.5$ GeV$^2$ using $\mu = 0.5$ GeV and $\mu = 0.3$ GeV,
  respectively. The normalization of the calculations are adjusted to
  fit the data.}
\end{figure}

The good agreement between the theory and the data shown in
Fig.~\ref{fig:dbarubar2},~\ref{fig:ssbar2}, and~\ref{fig:deltas} is
interpreted as evidence for the existence of the intrinsic light-quark
sea in the nucleons. The probabilities for the $|uudu\bar{u} \rangle$,
$|uudd\bar{d} \rangle$ and $|uuds\bar{s} \rangle$ Fock states are also
extracted as : ${\cal P}^{u \bar u}_5 = 0.122$, ${\cal P}^{d \bar d}_5
= 0.240$ and ${\cal P}^{s \bar s}_5 = 0.024$ for the initial scale
$\mu = 0.5~\rm{GeV}$, or ${\cal P}^{u \bar u}_5 = 0.162$, ${\cal P}^{d
  \bar d}_5 = 0.280$ and ${\cal P}^{s \bar s}_5 = 0.029$ for $\mu =
0.3~\rm{GeV}$. The results agree reasonably well with ${\cal P}^{u
  \bar u}_5 = 0.098$, ${\cal P}^{d \bar d}_5 = 0.216$ and ${\cal P}^{s
  \bar s}_5 = 0.057$ evaluated by the extended chiral constituent
quark model~\cite{an12}.

\subsubsection{Connected and disconnected sea}
\label{sec:CSDS}

\begin{figure}[H]
\centering
\subfigure[]
{{\includegraphics[width=0.2\textwidth]{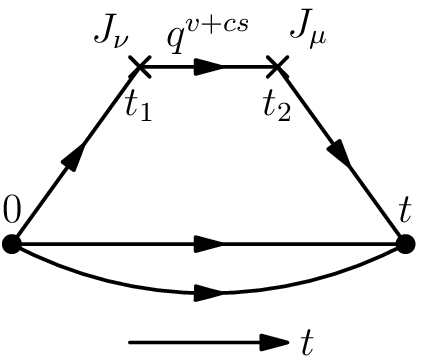}}
\label{val+CS}}
\subfigure[]
{{\includegraphics[width=0.2\textwidth]{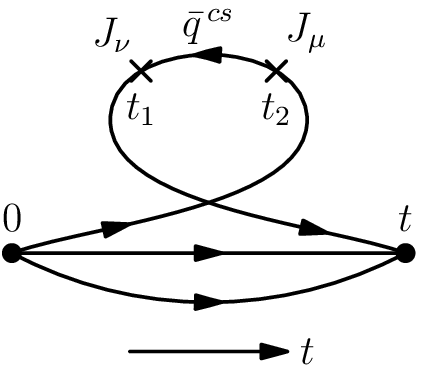}}
\label{CS}}
\subfigure[]
{{\includegraphics[width=0.2\textwidth]{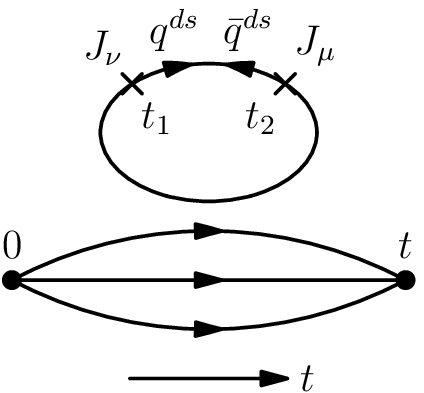}}
\label{DS}}
\caption{Three gauge invariant and topologically distinct diagrams
in the Euclidean path-integral
formalism of the nucleon hadronic tensor in the large momentum frame.
In between the currents
at $t_1$ and $t_2$, the parton degrees of freedom are
(a) the valence and CS partons $q^{v+cs}$, (b) the CS antipartons
$\bar{q}^{cs}$, and (c) the DS partons $q^{ds}$ and
antipartons $\bar{q}^{ds}$ with $q = u, d, s,$ and $c$. Only $u$
and $d$ are present in (a) and (b).}
\label{fig:CSDS}
\end{figure}

According to the Euclidean path-integral formalism of the hadronic
tensor, there are two distinct sources for the nucleon sea: the
connected sea (CS) and the disconnected sea (DS). The existence of the
connected sea and disconnected sea can be illustrated in three gauge
invariant and topologically distinct diagrams, as shown in
Fig.~\ref{fig:CSDS}. The various lines in Fig.~\ref{fig:CSDS}
represent the quark propagators from the source of the nucleon
interpolation field at time $t=0$ to the sink time at $t$ and the
currents are inserted at $t_1$ and $t_2$. These two components are
expected to have different $x$ distributions, as well as quark-flavor
dependence. Since there exists $u$ and $d$ valence quarks in the
nucleons, the $u$ and $d$ have both the CS and DS while $s$ and $c$
have only the DS. The small-$x$ behavior is scaled as $x^{-1/2}$ for
the valence and CS, and $x^{-1}$ for the DS.

It is clear from Fig.~\ref{fig:CSDS} that the two sources of the sea
quarks, CS and DS, have interesting quark-flavor dependence. For
example, while $u$ and $d$ have both CS and DS, $s$ and $c$ have only
DS. The small mass difference between the $u$ and $d$ quarks implies
that the DS cannot account for the large $\bar d / \bar u$
difference. Accordingly the flavor-nonsinglet quantity $x(\bar d(x) -
\bar u(x))$ is mainly contributed by the CS component. A lattice
calculation of the ratio $R$ of momentum fraction $\langle x \rangle$
for the strange and $u(d)$ in the disconnected insertion moment is
found to be 0.857(40)~\cite{doi08}. This result is utilized to derive
the DS component of $x (\bar u(x) + \bar d(x))$ as ${1 \over R} x(s(x)
+ \bar s(x))$ from HERMES's results of $x(s(x) + \bar
s(x))$~\cite{kfliu12}.

\begin{figure}[H]
\centering
\subfigure[]
{{\includegraphics[width=0.45\textwidth]{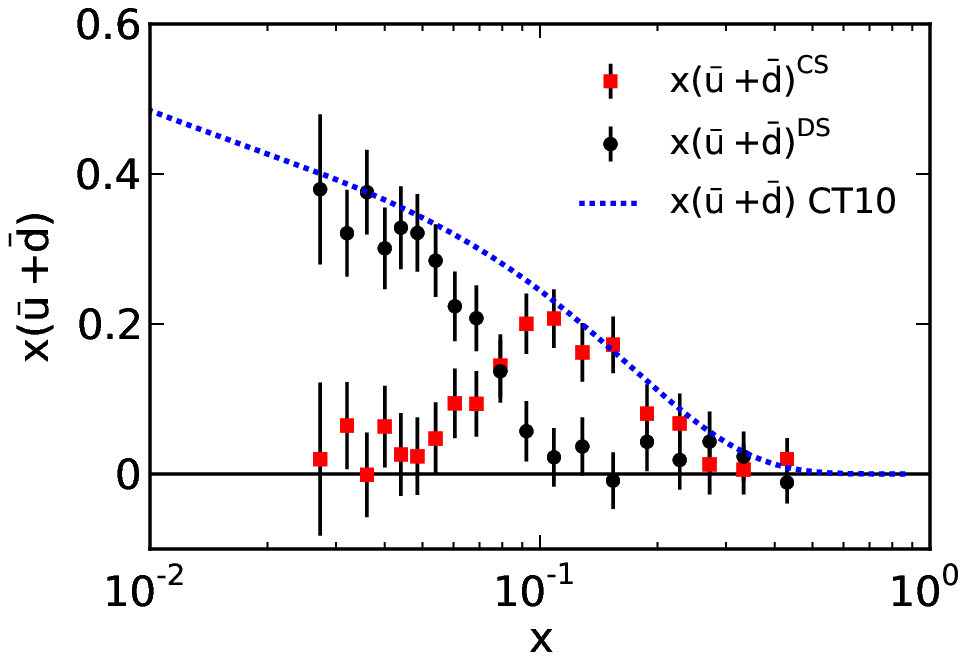}}
\label{fig:udbar}}
\subfigure[]
{{\includegraphics[width=0.45\textwidth]{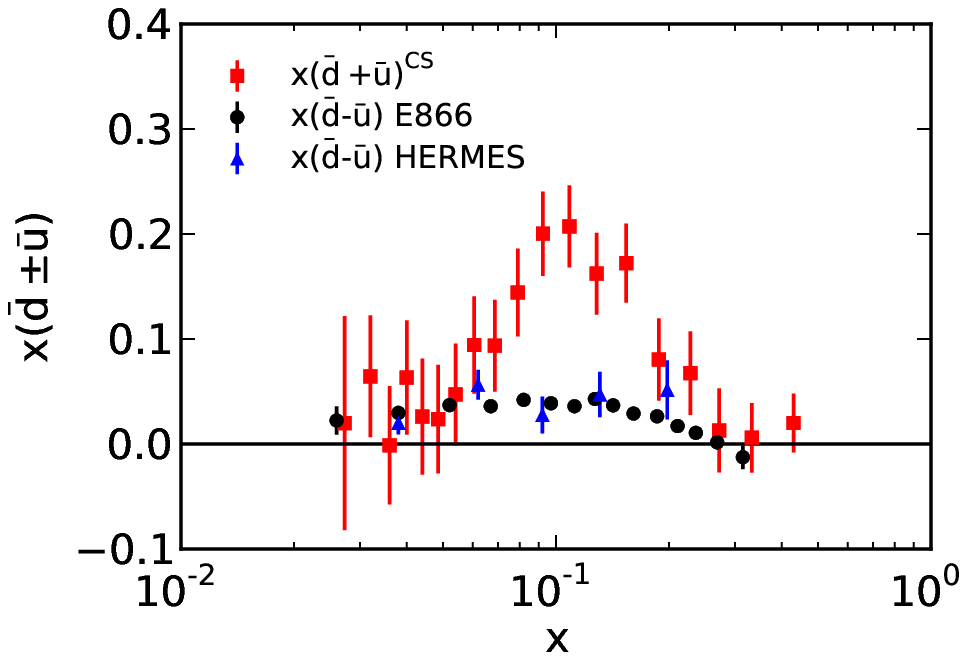}}
\label{fig:cspm}}
\caption{$x(\bar{u}^{cs}(x) + \bar{d}^{cs}(x))$ plotted together with
  (a) $x(\bar{u}(x) + \bar{d}(x))$ from CT10 and
  $\frac{1}{R}x(s(x)+\bar{s}(x))$ which is taken to be
  $x(\bar{u}^{ds}(x)+\bar{d}^{ds}(x))$ and (b) $x(\bar{d}(x) -
  \bar{u}(x))$ from E866 Drell-Yan experiment~\protect{\cite{E866}}
  and from SIDIS HERMES experiment~\protect{\cite{hermes98}}}
\end{figure}

The CS component of $u,d$ sea could be obtained by the subtraction of
the derived DS component from the CT10~\cite{CT10} results of $u,d$
sea as shown in Fig.~\ref{fig:udbar}. In Fig.~\ref{fig:cspm}, the
derived CS component of $u,d$ sea is plotted together with two data
sets of $x(\bar d(x) - \bar u(x))$ from E866 at $Q^2 = 54$ GeV$^2$ and
from HERMES at $<Q^2> = 2.3$ GeV$^2$. The valence-like shape of these
two distributions lends support to the lattice interpretation on the
origins of sea quarks. Since CS evolves like the valence, it is
suggested to have the CS and DS separately accommodated in the
extended QCD evolution to facilitate the global
fitting~\cite{kfliu12}.

\subsubsection{Comparison between two approaches}

The above two approaches take apparently different views of the origin
of sea quarks in interpreting the same data set and it is
interesting to compare them. The large $\bar d / \bar u$ difference
observed in the DIS and Drell-Yan experiments is considered to
originate primarily from the CS or intrinsic component of sea quark,
both of which are of non-perturbative nature. The structure of $x$
distribution is viewed as composed of both extrinsic (at small-$x$)
and intrinsic (at large-$x$) components. The CS component is mostly
associated with the intrinsic sea while the DS component contains both
intrinsic and extrinsic sea.

\subsection{Heavy Quarks Sea}

The existence of heavy quarks sea in the PDF of the proton at initial
scale other than those arising perturbatively through gluon splitting
in the DGLAP evolution, is a long standing issue in high-energy
physics. The charm structure functions $F_{2c}^{\gamma p}$ in the
relatively large-$x$ region measured by the EMC
collaboration~\cite{EMC1,EMC2,EMC3,EMC4,EMC5} were inconsistent with
the sole production of charm by photon-gluon
fusion~\cite{Vogt00}. This brought the speculation about the charm of
nonperturbative origin, so-called ``intrinsic charm'' (IC). Two
nonperturbative models of charm are suggested, the light-cone
five-quark BHPS model~\cite{Vogt00} and the virtual meson cloud model
for the proton wave function~\cite{pumplin06,hobbs14}.

This question is addressed quantitatively by the CTEQ Collaboration by
examining all relevant hard-scattering data in the global analysis of
CTEQ6.5c~\cite{pumplin07}. Three types of parametrization of charm
quarks originating from various scenarios were used at the initial
scales:
\begin{itemize} 

\item (a) the light-front five-quark BHPS model
\begin{equation}
c(x) = \bar c(x) = A x^2[6x(1+x)\ln x + (1-x)(1+10x+x^2)] 
\end{equation} 

\item (b) virtual meson-cloud model 
\begin{equation}
  c(x) = A x^{1.897}(1-x)^{6.095} \mbox{ and, } \bar c(x) = \bar A
  x^{2.511}(1-x)^{4.929}
\end{equation}
and 
\item (c) perturbatively sea-like, i.e. similar to that of the light quarks sea.
\begin{equation}
c(x) = \bar c(x) = A [\bar u(x) + \bar d(x)]
\end{equation} 

\end{itemize}
where $A$ and $\bar A$ are normalization constants. The
goodness-of-fit in the global analysis shows improvement with the
inclusion of the additional charm component at the initial scale.

\begin{figure}[H]
\resizebox{0.32\textwidth}{!}{
\includegraphics[clip=true,scale=0.32]{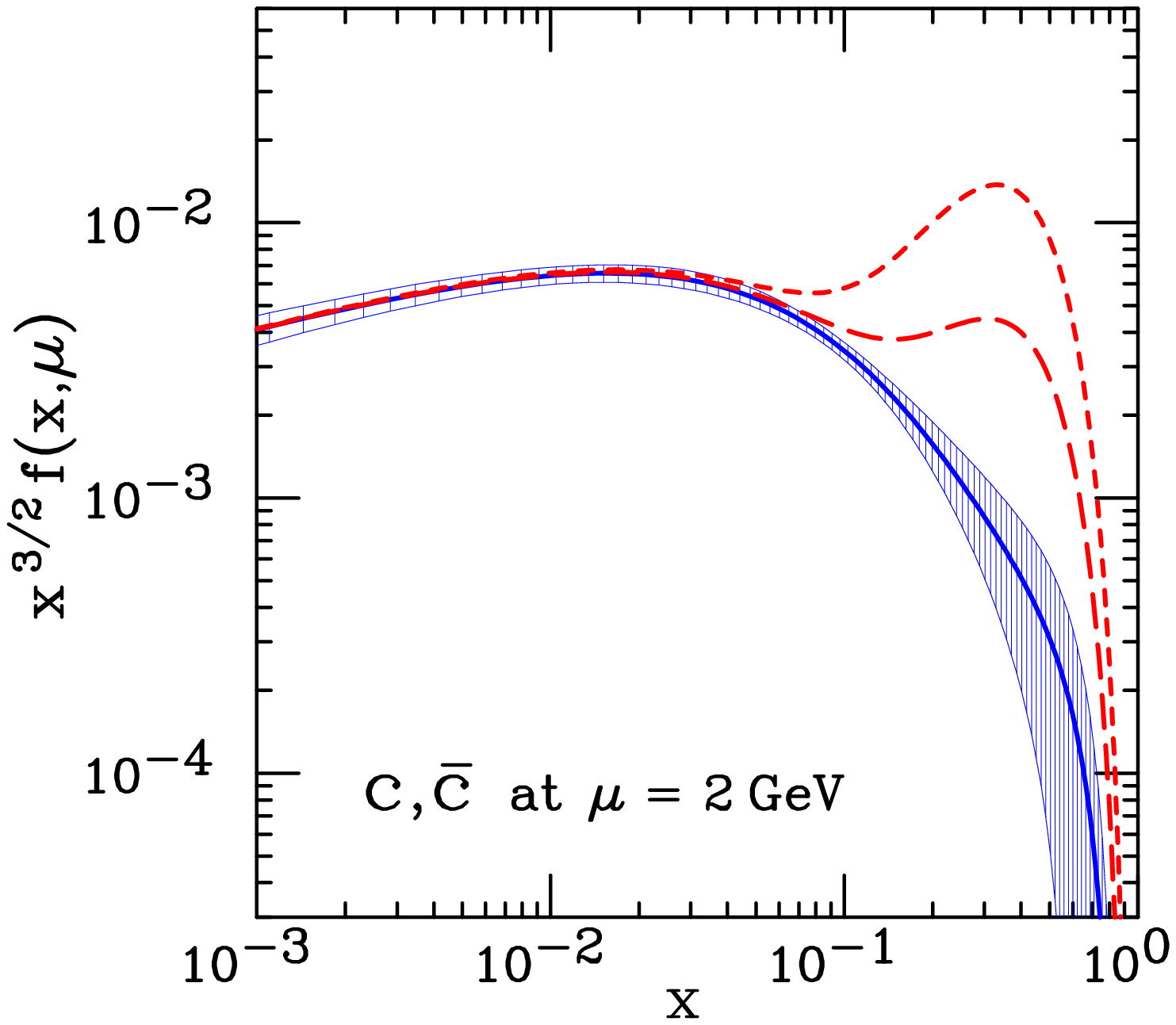}
}
\hfill
\resizebox{0.32\textwidth}{!}{
\includegraphics[clip=true,scale=0.32]{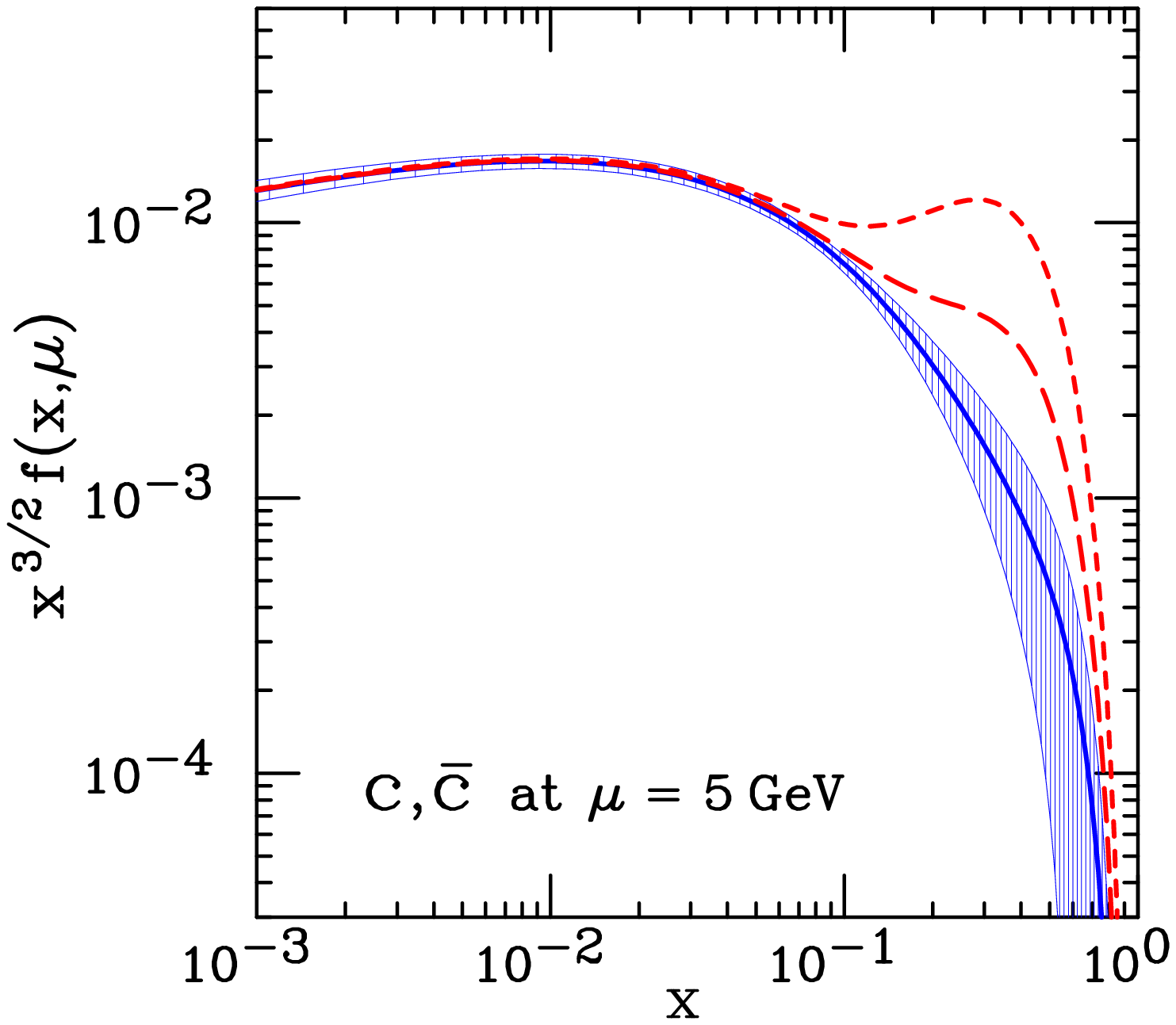}
}
\hfill
\resizebox{0.32\textwidth}{!}{
\includegraphics[clip=true,scale=0.32]{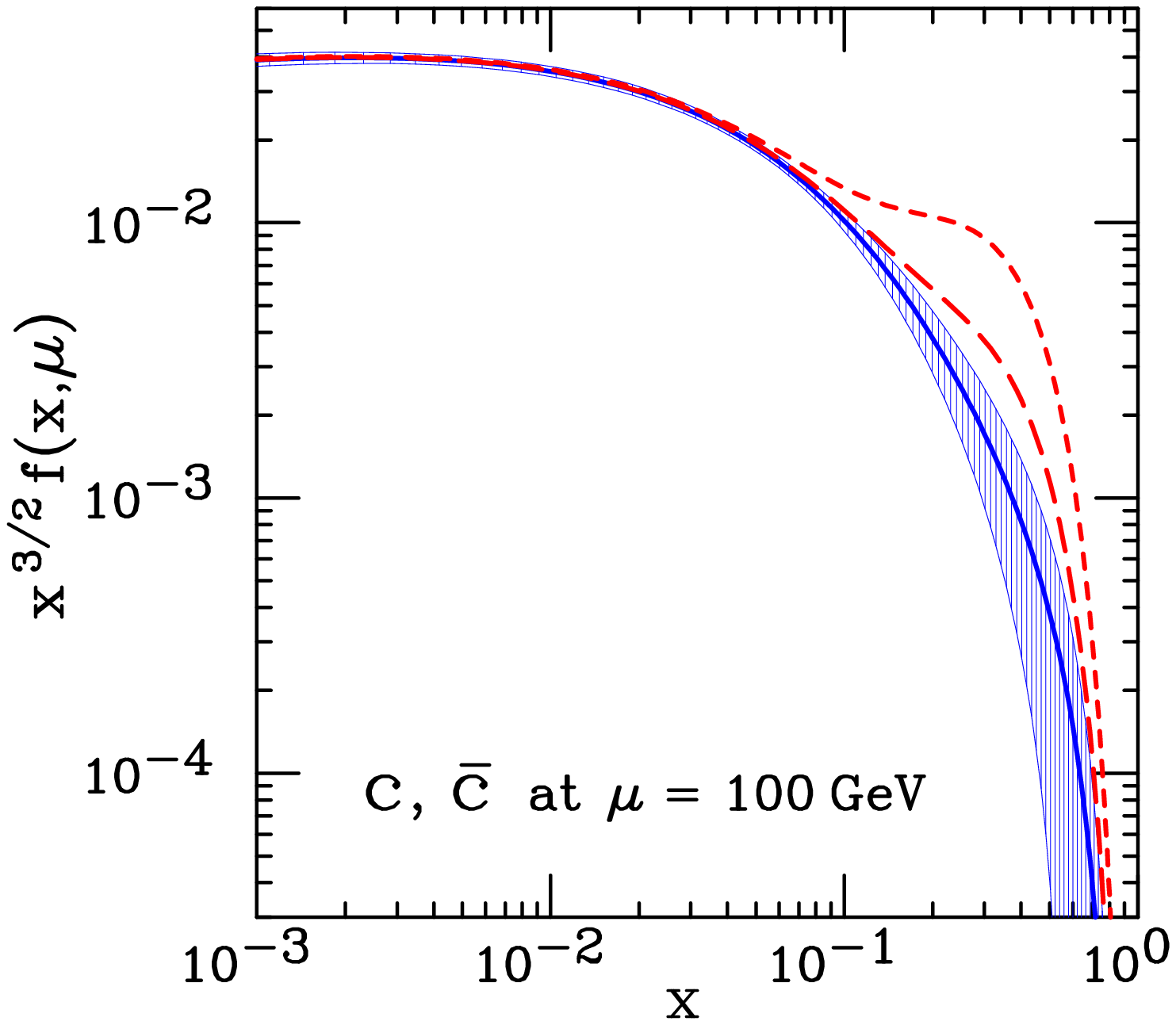}
} 
\caption{Charm quark distributions from the BHPS IC model. The three
  panels correspond to scales $\mu = 2$, $\mu = 5$, and $\mu = 100 \,
  \mathrm{GeV}$. The long-dash (short-dash) curve corresponds to
  $\langle x \rangle_{c + \bar{c}} = 0.57\%$ ($2.0\%$). The solid
  curve and shaded region show the central value and uncertainty from
  CTEQ6.5, which contains no IC. Figures
  from~\cite{pumplin07}.} \label{fig:IC_BHPS}
\end{figure}

\begin{figure}[H]
\resizebox{0.32\textwidth}{!}{
\includegraphics[clip=true,scale=0.32]{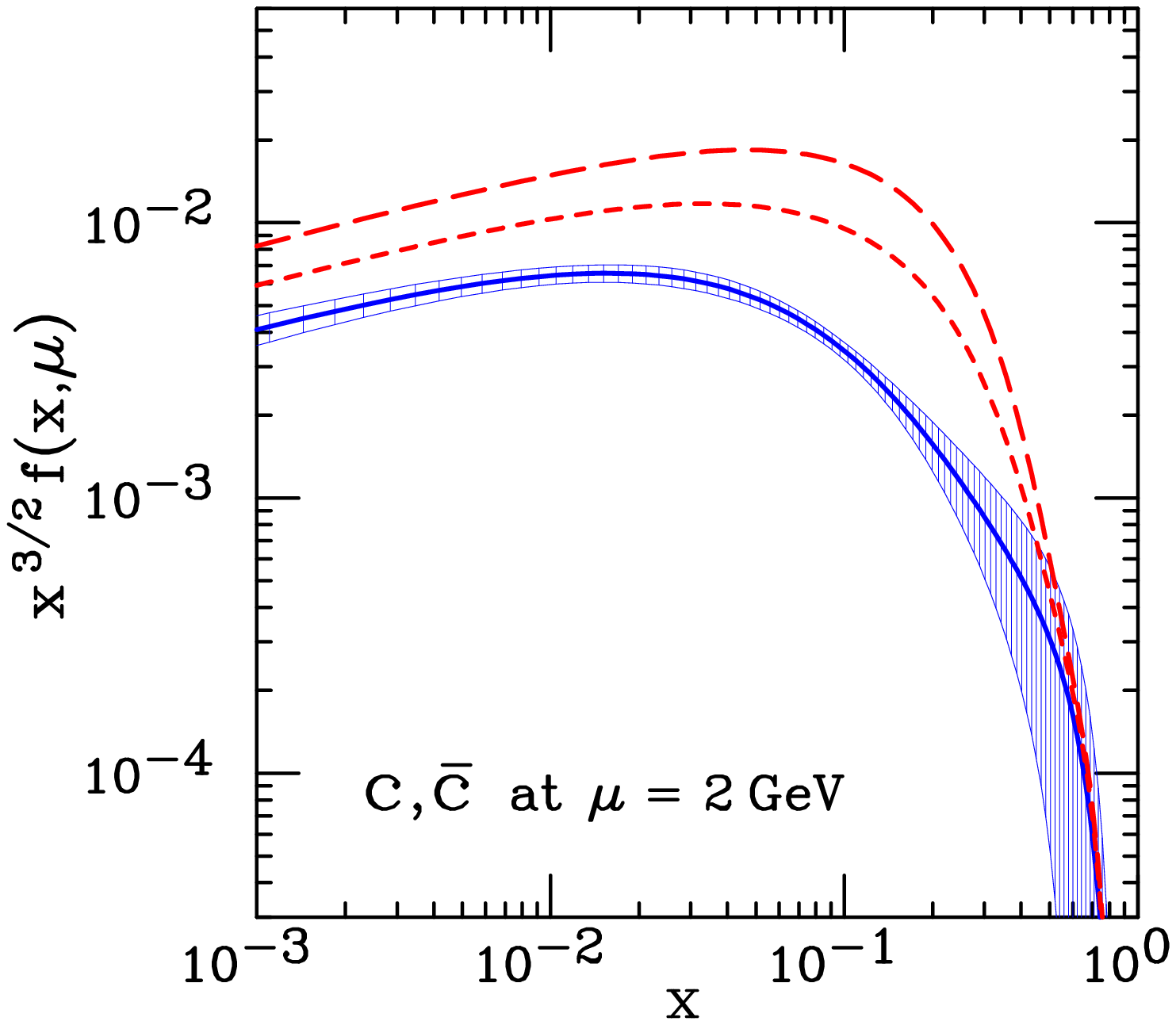}
}
\hfill
\resizebox{0.32\textwidth}{!}{
\includegraphics[clip=true,scale=0.32]{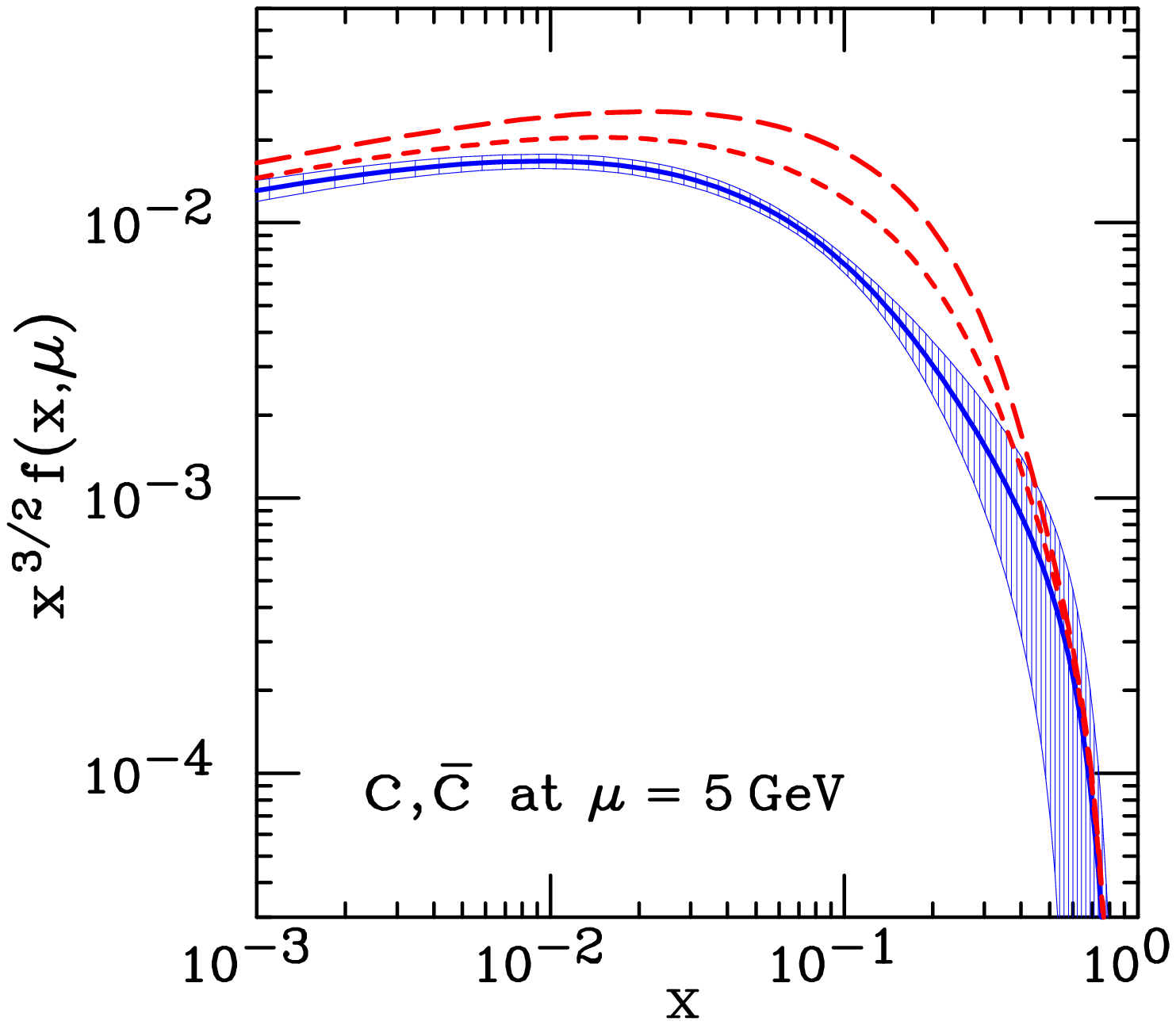}
}
\hfill
\resizebox{0.32\textwidth}{!}{
\includegraphics[clip=true,scale=0.32]{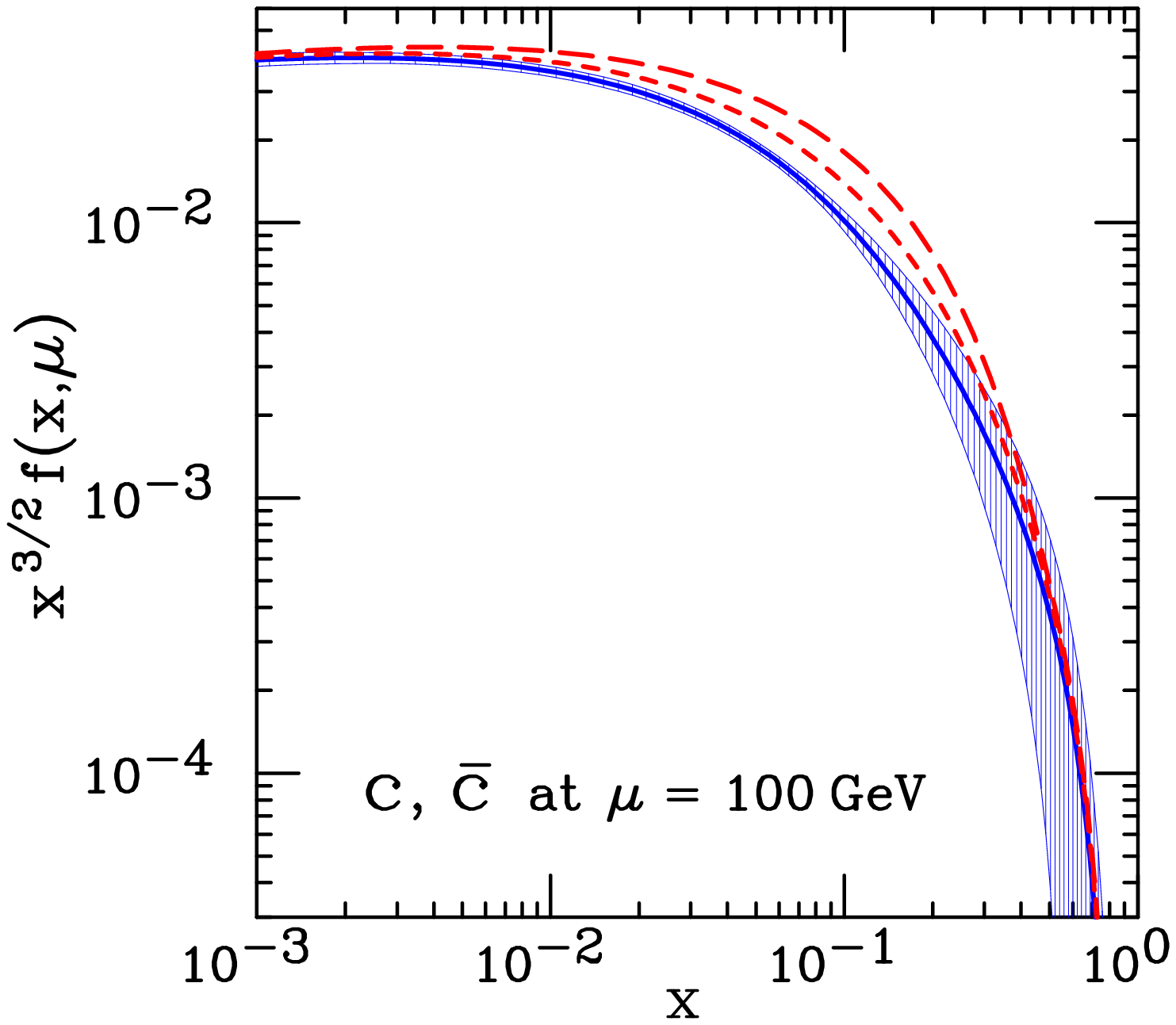}
} 
\caption{ Same as Fig.\,\ref{fig:IC_BHPS}, except for the sea-like
  scenario. The long-dash (short-dash) curves correspond to $\langle x
  \rangle_{c + \bar{c}} = 2.4\%$ ($1.1\%$). Figures
  from~\cite{pumplin07}.} \label{fig:IC_sealike}
\end{figure}

The magnitude of intrinsic charm is characterized by the momentum
fraction $\langle x \rangle_{c + \bar{c}} $ ($\equiv \int_0^1
x[c(x)+\bar c(x)]dx$) carried by charm at the initial scale $\mu$ =
1.3 GeV. For the shape suggested by non-perturbative light-cone BHPS
model, the data are consistent with a wide range of the intrinsic
charm magnitude, ranging from null to 2-3 times larger than the
estimate by the BHPS model as shown in Fig.~\ref{fig:IC_BHPS}. The
marginally allowed amount of IC is $\langle x \rangle_{c + \bar{c}} =$
0.020. A salient feature is that there could be a large enhancement of
charm at $x>0.1$ relative to the PDF assuming no IC even at a scale as
large as $\mu$ = 100 GeV. Fig.~\ref{fig:IC_sealike} shows the result
for an assumed shape of IC similar to other sea quarks; the maximum
amount of IC is $\langle x \rangle_{c + \bar{c}} =$ 0.024 and the
enhancement of charm shows up in a more broad range of $x$, $0.01 < x
<0.50$.

\begin{figure}[H]
\begin{center}
\includegraphics[width=0.3\textwidth]{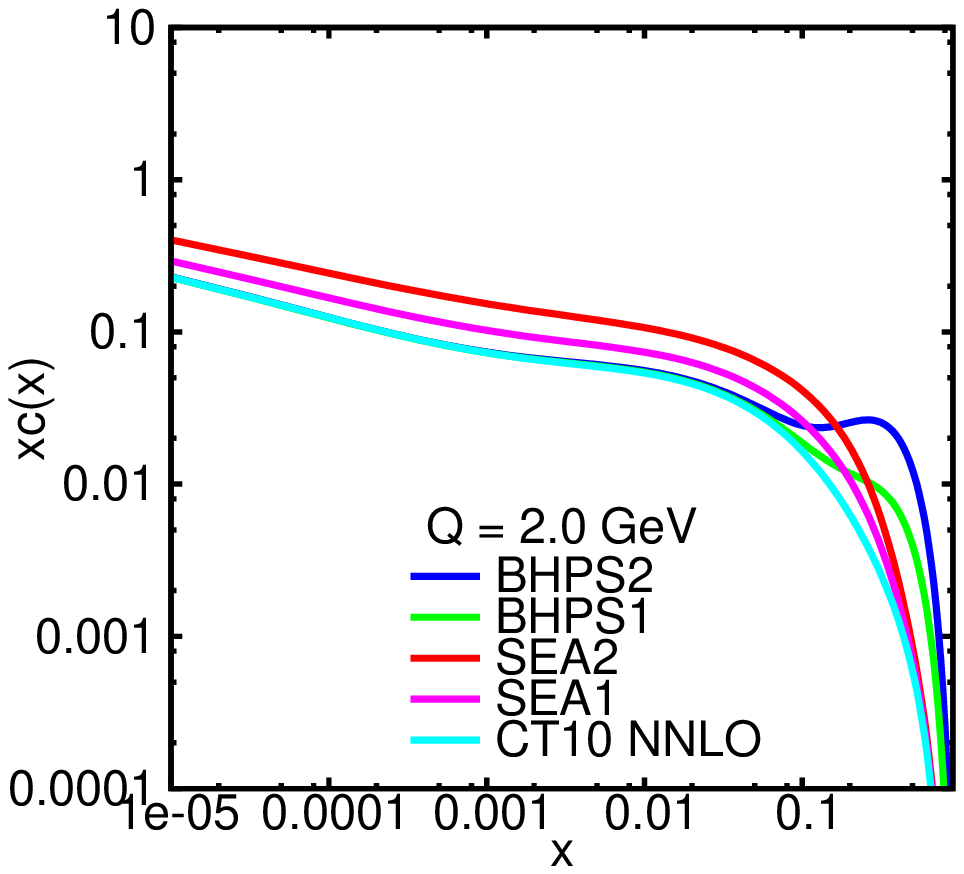}\hspace{.1in}
\includegraphics[width=0.3\textwidth]{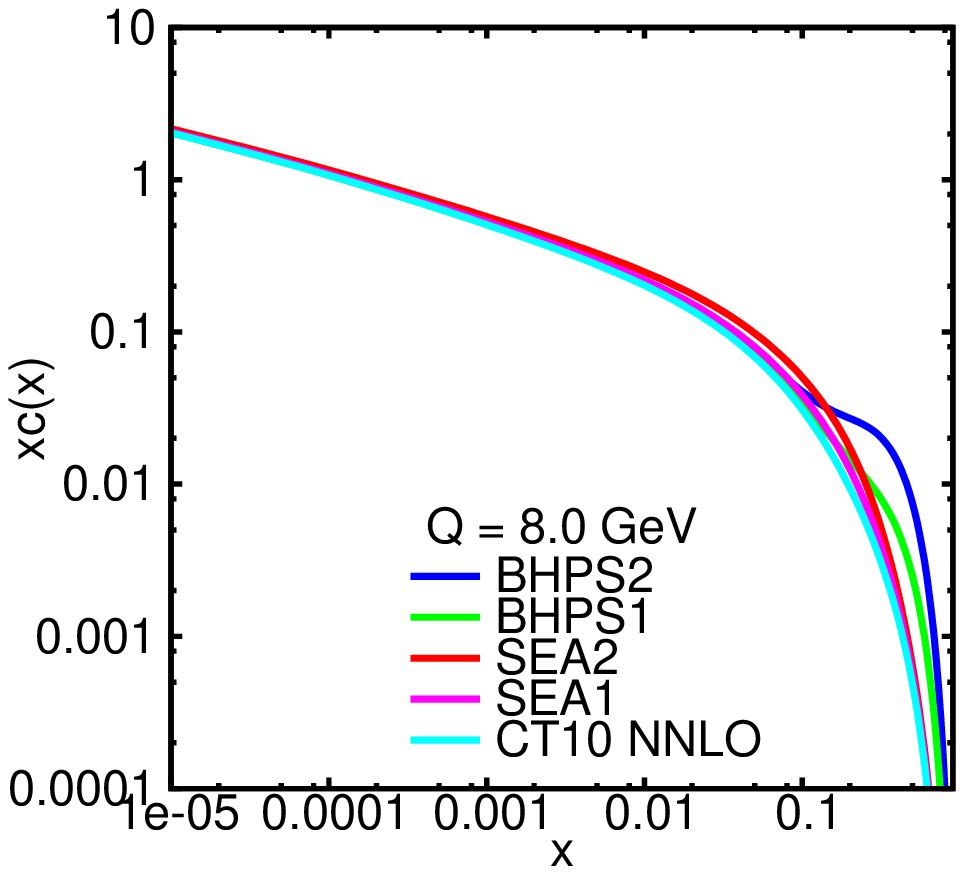}\hspace{.1in}
\includegraphics[width=0.3\textwidth]{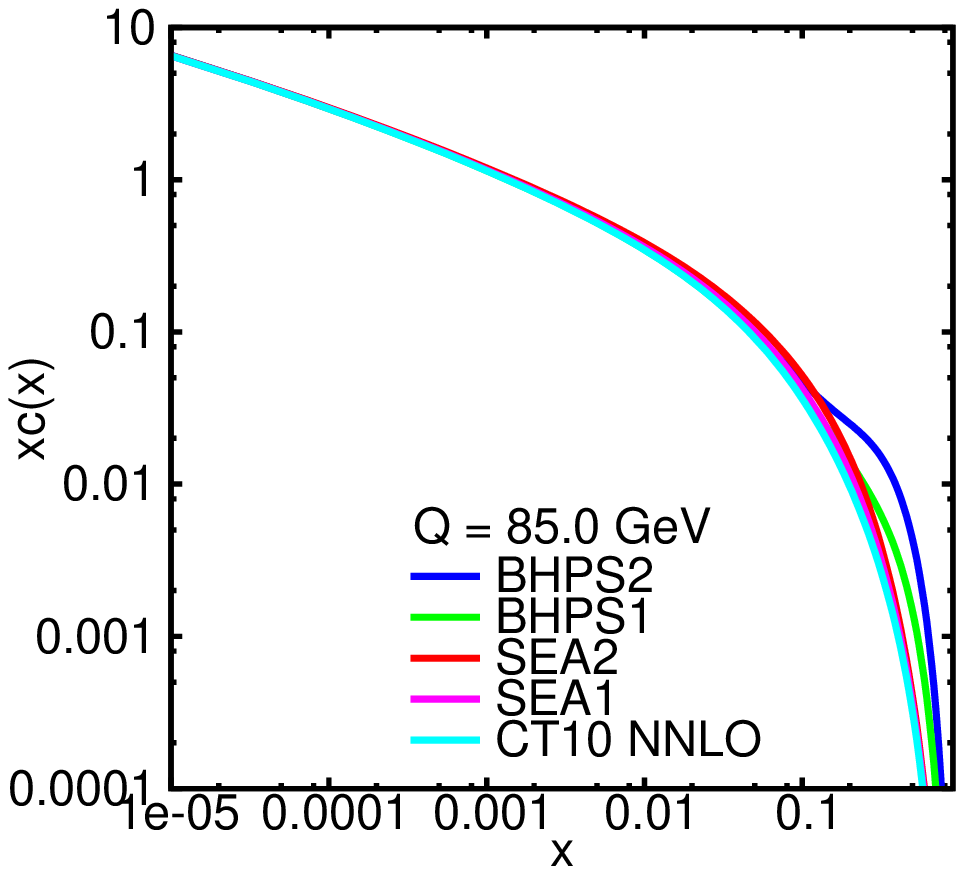}
\end{center}
\caption{Charm quark distribution $x\,c(x,Q)$ from the BHPS1 and BHPS2
  PDFs (which have $0.57\%$ and $2\%$ $\langle{x}\rangle_{\rm IC}$);
  from SEA1 and SEA2 PDFs (which have $0.57\%$ and $1.5\%$
  $\langle{x}\rangle_{\rm IC}$); and from CT10. The corresponding
  energy scale $Q$ of these three graphs is 2.0, 8.0, and 85 GeV
  respectively. Figures from~\cite{cteq-tea13}.}
\label{fig:cteq-tea}
\end{figure}

Very recently, CTEQ-TEA group reported an updated study of IC in the
structure function of the proton with CT10 NNLO global
analysis~\cite{cteq-tea13}. Besides the advances in the theoretical
approach, more relevant data set is included: the combined H1 and ZEUS
data for DIS~\cite{HERA1} and inclusive charm production~\cite{HERA2}
new data set are found to constrain the allowance of the sea-like IC
more strongly than other data. Figure~\ref{fig:cteq-tea} shows the
charm parton density distribution in the new global fits and the
conclusion is that a reasonable global analysis could still be done
without an IC component of proton but a small IC component with
$\langle x \rangle_{IC} \leq 2.5 \%$ for valence-like BHPS IC or
$\langle x \rangle_{IC} \leq 1.5 \%$ for sea-like IC cannot be ruled
out. Again, this calls for more relevant data to pin down the
existence of IC.

\section{Polarized Distributions of Sea Quarks}
\label{sec:polsea}

There are several important motivations for measuring and
understanding the helicity distributions of sea quarks in
the nucleons. First, the discovery of the spin puzzle at
SLAC and NMC showed clearly that the expectation from naive
SU(3) that quarks provide essentially all the spin of 
the nucleon is invalid. This led to extensive theoretical and
experimental efforts during the last several decades to understand the
decomposition of the proton's spin in terms of the quark/antiquark
spin, gluon spin, and quark/gluon orbital angular momenta.
While the recent polarized semi-inclusive DIS and 
$W$ production in polarized $pp$ collision suggest a relatively
small contribution of sea quarks to the proton's spin, the $x$ and 
flavor dependence of the antiquark helicity distributions are 
still poorly known. Second, as discussed below, many theoretical
models which are capable of explaining the flavor structure of
the unpolarized nucleon sea also have specific predictions on the
helicity distributions of the nucleon sea. While these models 
have very similar predictions for the unpolarized sea, they often
differ significantly on their predictions of the polarized sea.
Therefore, a stringent test of the various theoretical models 
could be provided by more precise information on the sea-quark
polarizations. Finally, the advent of the lattice QCD now offers the
possibility to calculate the $x$ distributions of the sea-quark
polarizations, and these calculations could be tested against the 
experimental data in the future.

In this Section we first discuss the subject of the helicity
distributions of the light-quark sea, namely $\Delta {\bar u}$ and
$\Delta {\bar d}$. We then examine the current knowledge on
the polarization of strange-quark sea, followed by a summary of
future prospects.

\subsection{Asymmetry Between $\Delta \bar u(x)$ and 
$\Delta \bar d(x)$}

The surprisingly large flavor asymmetry between the $\bar u(x)$
and $\bar d(x)$ naturally leads to the question whether the 
polarized $\bar u$ and $\bar d$ seas are also asymmetric. 
The flavor and the spin structures of the nucleon sea are 
closely connected in the sense that many theoretical models, 
originally proposed to explain the $\bar d / \bar u$ 
flavor asymmetry, also have specific implications for the spin 
structure of the nucleon sea. We now briefly summarize the 
predictions from various theoretical models on the flavor asymmetry
of the light-quark helicity distributions.

\subsubsection{Theoretical predictions on the asymmetry between 
$\Delta \bar u(x)$ and $\Delta \bar d(x)$}

\begin{itemize}

\item In the pion-cloud model, a pion is emitted from the proton in a p-wave
state to conserve parity. This reduces the spin projection of the
baryons ($N$ or $\Delta$) in the $\pi-N$ or $\pi - \Delta$ 
configurations along the initial proton spin direction. Therefore,
pion cloud would lead to a reduction of the proton's spin 
residing in the quark's spin in qualitative agreement with the
experimental results. This implies that a fraction of the proton's spin 
would reside in the orbital angular momentum of the pion-baryon 
configuration. Moreover, the antiquarks ($\bar u$ and $\bar d$)
in the pion-baryon configurations are unpolarized due to the spin-0
nature of pion. Therefore, the pion-cloud model would predict
that $\Delta \bar u = \Delta \bar d = 0$. An extension of the meson
cloud model to include the kaon-hyperon configuration would lead to the
conclusion that $\bar s$ quarks, which are present in the spin-0 kaon
cloud, are unpolarized. On the other hand, the strange quarks ($s$)
residing in the hyperons would be polarized. By extending the meson
cloud model to include a vector meson ($\rho$) cloud,
non-zero $\bar u, \bar d$ sea quark polarizations with
$\Delta \bar d - \Delta \bar u > 0$
were predicted~\cite{fries98,boreskov99,cao01,kumano01}.

\item In the chiral-quark model, a quark would undergo a spin flip 
upon an emission of a pseudoscalar meson 
($ u^\uparrow \to \pi^\circ (u \bar u, d \bar d) + 
u^\downarrow,~u^\uparrow \to 
\pi^+ (u \bar d) + d^\downarrow,~u^\uparrow \to K^+ + s^\downarrow$, etc.). 
This model predicts that antiquarks ($\bar u, \bar d, \bar s$) are unpolarized 
($\Delta \bar u = \Delta \bar d = \Delta \bar s = 0$) since they only 
reside in spin-0 Goldstone bosons. In contrast, the strange quarks ($s$) 
would have a negative polarization since the $u$ valence quarks 
in the proton are positively polarized and
the $u^\uparrow \to K^+ + s^\downarrow$ process would lead to an excess
of $s^\downarrow$. 

\item The Pauli-blocking model~\cite{steffens} implies that an excess of 
$q^\uparrow (q^\downarrow)$ valence quarks
would inhibit the creation of a pair of $q^\uparrow \bar q^\downarrow$
($q^\downarrow \bar q^\uparrow$) sea quarks. Since the polarization
of the $u$($d$) valence quarks in the proton is positive (negative), 
the Pauli-blocking 
model predicts a positive (negative) polarization for the $\bar u
(\bar d)$ sea $(\Delta \bar u > 0 > \Delta \bar d)$.

\item In the instanton model~\cite{dorokhov}, scattering of a 
valence quark off a nonperturbative fluctuation of the gluon field, i.e.
instanton, can result in a quark-antiquark pair. The instanton-induced 
interaction is described by the 't Hooft effective 
Lagrangian, which allows processes such as 
$u^\uparrow \to u^\downarrow d^\uparrow \bar d^\downarrow$, $d^\downarrow 
\to d^\uparrow u^\downarrow \bar u^\uparrow$. Since the flavor 
of the quark-antiquark pair
generated in this process is different from that of the initial valence
quark, the instanton model readily explains $\bar d > \bar u$. 
Furthermore, the correlation between the helicities of the sea quark 
and the valence quark
in the effective Lagrangian (i.e. $u^\uparrow$ leads to a
$\bar d^\downarrow$) predicts a positively (negatively) polarized
$\bar u (\bar d)$ sea. In particular, it predicts
a large $\Delta \bar u, \Delta \bar d$ flavor asymmetry with 
$\Delta \bar u > \Delta \bar d$, namely,
$\int_0^1 [\Delta \bar u(x) - \Delta \bar d(x)] dx = 
\frac{5}{3} \int_0^1 [\bar d(x) - \bar u(x)] dx$~\cite{dorokhov01}.

\item In the chiral-quark soliton model~\cite{diakonov96,wakamatsu98}, 
QCD at the large $N_c$ limit becomes an effective theory of 
mesons with the baryons appearing as solitons. Quarks are described 
by single particle wave 
functions, which are solutions of the Dirac equation in the field of
the background pions. In this model, 
the unpolarized isovector
distributions $\bar u(x) - \bar d(x)$ appear in next-to-leading order ($N_c$)
in a $1/N_c$ expansion,
while the polarized isovector distributions
$\Delta \bar u(x) - \Delta \bar d(x)$ appears in leading-order ($N_c^2$).
Therefore, this model predicts a large flavor asymmetry for the polarized sea
$[\Delta \bar u (x) - \Delta \bar d(x)] > [\bar d(x) - \bar u(x)]$.

\item In the statistical model~\cite{bourrely01,bhalerao00}, the 
nucleon is considered as a collection of
massless quarks, antiquarks, and gluons in thermal equilibrium within
a finite volume. The momentum distributions for quarks and
antiquarks follow a Fermi-Dirac function characterized
by a common temperature and a chemical potential $\mu$ which depends on
the flavor and helicity of the quarks. It can be shown that
\begin{equation}
\mu_{\bar q\uparrow} = - \mu_{q\downarrow};~\mu_{\bar q\downarrow} =
- \mu_{q\uparrow}.
\label{eq:helicity0}
\end{equation}
\noindent Equation~\ref{eq:helicity0}, together with the constraints 
of the valence quark
sum rules and inputs from polarized DIS experiments, can readily lead 
to the prediction that $\bar d > \bar u$ and $\Delta \bar u > 0 > 
\Delta \bar d$.
\end{itemize}
\begin{table}
\caption{Prediction of various theoretical models on the integral
$I_\Delta = \int_0^1 [\Delta \bar u(x) - \Delta \bar d(x)] dx$.}
\begin{center}
\label{tab:del_ubar-dbar}       
\begin{tabular}{lll}\hline\noalign
{\smallskip} Model & $I_\Delta$ prediction & Ref.  \\
\noalign{\smallskip}\hline\noalign{\smallskip}
Meson cloud ($\pi$-meson) & 0 & \cite{thomas,ehq} \\
Meson cloud ($\rho$-meson) & $\simeq -0.0007$ to $-0.027$ & \cite{fries98} \\
Meson cloud ($\pi-\rho$ interf.) & $= -6 \int_0^1 g^p(x) dx$ & 
\cite{boreskov99} \\
Meson cloud ($\rho$ and $\pi-\rho$ interf.) & $\simeq -0.004$ to $-0.033$ & 
\cite{cao01} \\
Meson cloud ($\rho$-meson) & $< 0$ & \cite{kumano01} \\
Meson cloud ($\pi-\sigma$ interf.) & $\simeq 0.12$ & \cite{fries02} \\
Pauli-blocking (bag-model) & $\simeq 0.09$ & \cite{cao01} \\
Pauli-blocking (ansatz) & $\simeq 0.3$ & \cite{gluck00} \\
Pauli-blocking & $= \frac{5}{3} \int_0^1 [\bar d(x) - \bar u(x)] dx \simeq 0.2$ 
& \cite{steffens02} \\
Chiral-quark soliton & 0.31 & \cite{dressler98} \\
Chiral-quark soliton & $\simeq \int_0^1 2x^{0.12} 
[\bar d(x) - \bar u(x)] dx$ & \cite{wakamatsu99} \\
Instanton & $= \frac{5}{3} \int_0^1 [\bar d(x) - \bar u(x)] dx \simeq 0.2$ 
& \cite{dorokhov01} \\
Statistical & $\simeq \int_0^1 [\bar d(x) - \bar u(x)] dx \simeq 0.12$ 
& \cite{bourrely01} \\
Statistical & $> \int_0^1 [\bar d(x) - \bar u(x)] dx > 0.12$ & 
\cite{bhalerao00} \\
\noalign{\smallskip}\hline
\end{tabular}
\end{center}
\end{table}

Predictions of various model calculations for $I_\Delta$,
the first moment of $\Delta \bar u(x) - \Delta \bar d(x)$, 
are listed in Table~\ref{tab:del_ubar-dbar}. 
While the meson cloud model gives small negative values for
$I_\Delta$, all other models predict a positive $I_\Delta$ with
a magnitude comparable or greater than the corresponding 
integral for unpolarized sea (recall that $\int_0^1 [\bar d(x) - 
\bar u(x)] dx \simeq 0.12$). 
Several calculations for the direct contribution of
$\rho$ meson cloud are in good agreement. However, the large $\pi - \rho$
interference effect reported in ~\cite{boreskov99} was not confirmed
in a later study~\cite{cao01}. It is worth noting that 
Ref.~\cite{fries02} 
considers $\pi - \sigma$ interference and predicts a large effect on 
$\Delta \bar d - \Delta \bar u$, with a sign opposite to other meson cloud
model calculations.

If the flavor asymmetry of the polarized sea is indeed as large as the
predictions of many models shown in Table~\ref{tab:del_ubar-dbar},
it would imply
that a significant fraction of the Bjorken sum,
$\int _0^1 [g^p_1(x) - g^n_1(x)] dx$,
comes from the flavor asymmetry of polarized nucleon sea.

\subsubsection{Experimental status on the asymmetry between
$\Delta \bar u(x)$ and $\Delta \bar d(x)$}

The experimental information on the helicity distributions of the partons
is mostly from polarized DIS experiments involving longitudinally 
polarized lepton beams and polarized targets. From the measured 
longitudinal ($A_\parallel$) and transverse ($A_\perp$) asymmetries
\begin{equation}
A_\parallel = \frac{d\sigma^{\rightarrow \Rightarrow} - d\sigma^{\rightarrow
\Leftarrow}}{d\sigma^{\rightarrow \Rightarrow} + d\sigma^{\rightarrow
\Leftarrow}}; A_\perp = \frac{d\sigma^{\rightarrow \Uparrow} - 
d\sigma^{\rightarrow \Uparrow}} {d\sigma^{\rightarrow \Uparrow} +
d\sigma^{\rightarrow \Uparrow}},
\end{equation}
the polarized structure function $g_1(x,Q^2)$ can be extracted.
In the parton model, $g_1(x,Q^2)$ at LO is given as 
\begin{equation}
g_1(x,Q^2) = \frac{1}{2} \sum_{i=1}^{n_f} e^2_i (\Delta q_i(x,Q^2)
+ \Delta \bar q_i(x,Q^2)),
\end{equation}
where $\Delta q_i(x,Q^2) = q^\uparrow_i (x,Q^2) - q^\downarrow_i(x,Q^2)$
and $\Delta \bar q_i(x,Q^2) = \bar q^\uparrow_i (x,Q^2) - 
\bar q^\downarrow_i(x,Q^2)$. At NLO, the $Q^2$ dependence of $g_1(x,Q^2)$ 
contains a contribution from $\Delta g (x,Q^2) = g^\uparrow (x,Q^2) 
- g^\downarrow (x,Q^2)$ through the term $\frac{\alpha_s}{2\pi} \Delta
C_g \otimes g$, and can be used to extract gluon polarization.

Extensive measurements of polarized DIS have been carried out at 
EMC~\cite{ashman89}, SMC~\cite{adeva98PRD}, 
SLAC~\cite{e142,e143,e154,e155a,e155b}, 
COMPASS~\cite{alexakhin07,alekseev10}, 
HERMES~\cite{ackerstaff99,airapetian07}, and JLAB~\cite{zheng04,clas06}.
These data have led to the determination of $\Delta u(x) + \Delta \bar u(x),
\Delta d(x) + \Delta \bar d(x), \Delta s(x) + \Delta \bar s(x)$, and
$\Delta g(x)$ in several global QCD analyses in the next-to-leading (NLO)
framework~\cite{grsv,bb02,bb10,lss07,lss10,acc06,acc09,soffer02,nnpdf13}.
These neutral current polarized DIS data on proton, deuteron, and
$^3$He allow the isospin separation of the polarized parton distributions 
(i.e., separating $\Delta u + \Delta \bar u$ from $\Delta d + \Delta \bar d$).
However, these data cannot disentangle the antiquark from the quark
helicity distributions. In order to extract the sea-quark polarizations,
several other experimental approaches have been considered. They include
the polarized semi-inclusive DIS, polarized Drell-Yan~\cite{moss93},
single-spin asymmetry in $W$-boson production in $pp$ 
collision~\cite{soffer93}, and charged-current DIS using neutrino
beam at neutrino factory~\cite{mangano01} or electron beam at
electron ion collider~\cite{boer11}. Among these various approaches, both 
the polarized SIDIS and the polarized $W$-boson production have been actively pursued in the last decades. In the following, we discuss their impact on
the determination of $\Delta \bar u(x)$ and $\Delta \bar d(x)$.

\begin{table}
\caption{Polarized SIDIS experiments and results on the first moments of
$\Delta \bar u(x)$, $\Delta \bar d(x)$, $\Delta \bar u(x) - \Delta \bar d(x)$
and $\Delta \bar u(x) + \Delta \bar d(x)$. The values of $\langle Q^2 \rangle$
for the SMC, HERMES, and COMPASS experiments are 10 GeV$^2$, 2.5 GeV$^2$,
and 3.0 GeV$^2$, respectively.}
\begin{center}
\begin{tabular}{lllcc}\hline\noalign
{\smallskip} Experiment & beam/target & det. part. 
& $\Delta \bar u$ 
& $\Delta \bar u - \Delta \bar d$ \\
{\smallskip} & & & ($\Delta \bar d$) & $(\Delta \bar u + \Delta \bar d)$ \\
\noalign{\smallskip}\hline\noalign{\smallskip}
EMC~\cite{ashman89} & $\mu$ on p & $h^\pm$ 
& $-$ & $-$ \\
SMC~\cite{adeva98} & $\mu^+$ on p, d & $h^\pm$ 
& $0.01 \pm 0.05 \pm 0.02$  & $-$ \\
 & & & ($0.01 \pm 0.14 \pm 0.12$) & \\
HERMES~\cite{hermes_pol_99} & $e^+$ on p, $^3$He & $h^\pm$ 
& $-0.01 \pm 0.02 \pm 0.03$ & $-$ \\ 
 & & & ($-0.02 \pm 0.03 \pm 0.04$) & \\
HERMES~\cite{hermes_pol_04} & $e^+$ on d & $\pi^\pm, K^\pm$ 
& $\sim 0$ & $\sim 0$ \\
HERMES~\cite{hermes_pol_05} & $e^+$ on p, d & $\pi^\pm, K^\pm$
& $-0.002 \pm 0.0036 \pm 0.023$ & $0.048 \pm 0.057 \pm 0.028$ \\ 
 & & & ($-0.054 \pm 0.033 \pm 0.011$) & \\
COMPASS~\cite{compass_pol_08} & $\mu$ on d & $h^\pm$ 
& $-$ & ($0.0 \pm 0.04 \pm 0.03$) \\
COMPASS~\cite{compass_pol_09} & $\mu$ on d & $\pi^\pm, K^\pm$
& $-$ & ($-0.04 \pm 0.03 \pm 0.01$) \\
COMPASS~\cite{compass_pol_10} & $\mu$ on p, d & $\pi^\pm, K^\pm$
& $0.02 \pm 0.02 \pm 0.01$ & $0.06 \pm 0.04 \pm 0.02$  \\
 & & & ($-0.05 \pm 0.03 \pm 0.02$) & ($-0.04 \pm 0.03 \pm 0.01$) \\
\noalign{\smallskip}\hline
\end{tabular}
\end{center}
\label{tab:del_u_sidis}       
\end{table}

The first polarized semi-inclusive DIS measurement was reported by the
EMC collaboration~\cite{ashman89}, followed by the SMC~\cite{adeva96,
adeva98} and the more recent extensive work at HERMES~\cite{hermes_pol_99,
hermes_pol_04,hermes_pol_05} and
COMPASS~\cite{compass_pol_08,compass_pol_09,compass_pol_10}.
As shown in Table~\ref{tab:del_u_sidis}, these polarized SIDIS experiments use
longitudinally polarized $e^\pm$, $\mu^\pm$ beams on a variety of 
longitudinally polarized $p, d,$ $^3$He targets. Either unidentified 
charged hadrons $(h^\pm)$ or identified $\pi^\pm$, $K^\pm$ are detected.
From the measured asymmetry $A^h_{meas}$, which is the asymmetry of the
normalized count rates when the beam and target polarization are
anti-aligned or aligned, the photo-absorption cross section asymmetry,
$A^h_1$, is determined as
\begin{equation}
A^h_1=\frac{\sigma^h_{1/2}-\sigma^h_{3/2}}{\sigma^h_{1/2}+\sigma^h_{3/2}},
\end{equation}
where $\sigma^h_{1/2(3/2)}$ refers to the semi-inclusive cross section for
producing hadron of type $h$ for photons with spin anti-parallel 
(parallel)
to the target nucleon spin. Since the spin-1 photon can only be absorbed
when the quark spin is in the direction opposite to the photon's spin
direction, the asymmetry between $\sigma^h_{1/2}$ and $\sigma^h_{3/2}$
is sensitive to the helicity distribution of the struck quark (or antiquark).
The charge and type of the detected hadron will further allow the
flavor separation of the quark helicity distribution. $A^h_1$ can be
written as follows:
\begin{equation}
A^h_1(x,Q^2,z)=\frac{\sum_q e^2_q \Delta q(x,Q^2) D^h_q(z,Q^2)}
{\sum_q e^2_q q(x,Q^2) D^h_q(z,Q^2)},
\end{equation}
where $D^h_q(z,Q^2)$ is the fragmentation function of a struck quark
with flavor $q$ hadronizing into a hadron $h$ carrying a fraction $z$ of
the virtual photon's energy.

\begin{figure}[H]
\includegraphics[width=1.0\textwidth]{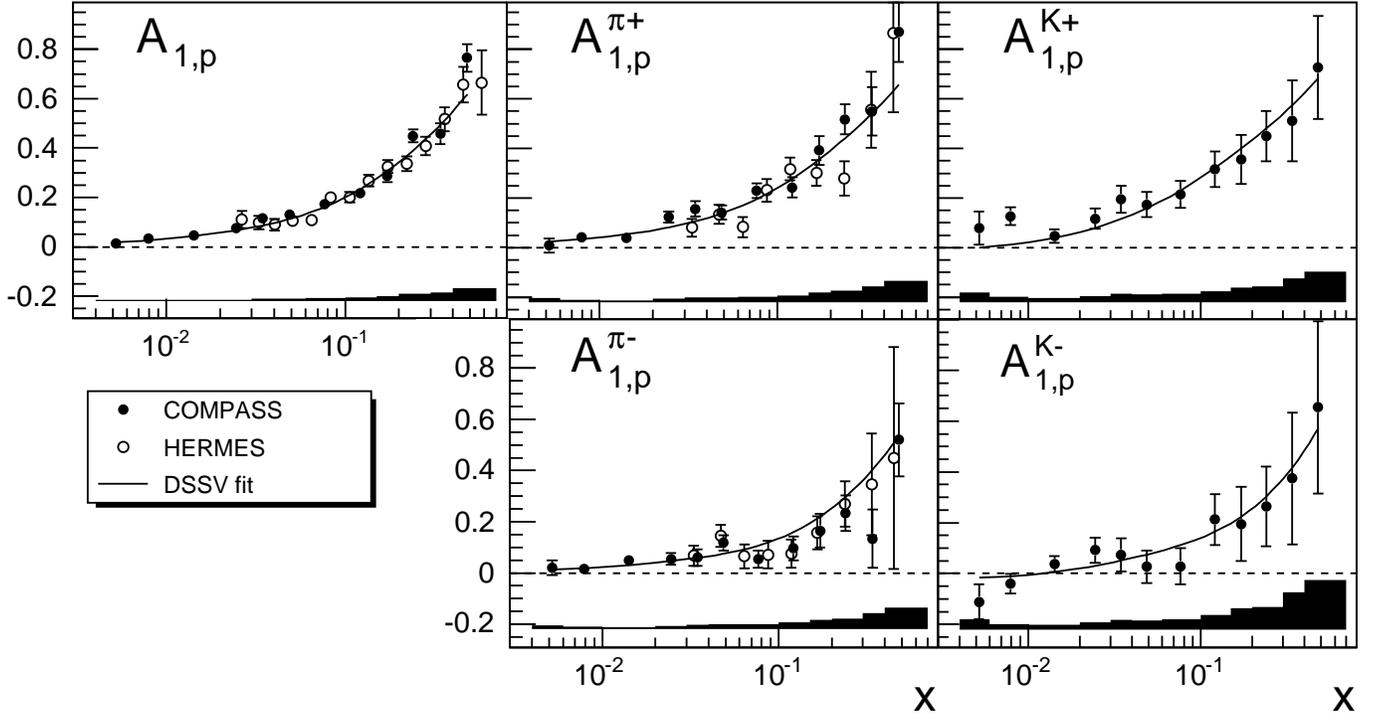}
\caption{The inclusive asymmetry $A_{1,p}$ and semi-inclusive asymmetries
$A^h_{1,p}$ from HERMES~\cite{airapetian07} and COMPASS~\cite{compass_pol_10}.
The curves are predictions of the DSSV parametrization~\cite{dssv}.}
\label{fig4.1}
\end{figure}

\begin{figure}[H]
\begin{center}
\includegraphics[width=0.9\textwidth]{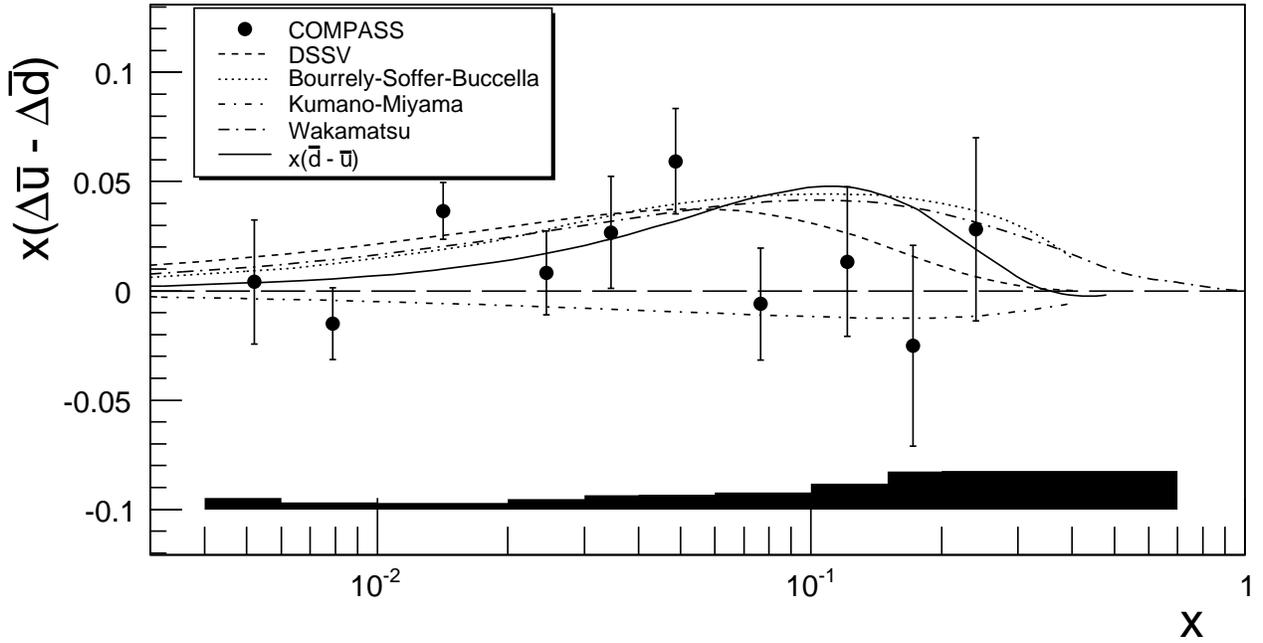}
\caption{Comparison of $x(\Delta \bar u - \Delta \bar d)$ extracted from
COMPASS~\cite{compass_pol_10} with the DSSV parametrization and 
various model predictions.}
\label{fig4.2}
\end{center}
\end{figure}

Figure~\ref{fig4.1} shows the $x$ dependence of the inclusive DIS asymmetry
$A_{1,p}$ and the semi-inclusive DIS asymmetries $A^{\pi^\pm,K^\pm}_{1,p}$
measured on polarized proton targets from HERMES and COMPASS. Very similar
trends, showing significant asymmetries at the large-$x$ valence quark 
region and a gradual fall-off as $x$ approaches the small-$x$ sea-quark
region, are observed for all asymmetries. This clearly indicates that 
the quark polarization is predominantly a valence-quark effect with a
relatively small role played by the sea quarks. Nevertheless,
the HERMES and COMPASS collaborations have 
reported their extractions of $\Delta \bar u(x)$ and 
$\Delta \bar d(x)$ using polarized semi-inclusive DIS data~\cite
{hermes_pol_05,compass_pol_10}. 
Although the early polarized SIDIS measurements were all consistent
with negligible $\bar u$ and $\bar d$ polarizations, the recent analysis
from COMPASS suggests a negative first moment for $\bar d(x)$ and
an intriguing $\Delta \bar u(x) - \Delta \bar d(x)$ distribution
shown in Fig.~\ref{fig4.2}. Although the accuracy of the measurement
is still quite limited, the data favor the models which predict
positive values for $\Delta \bar u(x) - \Delta \bar d(x)$, such as the
statistical model and the chiral-quark soliton model.

Figure~\ref{fig4.2} also shows the comparison between the extracted
values of $x(\Delta \bar u(x) - \Delta \bar d(x))$ and the DSSV
parametrization of polarized PDF~\cite{dssv}. Unlike the earlier
global fits to the polarized DIS data, the DSSV global fit also
included the polarized semi-inclusive DIS data as well as hadron
production data from polarized $pp$ collision at RHIC.  The DSSV
global fit also adopted the latest fragmentation functions~\cite{dss}
which provide a good description of the unpolarized SIDIS data.  Since
the earlier HERMES and COMPASS polarized SIDIS data were included in
the DSSV global fit, it is to be expected that the latest $\Delta \bar
u(x) - \Delta \bar d(x)$ result from COMPASS is well described by
DSSV.  Nevertheless, the uncertainties in the fragmentation function
are a main source of the systematic uncertainties in the extraction of
sea-quark polarization, as emphasized in
Ref.~\cite{compass_pol_10,leader12}.

\begin{figure}[htb]
\begin{center}
\includegraphics[width=0.7\textwidth]{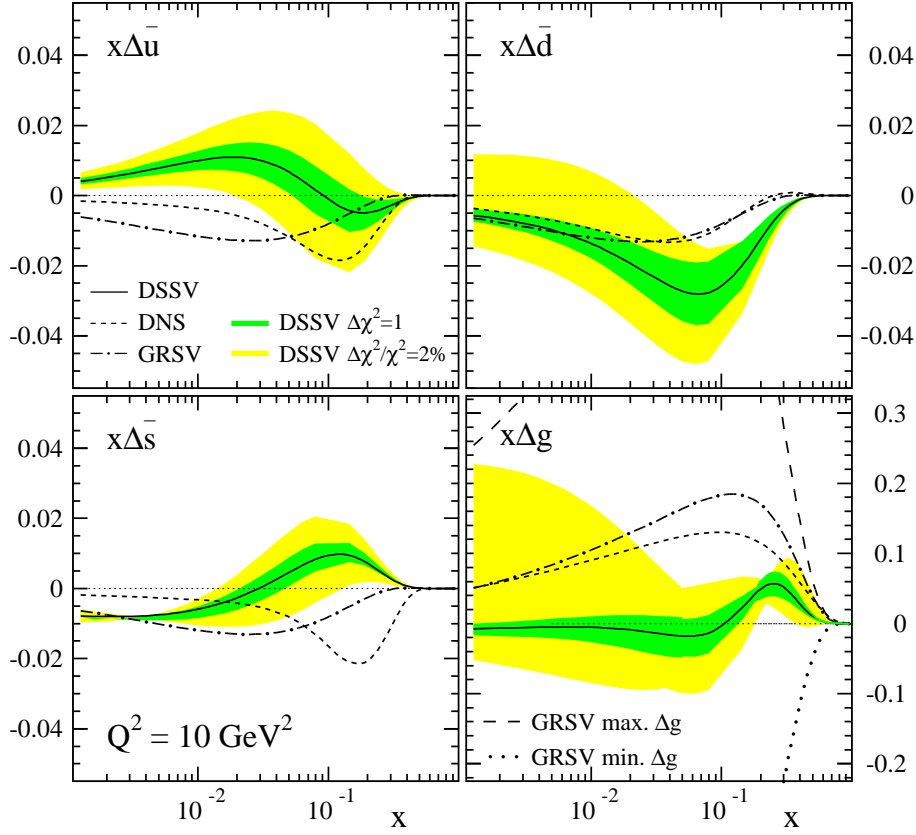}
\caption{The DSSV polarized sea and gluon distributions~\cite{dssv}.
The shaded bands correspond to $\Delta \chi^2 = 1$ and
$\Delta \chi^2/\chi^2 = 2\%$. Other polarized PDFs are also shown
for comparison.}
\label{fig4.3}
\end{center}
\end{figure}

Figure~\ref{fig4.3} shows the polarized sea and gluon distributions from 
the DSSV global fit. The assumption of an SU(3) symmetric polarized sea, $\Delta \bar u (x) = \Delta \bar d(x) = \Delta \bar s(x)$, is clearly at a variance
with the DSSV best-fit results. However, the error bands for $\Delta \bar u(x)$ 
and $\Delta \bar d(x)$ in Fig.~\ref{fig4.3} are still quite broad.
In particular, the narrower $\Delta \chi^2 = 1$ error band
gives the first moments of $\Delta \bar u = 0.028 \pm 0.021$ and
$\Delta \bar d = -0.089 \pm 0.029$. For the more conservative error band
($\Delta \chi^2/\chi^2 = 2\%$), both $\Delta \bar u$ and $\Delta \bar d$
are consistent with zero ($\Delta \bar u = 0.028 \pm 0.059$ and
$\Delta \bar d = -0.089 \pm 0.090$).

A unique experimental tool for measuring $\Delta \bar u$ and $\Delta \bar d$ 
sea-quark polarization is $W$ production
in polarized $pp$ collision at RHIC~\cite{soffer93,bunce00}.
As discussed by Bourrely and Soffer~\cite{soffer93}, three parity-violating
asymmetries in $W$ production are defined as

\begin{equation}
A_L=\frac{\sigma_+ - \sigma_-}{\sigma_+ + \sigma_-},~~
A^{PV}_{LL}=\frac{\sigma_{++}-\sigma_{--}}{\sigma_{++}+\sigma_{--}},~~
\bar A^{PV}_{LL} = \frac{\sigma_{+-}-\sigma_{-+}}{\sigma_{+-}+\sigma_{-+}},
\end{equation}
where $+,-$ in $A_{LL}$ refers to the proton beam helicity, while
$\sigma_+ = (\sigma_{++}+\sigma_{+-})/2$ and 
$\sigma_- = (\sigma_{-+}+\sigma_{--})/2$ in $A_L$. Ignoring the contributions
from strange and heavier quarks, the $W^+$ differential cross section can be 
written as

\begin{equation}
\frac{d\sigma}{dy}(W^+) = K\frac{\sqrt{2}\pi}{3} G_F x_1x_2 
\cos^2\theta_c [u(x_1) \bar d(x_2)+\bar d(x_1)u(x_2)].
\end{equation}

\begin{figure}[htb]
\centering
\begin{minipage}{0.6\textwidth}
\centering
\includegraphics[width=\textwidth]{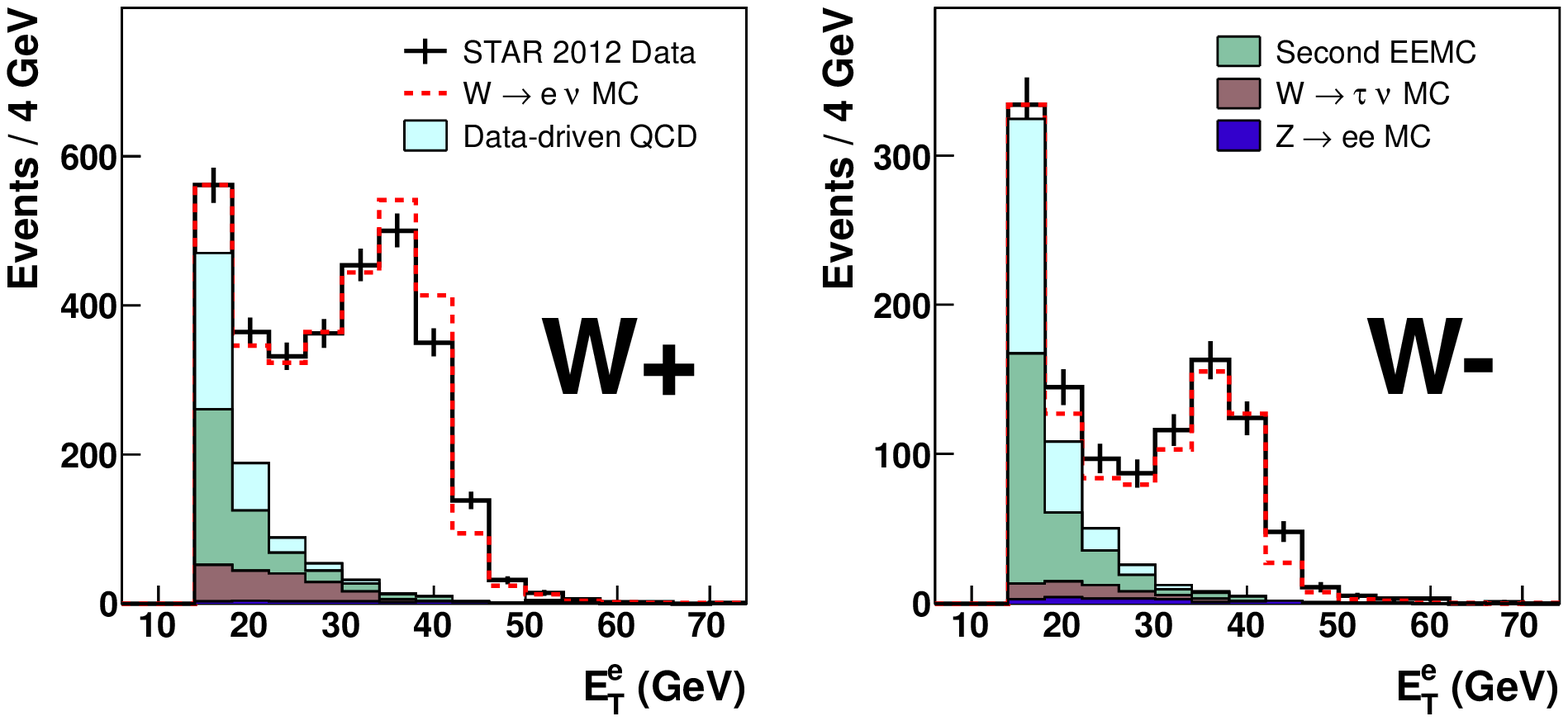}
\includegraphics[width=0.7\textwidth]{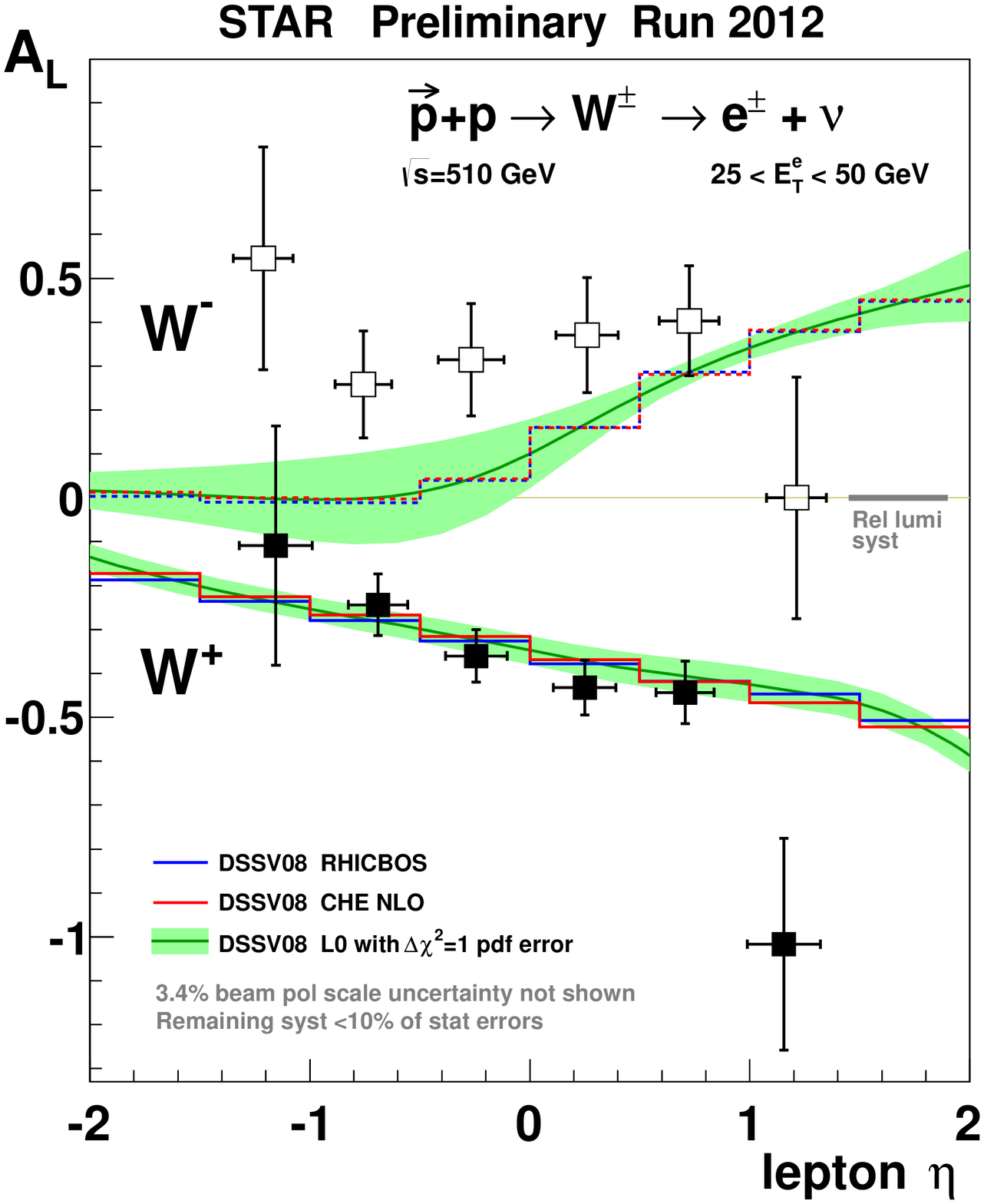}
\end{minipage}
\caption{Upper: (a): $W^+$ and $W^-$ candidate events from STAR. Lower: (b):
Single spin asymmetry $A_L$ from STAR~\cite{steven13}.}
\label{fig4.4}
\end{figure}

The factor $K$ takes into account the first-order QCD corrections, 
$K \simeq 1 + \frac{8\pi}{9} \alpha_s(Q^2) \sim 1.323$ at the $W$ mass
scale. Since $W$ couples to left-handed quarks and right-handed 
antiquarks, the expression for the four $W^+$ differential cross sections are
\begin{eqnarray}
\sigma_{++}\sim u^\downarrow(x_1) \bar d^\uparrow (x_2)
+ \bar d^\uparrow(x_1) u^\downarrow (x_2);~~
\sigma_{+-}\sim u^\downarrow(x_1) \bar d^\downarrow (x_2)
+ \bar d^\uparrow(x_1) u^\uparrow (x_2);\nonumber\\
\sigma_{-+}\sim u^\uparrow(x_1) \bar d^\uparrow (x_2)
+ \bar d^\downarrow(x_1) u^\downarrow (x_2);~~
\sigma_{--}\sim u^\uparrow(x_1) \bar d^\downarrow (x_2)
+ \bar d^\downarrow(x_1) u^\uparrow (x_2).
\label{eq:sig_W}
\end{eqnarray}
The superscripts $\uparrow,\downarrow$ refer to parton's helicity
parallel or anti-parallel to the proton's helicity. Eq.~\ref{eq:sig_W}
shows that the four beam-helicity dependent $W^+$ cross sections could
lead to the measurement of $u^\uparrow(x), u^\downarrow(x),
\bar d^\uparrow(x)$, and $\bar d^\downarrow(x)$. Similarly, the
four helicity dependent $W^-$ cross sections can determine
$d^\uparrow(x), d^\downarrow(x),
\bar u^\uparrow(x)$, and $\bar u^\downarrow(x)$.

The three parity-violating asymmetries as a function of $W^+$'s rapidity 
$y$ become~\cite{soffer93}
\begin{eqnarray}
A_L(y) & = & \frac{\Delta \bar d(x_1) u(x_2) - \Delta u(x_1) \bar d(x_2)}
{\bar d(x_1) u(x_2) + u(x_1) \bar d(x_2)}\nonumber\\
A^{PV}_{LL}(y) & = & \frac{[u(x_1)\Delta \bar d(x_2) - \Delta u(x_1)
\bar d(x_2)] - [\bar d(x_1) \Delta u(x_2) - \Delta \bar d(x_1) u(x_2)]}
{[\Delta u(x_1)\Delta \bar d(x_2) - u(x_1)
\bar d(x_2)] + [\Delta \bar d(x_1) \Delta u(x_2) - 
\bar d(x_1) u(x_2)]}\nonumber\\
\bar A^{PV}_{LL}(y) & = & \frac{[\bar d(x_1)\Delta u(x_2) + \Delta \bar d(x_1)
u(x_2)] - [u(x_1) \Delta \bar d(x_2) + \Delta u(x_1) \bar d(x_2)]}
{[\bar d(x_1) u(x_2) + \Delta \bar d(x_1)
\Delta u(x_2)] + [u(x_1) \Delta \bar d(x_2) + \Delta u(x_1) \Delta
\bar d(x_2)]}.
\label{eq:asym_W}
\end{eqnarray}
From Eq.~\ref{eq:asym_W}, it is clear that $A_L, A^{PV}_{LL}, \bar A^{PV}_{LL}$,
together with the unpolarized cross section $\sigma(y)$, would allow 
the extraction of $u(x), \Delta u(x), \bar d(x),$ and $\Delta \bar d(x)$
from $W^+$ production. Similarly, the three parity-violating observables and
$\sigma(y)$ for $W^-$ production would determine $d(x), \Delta d(x),
\bar u(x)$, and $\Delta \bar u(x)$, in principle. In practice, however,
it is the rapidity of the charged lepton ($l^\pm$) in the 
$W^\pm \to l^\pm \nu (\bar \nu)$ decay, rather than $W^\pm$ itself, 
which is measured experimentally. Moreover, as pointed out by Bourrely
and Soffer~\cite{soffer93}, under the reasonable assumption that 
$\Delta u \Delta \bar d \ll u \bar d$ 
(and $\Delta d \Delta \bar u \ll d \bar u$), both $A^{PV}_{LL} (y)$ 
and $\bar A^{PV}_{LL}(y)$ are related to the single spin asymmetry $A_L$:
\begin{equation}
A^{PV}_{LL}(y) = A_L(y) + A_L(-y);~~\bar A^{PV}_{LL}(y) = A_L(y)-A_L(-y).
\end{equation}
Effectively, there is only a single independent asymmetry observable, 
$A_L(y)$, in polarized $W$ production. From Eq.~\ref{eq:asym_W},
it is clear that at large forward rapidity region ($x_1 \gg x_2$),
\begin{equation}
A^{W^+}_L \sim -\frac{\Delta u(x_1)}{u(x_1)},~~A^{W^-}_L \sim 
-\frac{\Delta d(x_1)}{d(x_1)},
\end{equation}
while at large negative rapidity region ($x_1 \ll x_2$), 
\begin{equation}
A^{W^+}_L \sim \frac{\Delta \bar d(x_1)}{\bar d(x_1)},~~A^{W^-}_L 
\sim \frac{\Delta \bar u(x_1)}{\bar u(x_1)}.
\label{eq:w_AL}
\end{equation}
Eq.~\ref{eq:w_AL} shows that the most sensitive region for
determining the sea-quark polarization, $\Delta \bar u$ and $\Delta \bar d$,
is at large negative rapidity region. 
An important advantage of $W^\pm$ production for extracting $\Delta \bar u$
and $\Delta \bar d$ is that the uncertainty of fragmentation functions
encountered in polarized SIDIS is totally absent. The $W$ production mechanism
is well understood and theoretical calculations based on NLO (CHE)~\cite{che}
and resummation (RHICBOS)~\cite{rhicbos} are available.

First measurement of $A_L$ of $W^\pm$
production at RHIC has been reported by PHENIX~\cite{phenix11} and 
STAR~\cite{star11} using data collected in 2009 at $\sqrt{s} = 500$ GeV
with $\sim$ 10 pb$^{-1}$ integrated luminosity and an average beam polarization
of $30\%$.
A significantly higher integrated luminosity of 72 pb$^{-1}$ at
$\sqrt{s} = 510$ GeV and a beam polarization of 56\% during the 2012
run allowed an improved measurement of $A_L$. Preliminary results from
STAR~\cite{steven13} on the $W \to e \nu$ candidate events are shown in 
Fig.~\ref{fig4.4}. The measured $A_L$ over the pseudo rapidity region 
$|\eta| < 1.3$  are compared with theoretical predictions from calculations
using the DSSV proton helicity distributions. As discussed above, the
$A^{W^+}_L (A^{W^-}_L)$ at negative rapidity is sensitive to
$\Delta \bar d / \bar d (\Delta \bar u /\bar u)$. The good agreement
between data and prediction for $A^{W^+}_L$ supports the DSSV parametrization
for $\Delta \bar d(x)$. Interestingly, the preliminary STAR result for
$A^{W^-}_L$ suggests a significantly more positive $\Delta \bar u(x)$
than the DSSV parametrization. These new STAR results have prompted
a new DSSV++ global analysis reported in Ref.~\cite{elke}. A significant
shift from negative to positive values of the $\Delta \bar u$ moment
over $0.05 < x < 1$ is reported. This clearly illustrates the unique and
powerful constraints on $\Delta \bar u(x)$ ad $\Delta \bar d(x)$
provided by the polarized $W$ production data.

Very recently, RHIC completed a successful run in 2013, and the combined
$2012 + 2013$ integrated luminosity reaches $\sim 230$ pb$^{-1}$~\cite{xu13}.
Furthermore, the STAR forward calorimeter together with the PHENIX
forward muon spectrometer~\cite{oide14} extend the rapidity coverage to  
$-2.2 < \eta < 2.4$, allowing access to the kinematic region 
highly sensitive to 
$\Delta \bar u$ and $\Delta \bar d$. The significantly increased
integrated luminosity would also lead to measurements of
the sea-quark polarization through the double-spin asymmetries,
$A^{PV}_{LL}$ and $\bar A^{PV}_{LL}$, which could provide further 
consistency checks.

\subsection{Polarization of Strange Quark Sea: $\Delta s(x)$ and 
$\Delta \bar s(x)$}

The helicity structures of the strange quark sea, $\Delta s(x)$ and 
$\Delta \bar s(x)$, have attracted much theoretical and experimental
interest during the recent decades. The first indication for a sizable
polarization of the strange quark sea came from the measurement of
the first moment of the structure function $g_1(x,Q^2)$ from the
EMC experiment~\cite{ashman88}. In the scaling limit, 
\begin{equation}
\Gamma^p_1(Q^2) = \int^1_0 g^p_1 (x,Q^2) dx = \frac{1}{36} (4a_0+3a_3+a_8),
\end{equation}
where the three axial charges, $a_0,a_3,a_8$, are related to the first
moments of the quark helicity distributions as follows:
\begin{eqnarray}
a_0 & = & (\Delta u + \Delta \bar u) + (\Delta d + \Delta \bar d) +
(\Delta s + \Delta \bar s) \equiv \Delta \Sigma, \nonumber\\
a_3 & = & (\Delta u + \Delta \bar u) - 
(\Delta d + \Delta \bar d), \nonumber\\
a_8 & = & (\Delta u + \Delta \bar u) + (\Delta d + \Delta \bar d) -
2 (\Delta s + \Delta \bar s).
\label{eq:a_n}
\end{eqnarray}
The surprisingly small value of $\Delta \Sigma = 0.12 
\pm 0.094 (stat) \pm 0.138 (syst)$ at $Q^2 = 10.5$ GeV$^2$, 
obtained by the EMC 
Collaboration~\cite{ashman88,jaffe90}, shows that only a small 
fraction of the proton's spin is attributed to the quark's spin.
It also implies that $\Delta s + \Delta \bar s$ has a large
negative value~\cite{ellis95}. This is due to the fact that
$a_3$ and $a_8$ are related to the two constants $F$ and $D$
which can be determined from the weak decay of spin-1/2
baryon octet:
\begin{eqnarray}
a_3 & = & F + D = g_A/g_V = 1.269 \pm 0.003, \nonumber\\
a_8 & = & 3F - D = 0.586 \pm 0.031,
\label{eq:FD}
\end{eqnarray}
where $F=0.464 \pm 0.008$ and $D=0.866 \pm 0.008$ are obtained
from hyperon decay data assuming SU(3) flavor symmetry.
Eq.~\ref{eq:a_n} implies
\begin{equation}
\Delta s + \Delta \bar s = \frac{1}{3} (a_0 - a_8).
\end{equation}
The central values of $\Delta \Sigma = 0.12$ from EMC and 
$a_8 = 0.586$ from hyperon decay give 
$\Delta s + \Delta \bar s = -0.155$, a surprisingly large
negative value. As discussed later, a more precise later measurement
gives a central value of $a_0 = 0.330$, implying 
$\Delta s + \Delta \bar s = -0.085$. 
This corresponds to a surprisingly large strange-quark polarization 
in comparison with the lighter $\bar u$ and $\bar d$ quarks shown in 
Table~\ref{tab:del_u_sidis}.

The assumption of SU(3) flavor symmetry in hyperon decays, needed to
derive the expression for $a_8$ in Eq.~\ref{eq:FD},
was examined by Savage and Walden~\cite{savage97} who showed
that a violation of SU(3) flavor symmetry of up to 25\% may 
occur. Similar conclusion was found in a more recent 
work~\cite{bass10}. Nevertheless, recent lattice QCD 
calculations~\cite{lin09,erkol10,gockeler10} of hyperon axial
couplings support the assumption of SU(3) flavor symmetry.

The uncertainty in $\Delta S$, defined as $\Delta S = \Delta s +
\Delta \bar s$, also affects the calculation of
spin-dependent cross sections for dark-matter
scattering off nuclei~\cite{bottino02,ellis05,ellis08}.
The spin-dependent $(\chi + n)/(\chi + p)$ cross section ratio
where $\chi$ is the neutralino in the minimal supersymmetric
extension of the Standard Model (MSSM), is particularly sensitive
to the values of $\Delta S$. Assuming a $2\sigma$ variation for
$\Delta S = -0.09 \pm 0.03$, i.e., $-0.15 < \Delta S < -0.03$,
the $(\chi + n)/(\chi + p)$ cross section ratio would vary by a factor
of $2-3$~\cite{ellis08}.

In the following, we first summarize the status of various 
theoretical predictions on the strange quark polarization.
The experimental status as well as future prospects will
then be presented.

\subsubsection{Theoretical predictions on strange quark polarizations}

Various nucleon structure models, which have predictions at the confinement
scale on the
$\bar u(x) / \bar d(x)$ flavor asymmetry as well as
$\Delta \bar u(x) - \Delta \bar d(x)$, also have specific predictions
on $\Delta s(x)$ and $\Delta \bar s(x)$. In the simplest meson cloud
picture of proton consisting of the $K^+ + \Lambda$ Fock state, the
$\bar s$ quark residing in the spin-0 $K^+$ does not carry
any net spin, i.e., $\Delta \bar s = 0$. However, a net polarization
of $\Delta s(x)$ is expected since the spin of $s$ quark in $\Lambda$ is 
aligned with $\Lambda$'s spin. The relatively small coupling to the kaon 
cloud implies a small $\Delta s$. 
By taking into account the $K^* (S=1)$ cloud, Cao and Signal~\cite{cao03PRD}
found that $\Delta s$ and $\Delta \bar s$ are both non-zero, but
small ($\Delta s + \Delta \bar s = 0.01$).

In the chiral quark model~\cite{eichten92,cheng95,cheng98}, the
$u^\uparrow \to K^+ s^\downarrow$ is the dominant process resulting
in $\Delta s < 0$ and $\Delta \bar s = 0$. Using the simplest
version of the chiral quark model including the Goldstone boson
octet ($\pi,K,\eta$)~\cite{eichten92}, the moment of $\Delta s(x)$ is
related to the moment of $\bar u(x) - \bar d(x)$, namely,
$\Delta s = 3/2 (\bar u - \bar d) < 0$. By including the contribution
from $\eta^\prime$ and the axial U(1) breaking correction, a prediction
of $\Delta s = -0.10$ was obtained~\cite{cheng95}. Allowing SU(3)
breaking, a range of $-0.10 < \Delta s < -0.05$ was finally 
predicted~\cite{cheng98}.

\begin{table}[H]
\caption{First moment of $\Delta S$ ($\Delta S = \int^1_0 [\Delta s(x) +
\Delta \bar s(x)] dx$) from various lattice QCD calculations.}
\begin{center}
\begin{tabular}{ccccccc}\hline\noalign
{\smallskip} Reference & \cite{dong95,deka13} & \cite{gusken99} 
& \cite{babich12} 
& \cite{bali12} & \cite{engelhardt12} & \cite{etmc14} \\
$\Delta S$ & -0.12 (1) & -0.12 (7) & -0.019 (11) & -0.020 (11) 
& -0.031 (17) & -0.0227 (34)\\
\noalign{\smallskip}\hline
\end{tabular}
\end{center}
\label{tab:lattice_dels}       
\end{table}

Several lattice QCD calculations for $\Delta S$ have also been
reported.  Since strange quarks are not present in the valence
component of the nucleon, $\Delta S$ only involves the quark-line
disconnected diagrams.  Pioneering lattice calculations of
disconnected matrix elements show large negative values of $\Delta S =
-0.12 \pm 0.01$~\cite{dong95} or $\Delta S = -0.12 \pm
0.07$~\cite{gusken99} consistent with the values deduced from the EMC
polarized DIS experiment. More recent lattice calculations, however,
found much smaller values for $\Delta S$, as shown in
Table~\ref{tab:lattice_dels}. The discrepancy is likely due to the use
of quenched action in the early work~\cite{dong95,gusken99}.  A full
calculation of all the contributions to the proton spin, including the
spin of quarks (antiquarks) and gluons and the orbital angular momenta
of quarks (antiquarks), was recently reported~\cite{deka13} using a
quenched action. An interesting finding of this recent work is that
the orbital angular momentum (OAM) of strange quarks has a positive
value ($2L_s = 0.14 \pm 0.01$), which nearly cancels the spin part of
the strange quarks ($\Delta S = -0.12 \pm 0.01$). The net contribution
of the strange quarks to the proton's spin is $0.02 \pm 0.01$, which
is close to the values obtained by the recent unquenched calculations
for $\Delta S$, as shown in Table~\ref{tab:lattice_dels}. Note that
the OAM for strange quarks only originates from the disconnected
diagram.  The interplay between the roles of OAM and spin for the
strange quarks is an interesting topic deserving further
studies~\cite{syritsyn14}.

An inherent uncertainty in the extraction of $\Delta S$ from the 
inclusive or semi-inclusive polarized DIS is the contribution from 
the unmeasured small-$x$ region. An alternative method which bypasses this
uncertainty is to measure the strange quark's contribution to the 
axial form factor, $G^s_A(Q^2)$, from $\nu p$ and $\bar \nu p$ elastic
scattering cross sections. The value of $\Delta S$ can be
deduced by extrapolating $G^s_A(Q^2)$ to $Q^2 = 0$, namely,
$\Delta S = G^s_A(Q^2 = 0)$. At low $Q^2$, the $\nu p$ elastic cross section
is dominated by the axial form factor:
\begin{equation}
\frac{d\sigma}{dQ^2} (Q^2 \to 0) \sim \frac{G^2_F}{32\pi} 
\frac{M^2_p} {E_\nu^2} [(G^Z_A)^2 + (1 - 4 \sin^2\theta_W)^2],
\end{equation}
where
\begin{equation}
G^Z_A = \frac{1}{2} (-G^u_A + G^d_A + G^s_A).
\end{equation}
The $u$ and $d$ contribution to $G^Z_A$ is well determined from neutron
$\beta$ decay, $G^\mu_A - G^d_A = g_A = 1.269 \pm 0.003$. Several
model calculations for $G^s_A(Q^2)$ using chiral quark soliton
model~\cite{silva05,goeke07}, five-quark model~\cite{an06,riska06},
chiral quark model~\cite{lyubovitskij02}, and SU(3) skyrme 
model~\cite{park92} also exist. Table~\ref{tab:ff_dels} lists the prediction
of these models on $\Delta S = G^s_A(Q^2 = 0)$. It should be cautioned
that the value of $\Delta S$ deduced from $G^s_A(Q^2 = 0)$ corresponds 
to $Q^2=0$, while $\Delta S$ extracted from polarized semi-inclusive
DIS corresponds to $Q^2$ of several GeV$^2$. It is not obvious how the
value of $\Delta S$ obtained at $Q^2 = 0$ can be compared with
that obtained at larger $Q^2$ values.
\begin{table}[H]
\caption{Predictions of $\Delta S = G^s_A(Q^2 = 0)$ from various models.}
\begin{center}
\begin{tabular}{ccccc}\hline\noalign
{\smallskip} Reference & \cite{silva05,goeke07} & \cite{an06,riska06}
& \cite{lyubovitskij02} & \cite{park92} \\
$\Delta S$ & -0.075 & $-0.05 \pm 0.02$ & $-0.0052 \pm 0.0015$ 
& -0.03 \\
\noalign{\smallskip}\hline
\end{tabular}
\end{center}
\label{tab:ff_dels}       
\end{table}

\subsubsection{Experimental status on strange quark polarizations}

There are several experimental tools for probing the helicity distributions
of strange quarks. They include the polarized inclusive DIS, the polarized
semi-inclusive DIS, neutrino elastic scattering, hyperon weak decays, and
the longitudinal spin transfer $D_{LL}$ of $\Lambda$ and $\bar \Lambda$
production in polarized $pp$ collision. We briefly summarize
the current status on the extraction of strange quark polarization,
as well as the impact of these data on the parametrization of recent
polarized PDFs.

As discussed in Section 4.1.2, extensive measurement of polarized inclusive
DIS has led to the determination of $\Delta u(x) + \Delta \bar u(x)$,
$\Delta d(x) + \Delta \bar d(x)$, $\Delta s(x) + \Delta \bar s(x)$, and
$\Delta g(x)$ in several QCD analyses. Based on their measurements of
$g^p_1(x,Q^2)$ and $g^d_1(x,Q^2)$, the HERMES Collaboration determined
the first moment of the strange quark helicity distribution to be
$\Delta s + \Delta \bar s = -0.085 \pm 0.018$ 
at $Q^2 = 5$ GeV$^2$~\cite{airapetian07}
in the NNLO analysis.
From a measurement of $g^d_1(x,Q^2)$ alone, the COMPASS Collaboration
extracted $\Delta s + \Delta \bar s = -0.08 \pm 0.01(stat) 
\pm 0.02(syst)$ at $Q^2 = 3$ GeV$^2$~\cite{alexakhin07}. 
While these are very important
results, the assumption of SU(3) symmetry was adopted in the analysis.
Moreover, as discussed earlier, the inclusive DIS data do not facilitate 
the separation of $\Delta s$ from $\Delta \bar s$.

\begin{figure}[H]
\begin{minipage}{0.50\textwidth}
\includegraphics[width=\textwidth]{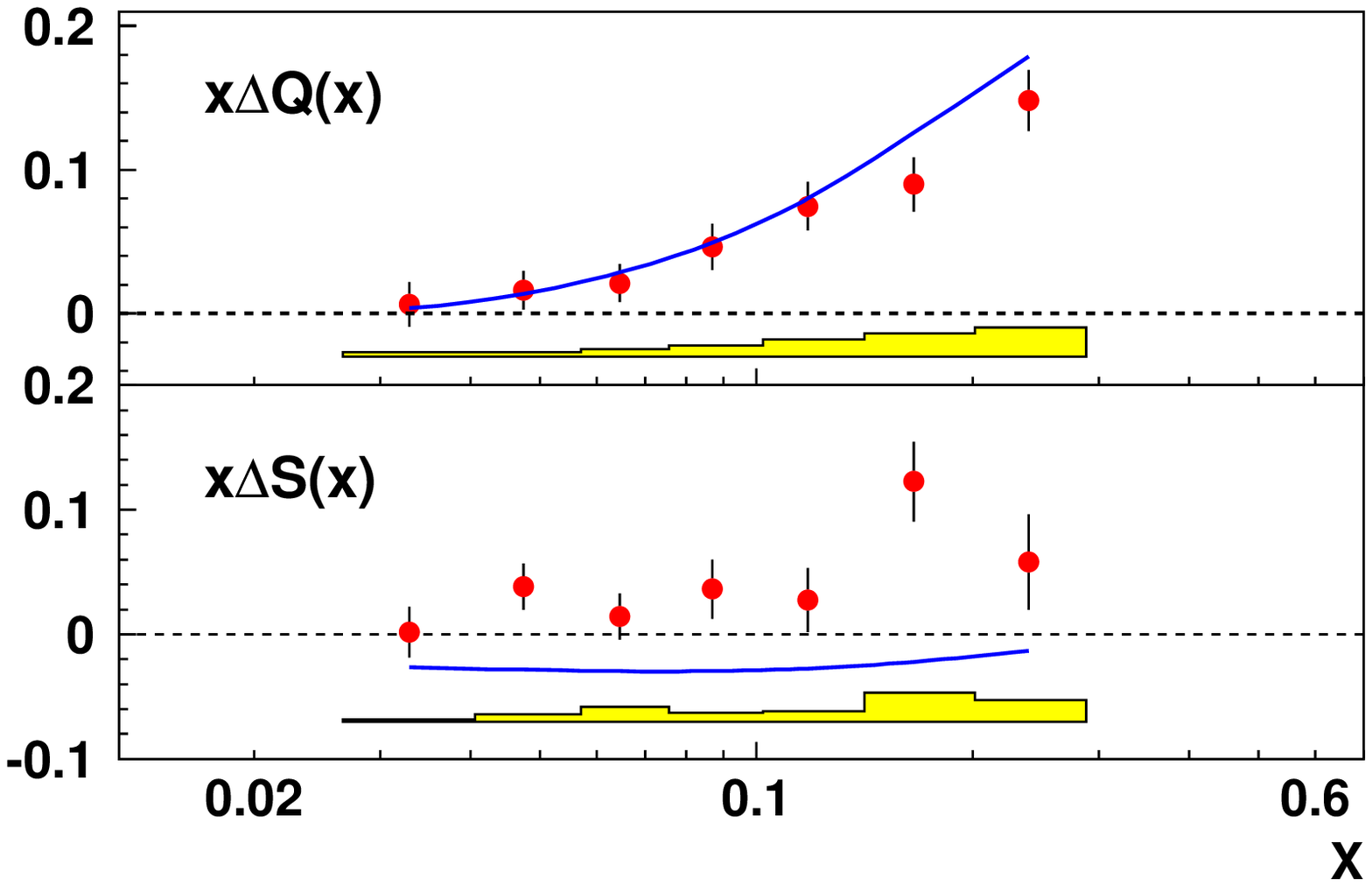}
\end{minipage}
\begin{minipage}{0.50\textwidth}
\includegraphics[width=\textwidth]{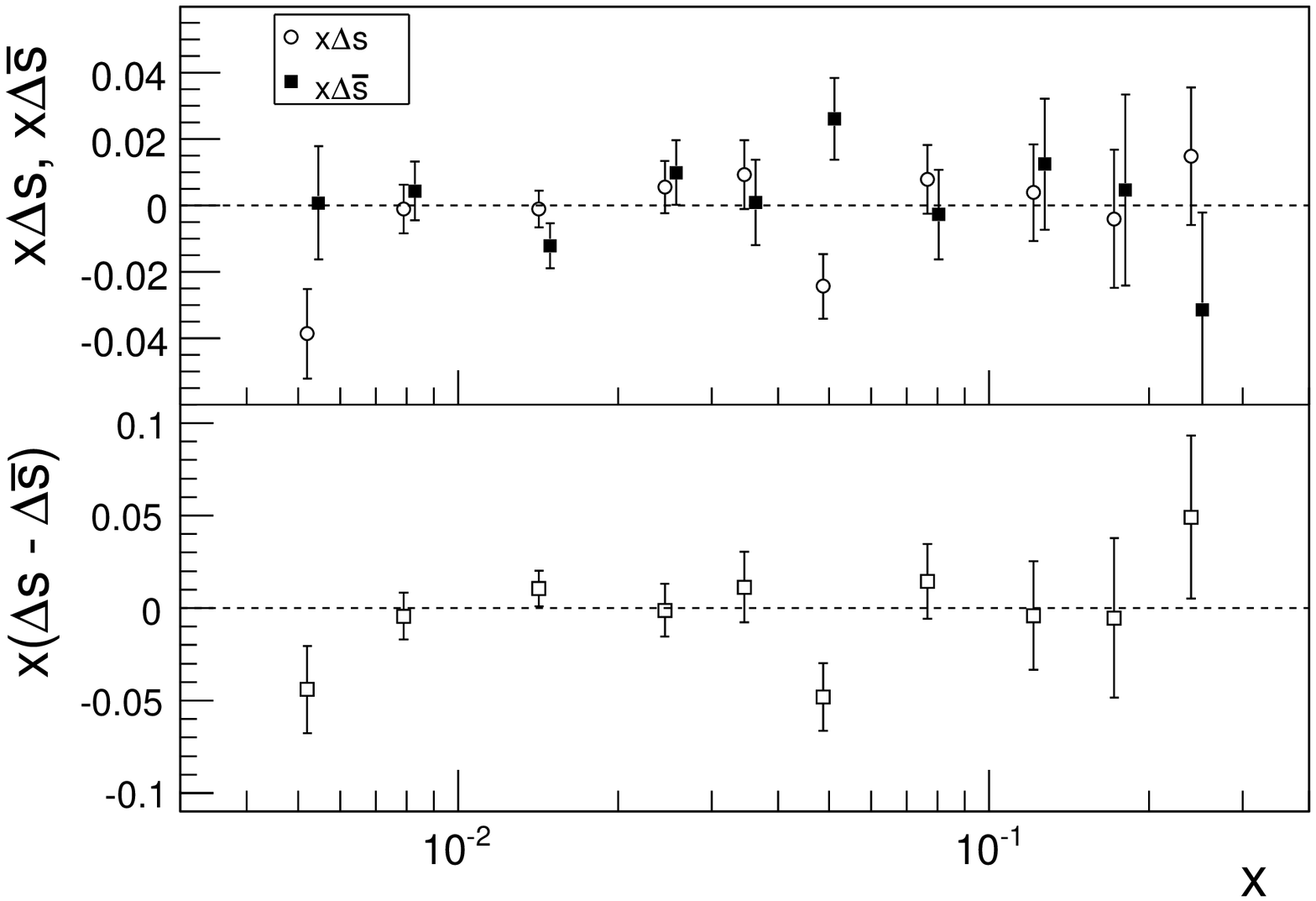}
\end{minipage}
\caption{Left (a): Nonstrange and strange quark helicity distributions
at $Q^2=2.5$ GeV$^2$ from HERMES~\cite{hermes08}. The curves are from the
global fit of Leader et al.~\cite{leader06}.
Right (b): Extraction of $x \Delta s(x)$, $x \Delta \bar s(x)$,
and $x (\Delta s(x) -\Delta \bar s(x))$ from 
COMPASS~\cite{compass_pol_10}.}
\label{hermes_del_s}
\end{figure}

From polarized semi-inclusive kaon production data, both the HERMES and
the COMPASS Collaborations have also reported the extraction of 
the $x$-dependence of the $\Delta S(x) \equiv \Delta s(x) + \Delta \bar s(x)$
distributions. Figure~\ref{hermes_del_s} shows the HERMES result on the 
nonstrange and strange quark helicity distributions at 
$Q^2 = 2.5$ GeV$^2$~\cite{hermes08}. While the nonstrange
quark helicity distribution 
($\Delta Q(x) \equiv \Delta u(x) + \Delta \bar u(x) + \Delta d(x) 
+ \Delta \bar d(x)$) is in good agreement with the result obtained
by Leader et al.~\cite{leader06} from their global fit to polarized
inclusive DIS data, the strange quark helicity distribution ($\Delta S(x)$)
favors positive values and differs from the inclusive DIS result.
This apparent tension between the extraction of $\Delta S(x)$ using
inclusive DIS versus semi-inclusive DIS was also observed by the COMPASS
experiment. Figure~\ref{hermes_del_s} shows the 
COMPASS extraction~\cite{compass_pol_10}
of $\Delta s(x)$, $\Delta \bar s(x)$ and $\Delta s(x) - \Delta \bar s(x)$.
This represents the first flavor separation between $\Delta s(x)$ and
$\Delta \bar s(x)$, demonstrating a unique capability of polarized 
semi-inclusive DIS reaction. There is an apparent discrepancy of 
$\Delta S(x)$ extracted from inclusive DIS versus semi-inclusive DIS.
This discrepancy could
reflect the uncertainty of the kaon fragmentation functions used in
the SIDIS analysis which is absent in the inclusive DIS 
analysis~\cite{compass_pol_10,leader11}. 
Indeed, it was shown~\cite{leader11} that a different choice of the 
kaon fragmentation function~\cite{hkns} would greatly reduce this
discrepancy. Another possible source of this discrepancy could be the
assumption of SU(3) symmetry in the inclusive DIS analysis. Several
attempts to fit both the inclusive DIS and the semi-inclusive DIS
data have been carried out~\cite{dssv,leader11,dns}. One example of
such a combined fit~\cite{dssv} is shown in 
Fig.~\ref{fig4.3}, in which the $\Delta S(x)$
distribution contains a node at small $x$. This allows positive values
of $\Delta S(x)$ at the $x$ region probed by semi-inclusive DIS, while
agreeing with the negative value of the moment of $\Delta S(x)$ favored
by the DIS analysis. A very recent report from HERMES~\cite{hermes2013} 
showed that the updated kaon multiplicity analysis~\cite{hermes2013a} does
not lead to any significant difference in the extraction of $\Delta S(x)$.
Clearly, more precise data on semi-inclusive DIS, as well as accurate
determination of the kaon fragmentation function, are needed for 
an accurate extraction of the strange quark helicity distribution.


The neutral-current neutrino elastic scattering reactions,
$\nu_\mu + p \to \nu_\mu + p$ and $\bar \nu_\mu + p \to \bar \nu_\mu + p$,
were measured sometime ago in the E734 
experiment~\cite{ahrens87} at BNL. This experiment led to an extraction 
of the weak mixing angle $\sin^2 \theta_W$, as well as the axial-vector
form factor $G_A(Q^2)$. The E734 analysis suggested an additional
contribution to $G_A(Q^2)$ possibly originating from the heavy-quark
currents. A later analysis by Garvey, Louis, and White~\cite{garvey93}
showed that the strange axial form factor $G_1^s(Q^2)$ can be extracted,
provided that strange vector form factors are better known. An extensive
effort to determine the strange quark vector form factors has been conducted
in several parity-violating electron nucleon scattering experiments at 
Bates, JLab, and Mainz (see Ref.~\cite{armstrong12} for a recent review). This allows a combined fit~\cite{pate04,pate08} to the E734 
neutrino scattering data
and the parity-violating electron scattering data, leading to the
determination of strange vector ($G^s_E$ and $G^s_M$) and axial
($G^s_A$) form factors. The extracted
values of $G^s_A(Q^2)$ suggest that $\Delta S = G^s_A (Q^2 = 0)$
is negative, $\Delta S = -0.30 \pm 0.42$, consistent with 
the result from polarized DIS experiments.
The accuracy of this independent measurement of $\Delta S$ could be
significantly improved~\cite{pate14} when the data from MicroBooNE
become available. The possibility of utilizing intense neutrino beam from
pion decays has also been considered~\cite{pagliaroli13}.

In polarized $pp$ collision, the longitudinal spin transfer $D_{LL}$ for 
$\Lambda$ and $\bar \Lambda$ production is sensitive to the helicity
distributions of $s$ and $\bar s$~\cite{florian98,xu06}. A first measurement
of $D_{LL}$ at mid-rapidity region for $\Lambda$ and $\bar \Lambda$ was 
reported by the STAR Collaboration~\cite{abelev09}. More recently, an 
eight-fold increase in data sample was collected at 200 GeV~\cite{xu13}.
Preliminary result suggests a positive value of $D_{LL}$ for both
$\Lambda$ and $\bar \Lambda$ at a $p_T$ of 6 GeV/c. While the extraction
of $\Delta s(x)$ and $\Delta \bar s(x)$ is likely to depend on the 
poorly known fragmentation functions, this process could be an interesting 
new tool for probing the strange quark helicity distributions.


\section{Transverse Structure of the Nucleon Sea}
\label{sec:TMDsea}

In addition to the spin-independent and the helicity distributions
of sea quarks discussed so far, there are also other novel sea
quark distributions involving transverse degrees of freedom for the
nucleon or quarks. These transverse degrees of freedom include the
transverse spin of the nucleon and the quark, as well as the transverse
momentum of the quark. There has been intense experimental effort in the
last decade to measure these novel parton distributions using lepton
and hadron beams. These novel distributions invoking the transverse
degrees of freedom potentially offer new insights on the nucleon
structure. They also provide stringent tests for various models on
nucleon structure. Moreover, the progress in lattice QCD also allows
comparisons between the lattice results with the experiments. 

From the three transverse quantities, namely, the nucleon's transverse
spin ($\vec S^N_\perp$), the quark's transverse spin ($\vec s^q_\perp$),
and the quark's transverse momentum ($\vec k^q_\perp$), three different
correlations could be formed. The correlation between the quark's
and the nucleon's transverse spins leads to the ``transversity" distribution.
The correlation between quark's transverse momentum and the nucleon's
transverse spin is the Sivers function. The Boer-Mulders function
corresponds to the correlation between the quark's transverse spin and
its transverse momentum. Among the various novel parton distributions,
the bulk of recent progress centered on these three distributions.
In the remainder of this Section we discuss the recent progress related to
these three distributions, focusing on their sea-quark components.

\subsection{Transversity Sea}

The nucleon unpolarized and polarized quark distributions, 
$q(x,Q^2)$ and $\Delta q(x,Q^2)$, are now rather well known. 
In contrast, a third quark distribution, called 
transversity ($\delta q(x,Q^2)$ or $h^q_1(x,Q^2)$),
was only beginning to be measured during the last decade. 
The transversity distribution, which can be described in the 
quark-parton model as the net transverse polarization of quarks 
in a transversely polarized nucleon~\cite{barone02}, has many 
interesting properties:

\begin{itemize}

\item In the non-relativistic quark model, where boosts and rotations 
commute, the transversity distributions are identical to the 
helicity distributions. The differences between the transversity and
the helicity distributions would reflect the relativistic nature
of the quarks inside the nucleon.

\item The transversity distributions have a valence-like 
behavior. Since the transversity of gluons in a nucleon does not exist, 
the transversity distributions are expected to follow a simple evolution 
as a flavor non-singlet quantity~\cite{bourrely98}.

\item The transversity distributions are predicted to obey some inequality
relations. 
The first, $|\delta q(x)| \leq q(x)$, follows from its interpretation as a 
difference of probabilities. The second (Soffer's bound) has its origins 
in the positivity of helicity amplitudes~\cite{soffer95}, 
$|h^q_1(x)| \leq (f^q_1(x)+g^q_1(x))/2$. As discussed below, other inequalities
can be obtained in the limit of large $N_c$.

\item The lowest moment of valence-quark transversity distribution 
measures a simple local operator analogous 
to the axial charge, known as the ``tensor charge". The tensor charge has
been calculated in various theoretical models and in lattice QCD.

\end{itemize}

In the following, we focus our discussion on the sea-quark transversity
distribution. We adopt the notation of $\delta q(x)$ and $\delta \bar q(x)$
for the quark and antiquark transversity distributions. While various
models predict similar behaviors for the valence-quark transversity
distributions, the predictions on the sea-quark transversity are sensitive
to details of the models. 

In the non-relativistic limit, one expects the same helicity and transversity
distributions. Therefore, one would expect similar flavor structure for
the sea-quark transversity and helicity distributions. However, model
calculations for the sea-quark transversity have distinct predictions
and do not necessarily follow this expectation. In a chiral-quark soliton
model calculation~\cite{schweitzer01} in the large $N_c$ limit, it was
predicted that $\delta \bar u(x) - \delta \bar d(x) < 0$, which is
opposite to the case for helicity, namely, 
$\Delta \bar u(x) - \Delta \bar d(x) > 0$. A similar result was obtained
by Wakamatsu~\cite{wakamatsu07}, who predicts the first moments of
$\delta \bar u = -0.05$ and $\delta \bar d = 0.08$ at $Q^2 = 0.36$ GeV$^2$. 
In a statistical
model approach~\cite{bourrely09}, $\delta \bar q(x) = \kappa \Delta \bar q(x)$
is predicted with an estimated value of 0.6 for $\kappa$. This leads to a
positive $\delta \bar u(x)$ and a negative $\delta \bar d(x)$, and hence
$\delta \bar u(x) - \delta \bar d(x) > 0$, which is
opposite to the predictions of the chiral-quark 
soliton model~\cite{schweitzer01,wakamatsu07}.
Finally, the very recent attempt to calculate the isovector 
$\delta \bar u(x) - \delta \bar d(x)$ in lattice QCD found~\cite{lin13} 
large negative values in qualitative agreement with the prediction of the
chiral quark soliton model. It would be very interesting to measure the
sign and magnitude of $\delta \bar u(x)$ and $\delta \bar d(x)$
to test the conflicting predictions from various models. 


\begin{figure}[htb]
\centering
\begin{minipage}{0.45\textwidth}
\includegraphics[width=\textwidth]{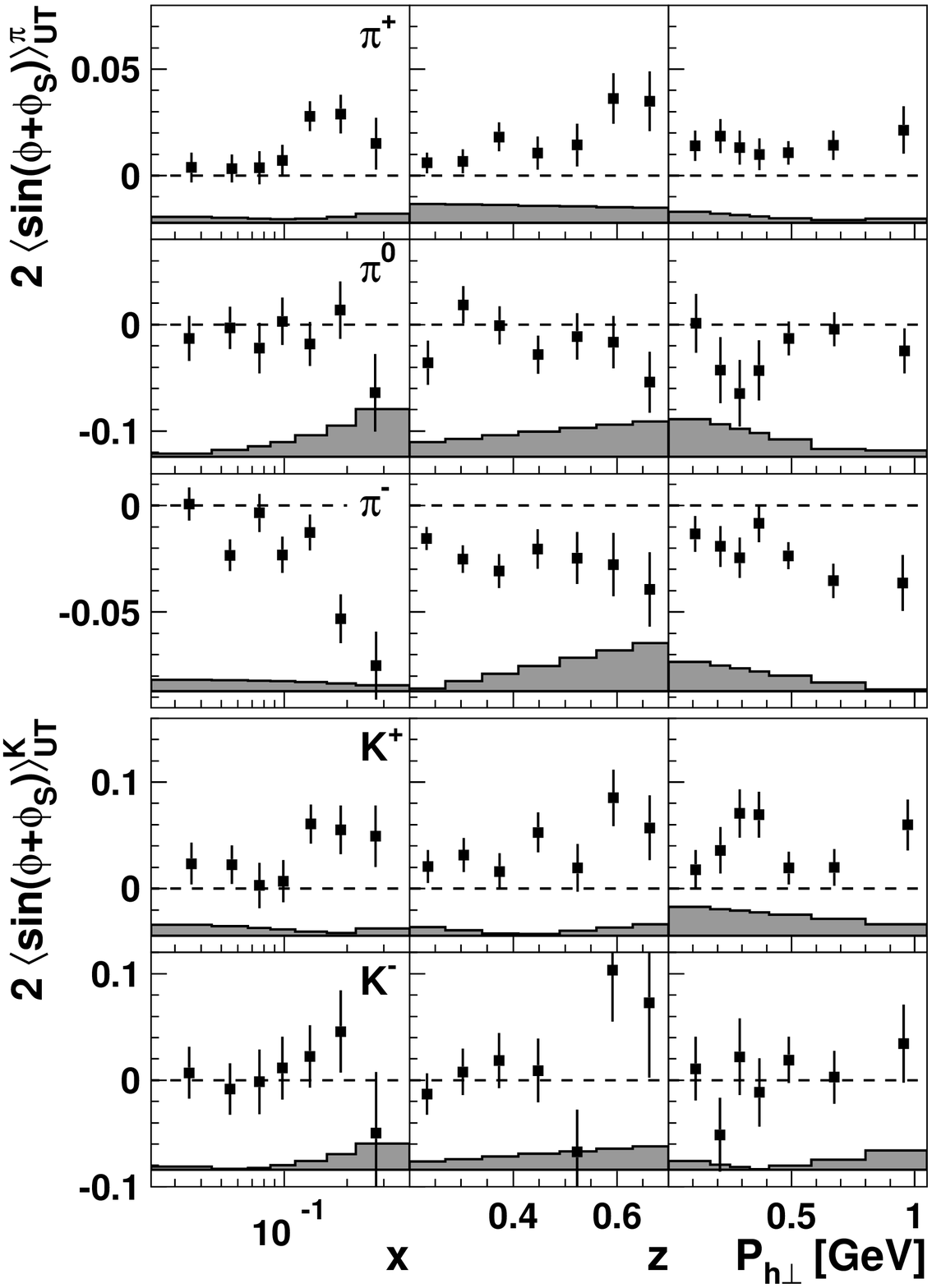}
\end{minipage}
\begin{minipage}{0.50\textwidth}
\includegraphics[width=\textwidth]{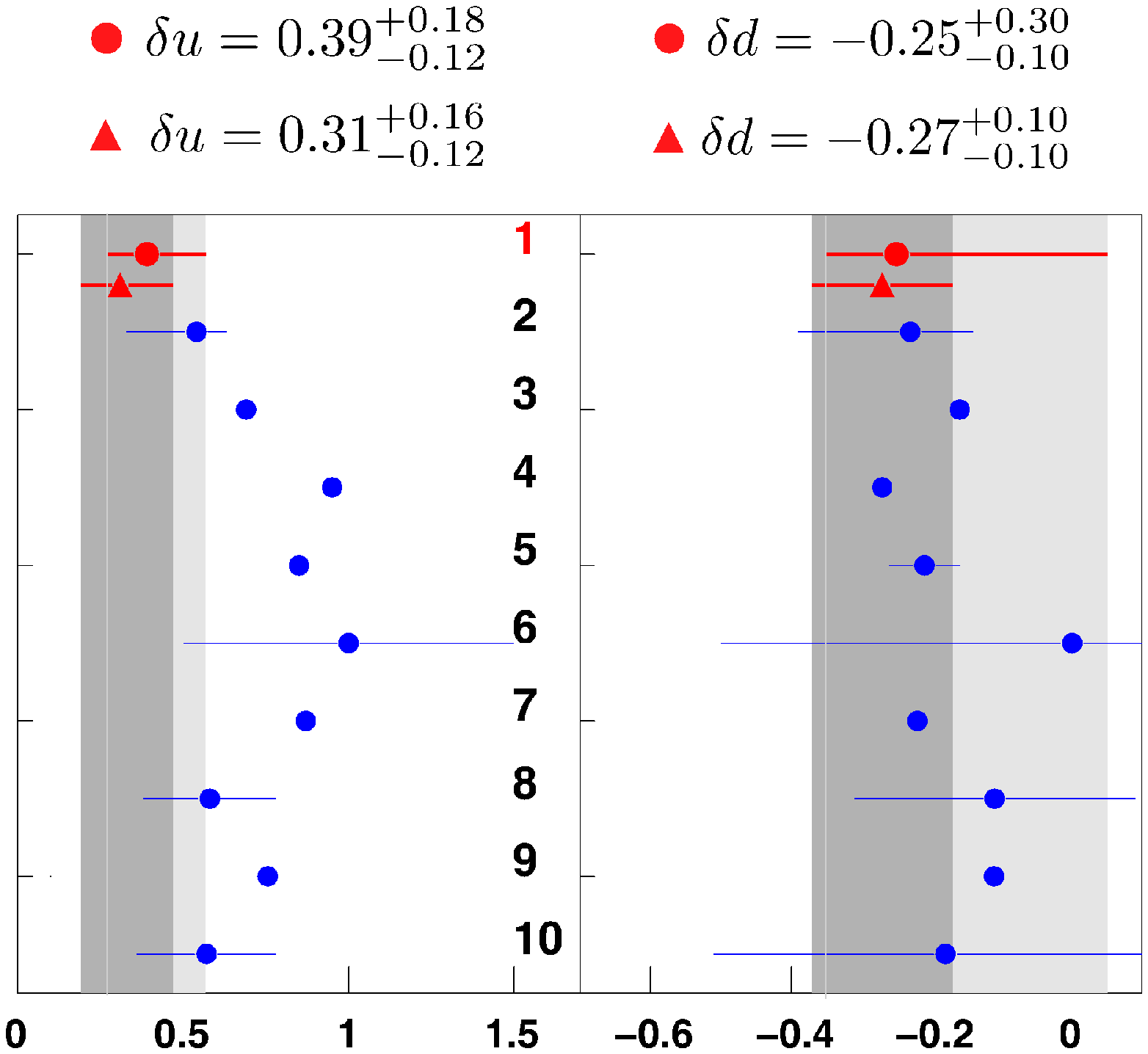}
\end{minipage}
\caption{Left (a): Amplitudes for the $\sin (\phi_h + \phi_s)$ azimuthal
dependence in semi-inclusive DIS measured by the HERMES Collaboration.
From~\cite{hermes_10}. Right (b): Tensor charge $\delta u$ and $\delta d$
obtained from the fit to the semi-inclusive DIS data at $Q^2 = 2.41$ GeV$^2$
in comparison with
various calculations~\cite{anselmino13}.}
\label{fig:trans}
\end{figure}

Due to the chiral-odd nature of the transversity distribution, 
it cannot be measured in inclusive DIS experiments. In order to 
measure $\delta q(x,Q^2)$, an additional chiral-odd
object is required. For example, the double spin asymmetry, $A_{TT}$, 
for Drell-Yan cross section in transversely polarized $pp$ collision, 
is sensitive to transversity since $A_{TT} \sim \Sigma_i e^2_i \delta q_i (x_1)
\delta \bar q_i (x_2)$~\cite{ralston}. Such a measurement could 
be carried out at 
RHIC~\cite{bunce00}, although the anticipated effect is small, on the order 
of $1 - 2$\%, due to the presumably small size of the sea-quark transversity
distributions. However, significantly larger values of $A_{TT}$ are expected
if the sea quark transversity is as large as what the lattice QCD 
predicts~\cite{lin13}.
An interesting proposal is to produce polarized antiproton
beam at the FAIR facility~\cite{fair}. This would allow the measurement 
of $A_{TT}$
for the Drell-Yan process in $\bar p p$ collision, involving the
valence transversity distributions in $\bar p$ and $p$. Much larger
values of $A_{TT}$ are expected~\cite{fair} for $\bar p p$ collision than
those of $pp$ collision.

Several other methods for measuring transversity have been 
proposed. In particular, Collins suggested~\cite{collins93} that a novel 
chiral-odd fragmentation
function, called Collins function ($H^\perp_1$), in 
conjunction with the chiral-odd 
transversity distribution, would lead to a single-spin azimuthal asymmetry 
in semi-inclusive meson production. Using a transversely polarized target,
an azimuthal angular modulation of $\sin (\phi_h+\phi_s)$ would be
proportional to $h_1 \otimes H^{\perp}_1$, which is the convolution 
of the transversity and the Collins fragmentation function.

The Collins fragmentation function can be extracted from the $\cos 2 \phi$
azimuthal dependence in the $e^+ e^- \to h^+ h^-$ reactions~\cite{boer97}.
The observation of a sizable Collins fragmentation function at 
Belle~\cite{seidl06} provided a crucial input for extracting the transversity
distribution from the detection of $\sin (\phi_h+\phi_s)$ angular
modulation in polarized SIDIS.

Major efforts to measure the transversity distributions via SIDIS on
transversely polarized targets have been carried out at 
HERMES~\cite{hermes_05,hermes_10}, 
COMPASS~\cite{compass_07,compass_12,compass_13}, and JLab~\cite{qian11,zhao14}
during the last decade. An example of the results obtained from these
experiments is shown in Fig.~\ref{fig:trans} (a). The measurement from
HERMES~\cite{hermes_10} using a transversely polarized hydrogen target shows
non-zero $\sin(\phi +\phi_s)$ amplitudes for charged pions and $K^+$.
In contrast, the amplitudes for $K^-$ are consistent with zero. Since 
the valence quarks in $K^-$ have flavors different from the valence
quarks in the nucleons, this suggests a small sea quark transversity.
Similar results were also reported by the COMPASS 
Collaboration~\cite{compass_13}.

Global analysis of the HERMES, COMPASS, and Belle data has led to the
extraction of the transversity distributions of the $u$ and $d$
quarks. In the most recent analysis~\cite{anselmino13} the sea
quark transversity was assumed to be zero. The salient feature of 
the results from this global analysis is that the $u(d)$ quark
transversity is positive (negative), just like the $u(d)$ helicity
distributions. However, the magnitudes of the transversity distributions
are smaller than the corresponding helicity distributions. 
Figure~\ref{fig:trans} (b) shows the tensor charges $\delta u$ and 
$\delta d$, where 
$\delta q \equiv \int^1_0 [\delta q(x) - \delta \bar q(x)] dx$, 
extracted from this analysis. Predictions from various models and lattice
QCD are also shown. It is noted that the central values for the 
tensor charges are $\delta u = 0.31$ and $\delta d = -0.27$,
which are significantly smaller in magnitude than the corresponding
values for the axial charges, $\Delta u = 0.787$ and $\Delta d = -0.319$.
Figure~\ref{fig:trans} (b) indicates that the extracted $\delta u$ is 
significantly smaller than the model predictions. The neglect of the
sea quark transversity in the analysis could introduce significant
systematic uncertainties in the determination of the tensor charges
$\delta u$ and $\delta d$. It is worth noting that the recent result
from JLab~\cite{zhao14} shows a large negative $\sin (\phi + \phi_s)$
amplitude for $K^-$ production on a transversely polarized $^3$He target,
hinting a sizable sea quark transversity. Future high-statistics
measurements proposed at the 12 GeV JLab upgrade~\cite{gao11}, as well
as measurements of $A_{TT}$ in Drell-Yan process at RHIC are required
to pin down the role of sea quarks in the transversity distributions.

\subsection{Sivers Sea}

It was suggested by Sivers~\cite{sivers90} that
correlations between the transverse spin of the target nucleon and the
transverse momentum of the unpolarized quark could lead to single-spin
asymmetries in various processes. This correlation is expressed in
terms of the ``Sivers Function'', which is an example of
transverse-momentum-dependent (TMD) parton distribution functions. 
Note that different notations have been adopted for Sivers function,
either $f^{\perp}_{1T}(x,k_\perp)$ or $\Delta^N (x,k_\perp)$.
As a time-reversal odd object, the Sivers function requires
initial/final state interactions via a soft gluon. As
shown in~\cite{brodsky02,collins02,ji02}, such interactions
are incorporated in a natural fashion by the gauge link that is
required for a gauge-invariant definition of the TMD parton
distribution. An unambiguous measurement of Sivers function would be
very valuable for understanding the nature of the TMD parton
distributions.

An important feature of the Sivers function is that it is related to
the forward scattering amplitude of \(N^{\Rightarrow}q\rightarrow
N^{\Leftarrow}q \) where the helicity of the target nucleon is
flipped. The helicity flip of the nucleon must involve the orbital
angular momentum of the unpolarized quark. Therefore, the Sivers
function is connected to the angular momentum of the quark.
In the meson-cloud model, a fraction of the nucleon's momentum
resides in the orbiting mesons. Since the mesons contain valence
antiquarks, it is natural to expect that antiquarks carry non-zero
orbital angular momentum~\cite{thomas08,garvey10}. Using the chiral quark
soliton model, Wakamatsu has shown the striking result that $\bar u$ and
$\bar d$ have dominant contributions to the proton's orbital angular
momentum~\cite{wakamatsu10}. The recent lattice 
calculation~\cite{deka13} also found a significant fraction of
proton's spin coming from the $\bar u$ and $\bar d$ orbital 
angular momentum. All of these theoretical studies suggest that 
the Sivers functions for sea quarks could be sizable and measurable.

\begin{figure}[H]
\begin{center}
\begin{minipage}{0.42\textwidth}
\includegraphics[width=\textwidth]{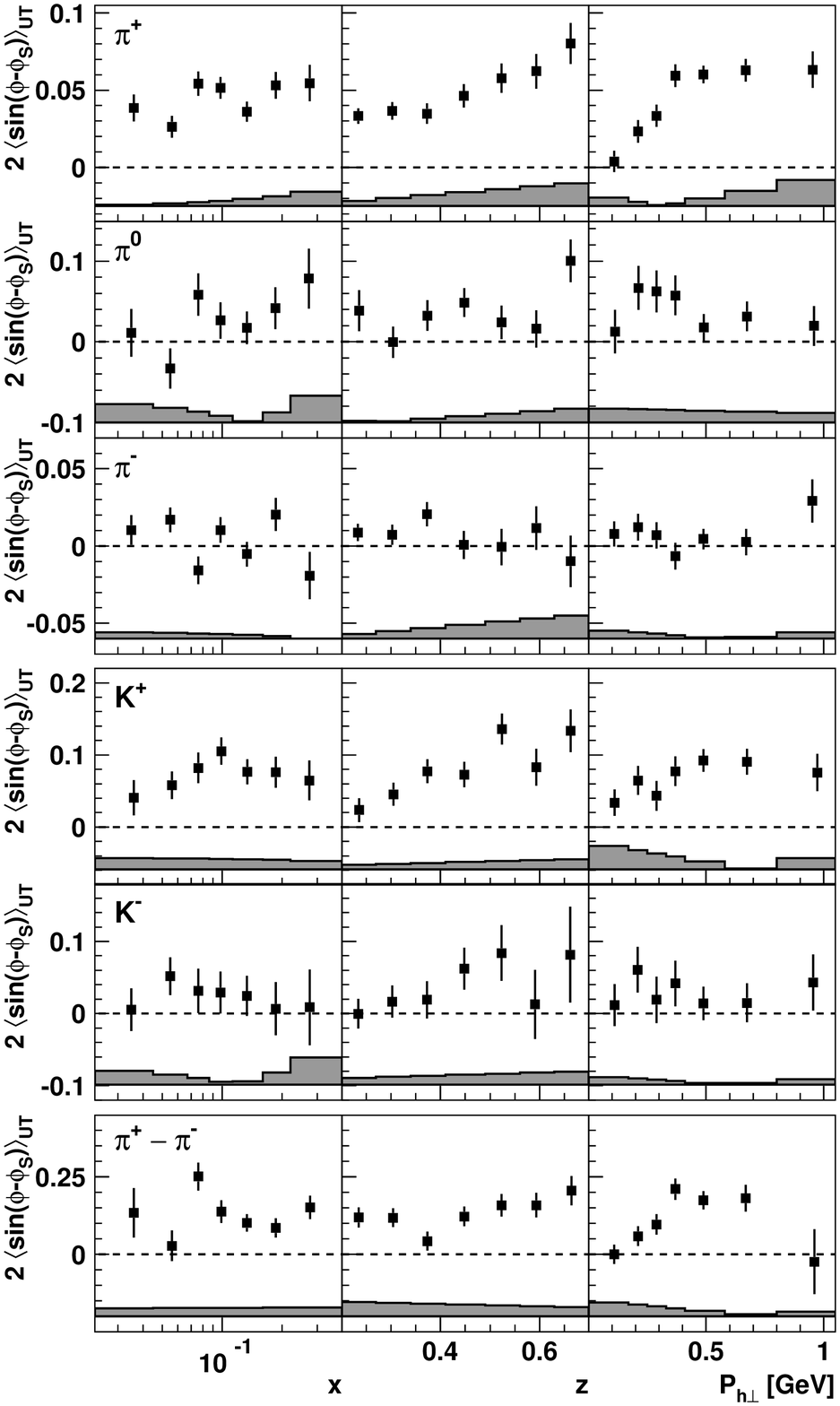}
\end{minipage}
\begin{minipage}{0.57\textwidth}
\includegraphics[width=\textwidth,bb= 10 150 580 660]{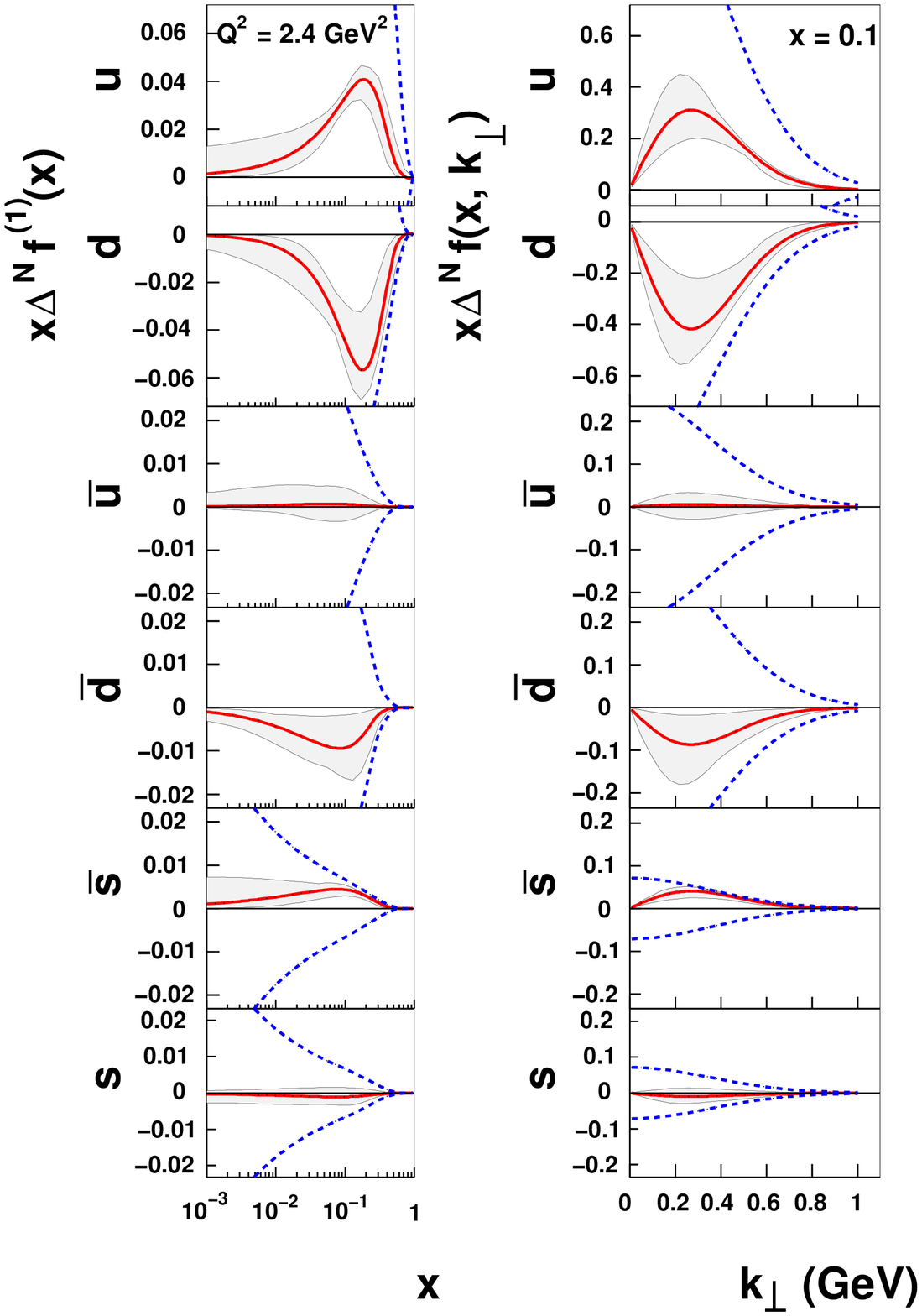}
\end{minipage}
\end{center}
\caption{Left (a): Amplitudes for the $\sin (\phi_h - \phi_s)$ azimuthal
dependence in semi-inclusive DIS measured by the HERMES
Collaboration~\cite{hermes_sivers}. Right (b): Sivers functions for 
quarks and antiquarks extracted 
from global analysis of semi-inclusive DIS data~\cite{anselmino09}.}
\label{sivers_fit}
\end{figure}

A first experimental indication for a sizable sea quark
Sivers function came from the HERMES measurement of 
a larger amplitude for the 
$\sin (\phi_h - \phi_s)$ Sivers moment in $K^+$ 
relative to $\pi^+$ production~\cite{markus}.
Other measurements of the Sivers moment have been reported by the
HERMES~\cite{hermes_sivers} and 
COMPASS~\cite{compass_sivers_1,compass_sivers_2} Collaborations. 
An example is shown in Figure~\ref{sivers_fit} (a) from the HERMES
Collaboration~\cite{hermes_sivers}. A global analysis to extract both 
the quark and antiquark 
Sivers functions was performed~\cite{anselmino09}. Figure~\ref{sivers_fit} (b)
shows the results on the extracted Sivers functions for the $u$, $d$, 
and various sea quarks. 
Another global analysis to extract the Sivers function, using more recent 
data from COMPASS, was recently reported~\cite{sun13}. Nonzero $\bar u$
and $\bar d$ Sivers functions are also obtained. 
Although the global analysis suggests non-zero
sea-quark Sivers functions, especially for $\bar d$, to account for
the large Sivers moment observed for $K^+$, much better statistical
accuracy~\cite{gao11} is required to draw definitive conclusions.

Sivers functions could also be studied in hadron-hadron collisions.
The observed large single spin asymmetries in the
production of $\bar p$ and $K^-$ from the polarized 
$pp$ collisions~\cite{brahms09} might
hint non-vanishing sea-quark Sivers distributions.
Polarized $pp$ Drell-Yan experiment provides a clean probe to extract sea quark
Sivers distributions without the ambiguities of parton fragmentation. A
proposal~\cite{jiang13} of replacing the target of the Fermilab E906/SeaQuest
DY experiment by a transversely-polarized $NH_3$ target to measure the sea quark
Sivers function  has been approved.

\subsection{Boer-Mulders Sea}

The Boer-Mulders function~\cite{boer98},
$h^\perp_1(x,k_\perp)$, is another example of a T-odd TMD. It 
signifies the correlation
between $\vec k_\perp$ and the quark transverse spin, $\vec s_\perp$, 
in an unpolarized
nucleon. As a 
chiral-odd analog of the Sivers function, the Boer-Mulders function 
also owes its existence to the presence of initial/final state
interactions~\cite{brodsky02}. 
Such interactions are incorporated in a natural fashion by the gauge
link that is required for a gauge-invariant definition of
the TMDs~\cite{collins02}.

The flavor and $x$ dependence of the  
Boer-Mulders functions have been calculated using various models.
In the quark-diquark model, the Boer-Mulders functions are 
shown to be identical to
the Sivers functions when the scalar diquark configuration alone
is considered~\cite{boer03,gamberg03}. However, calculations
taking into account both the axial-vector and 
the scalar configurations found significant
differences in the flavor dependence between the Sivers and Boer-Mulders
functions~\cite{gamberg08}.
In particular, while the $u$ and $d$ valence quark Boer-Mulders
functions are predicted
to be both negative, the Sivers function is negative for the $u$
and positive for the $d$ valence quarks. Other model calculations
utilizing the MIT bag model~\cite{yuan03}, the large-$N_c$ 
model~\cite{pobylitsa03}, the relativistic constituent
quark model~\cite{pasquini07}, as well as lattice QCD~\cite{gockeler07}
all predict negative signs for the $u$ and $d$ valence Boer-Mulders
functions. It is interesting to test their prediction
that the $u$ and $d$ Boer-Mulders functions both have negative signs.

The Sivers functions do not exist for spin-zero hadrons, such as pions,
since it corresponds to the correlation between
quark's $\vec k_\perp$ and hadron's spin. On the other
hand, the Boer-Mulders functions can exist for pions, since they do not 
depend on hadron's spin. Calculations for the pion's
valence-quark Boer-Mulders functions
using the quark-spectator-antiquark model
predict~\cite{ma_pion} a negative sign,
just like the $u$ and $d$ quark Boer-Mulders functions of the 
nucleons. Using the bag model, the valence 
Boer-Mulders functions for mesons and nucleons
were predicted~\cite{burkardt08} to have similar magnitude with the same
signs. This prediction of a universal
behavior of the Boer-Mulders functions for pions and nucleons awaits
experimental confirmation.

For nucleon's antiquark Boer-Mulders functions there
exists only one model calculation so far. It was pointed out~\cite{ma_sea}
that the meson cloud could contribute to nucleon's sea-quark 
Boer-Mulders functions. As discussed in Section 3, the
meson cloud as an important source for sea quarks in the
nucleons was evidenced by the large $\bar d / \bar u$ flavor
asymmetry observed in DIS and Drell-Yan experiments.
A significant fraction of the nucleon's antiquark sea comes from 
the meson cloud. This suggests
that pion cloud can contribute to the nucleon's
antiquark Boer-Mulders functions~\cite{ma_sea}. An interesting implication is
that nucleon's antiquark Boer-Mulders functions would have negative signs, just
like Boer-Mulders functions for pion's valence quarks.

The gauge-link operator leads to a remarkable prediction~\cite{collins02}
that the T-odd Sivers and Boer-Mulders functions are process dependent, namely,
they must have opposite signs depending on whether they are involved
in the spacelike SIDIS or the timelike
Drell-Yan process. An experimental verification of the sign-reversal
prediction of the Sivers and Boer-Mulders functions would provide an important
test of QCD at the confinement scale, and represents a significant step
towards understanding the properties of these novel TMDs.

The Boer-Mulders functions can be extracted from the azimuthal angular
distribution of hadrons produced in unpolarized SIDIS.
At leading twist, the $\cos 2\phi$ term is proportional to the product
of the Boer-Mulders function $h^\perp_1$ and the Collins fragmentation
function $H^\perp_1$. The azimuthal angle $\phi$ refers to the angle between
the hadron plane and the lepton plane. At the low $p_T$ region, the
$\langle \cos 2 \phi \rangle$ moment has been measured by the
HERMES~\cite{giordano09,hermes13} and COMPASS~\cite{kafer08,bressan09}
Collaborations. An analysis of these $\langle \cos 2 \phi \rangle$
data for pion SIDIS, taking into account the higher-twist 
Cahn effect~\cite{cahn78}, was performed~\cite{barone10a} by assuming
the functional form for the Boer-Mulders function as
\begin{equation}
h_1^{\perp q} (x,k_\perp^2) = \lambda_q f_{1T}^{\perp q} (x, k_\perp^2),
\label{eq:bmfit}
\end{equation}
\noindent where $q$ refers to the quark flavor and $h_1^{\perp q}$ and
$f_{1T}^{\perp q}$ are the Boer-Mulders and Sivers functions, respectively.
Equation~\ref{eq:bmfit} assumes the same $x$ and $k_\perp^2$ dependence 
for the Boer-Mulders and Sivers functions with the proportionality factor
$\lambda_q$ determined by data. The pion SIDIS
data are not yet able to constrain the antiquark Boer-Mulders functions,
which are expected to be dominated by their
quark counterparts. Therefore, the antiquark Boer-Mulders functions
were assumed to be equal in magnitude to the corresponding
Sivers function with a negative sign, namely,
\begin{equation}
h_1^{\perp \bar q} (x,k_\perp^2) = -|f_{1T}^{\perp \bar q} (x, k_\perp^2)|.
\label{eq:bmfita}
\end{equation}
\noindent The Sivers functions determined from a fit~\cite{anselmino09} to
the polarized SIDIS data together with the Collins function from
Ref.~\cite{anselmino09b}
were used. The best-fit values are $\lambda_u = 2.0 \pm 0.1$
and $\lambda_d = -1.111 \pm 0.001$. Since the Sivers function for $u (d)$
is negative (positive), these values imply that $h_1^{\perp u}$ and
$h_1^{\perp d}$ are both negative in agreement with the theoretical
expectation. It should be cautioned that the extracted signs of the 
Boer-Mulders functions depend on the signs of the Collins functions adopted in
the analysis. Although the signs chosen for the Collins functions
are based on plausible arguments, some uncertainties do remain
in determining the signs of the Boer-Mulders functions.

\begin{figure}[H]
\begin{center}
\includegraphics[width=\textwidth]{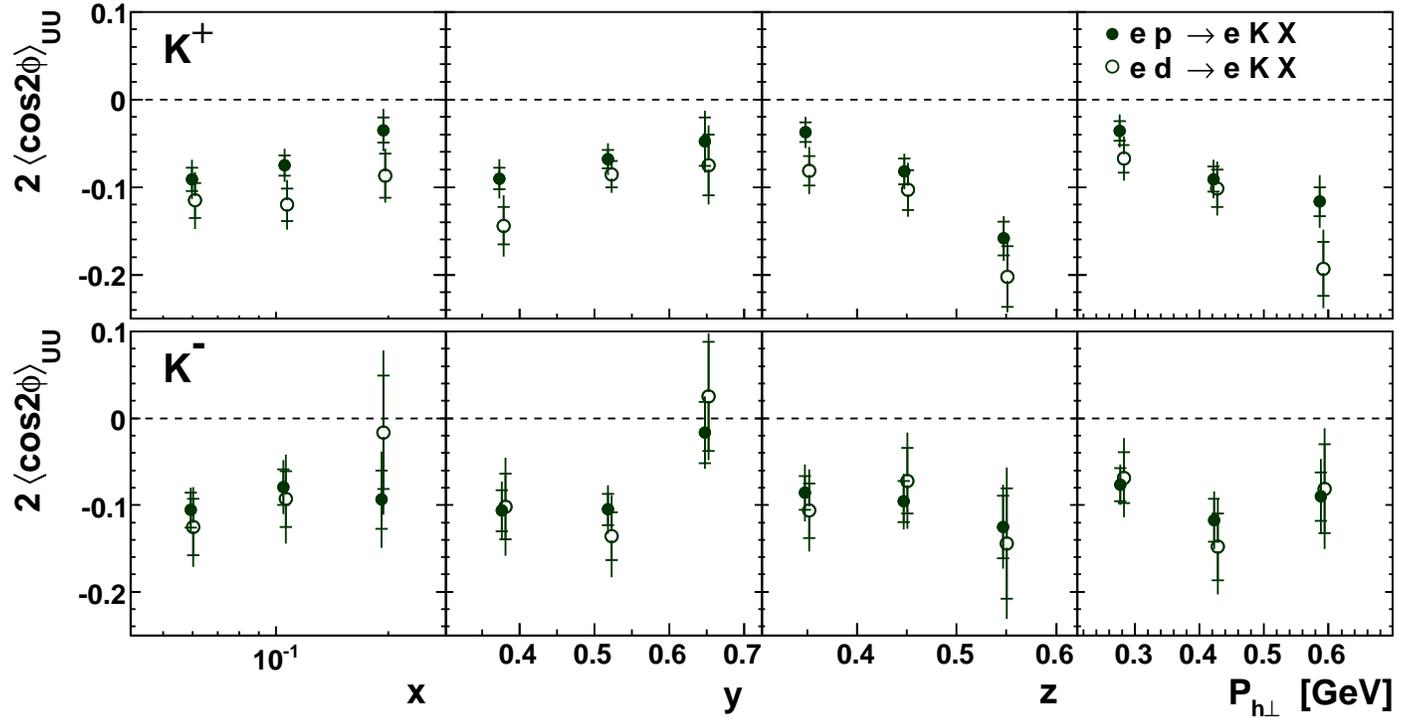}
\end{center}
\caption{$\cos 2 \phi$ amplitudes for $K^+$ and $K^-$ in unpolarized
SIDIS measurement by the HERMES Collaboration~\cite{hermes13}.}
\label{hermes_bm}
\end{figure}

The HERMES collaboration recently reported~\cite{hermes13} results on the
azimuthal $\cos 2 \phi$ modulations for $\pi^{\pm}$, $K^{\pm}$,
and unidentified hadrons in unpolarized $e+p$ and $e+d$ SIDIS.
The $K^{\pm}$ and unidentified hadron data were not included in the
earlier work~\cite{barone10a} to extract nucleon Boer-Mulders functions.
These new HERMES data could lead to a more precise extraction of the
valence Boer-Mulders functions. In addition, these data are sensitive to
the sea-quark Boer-Mulders functions. In particular, the $\cos 2 \phi$
moments for $K^-$ production are observed to be large
and negative~\cite{hermes13}, as shown in Fig.~\ref{hermes_bm}. 
Since the valence quark content of $K^-$,
$s \bar u$, is distinct from that of the target nucleons, the large
negative $K^-$ $\cos 2 \phi$ moments suggest sizable sea-quark 
Boer-Mulders functions.
An extension of the global fit in Ref.~\cite{barone10a} to include the
new $K^\pm$ data would be very valuable and could determine the sea-quark
Boer-Mulders functions in SIDIS.

The Boer-Mulders functions can be extracted~\cite{boer99}
from the azimuthal angular distributions in the unpolarized Drell-Yan
process, $h_1 h_2 \to l^+ l^- x$. The general expression for the Drell-Yan
angular distribution is~\cite{lam78}
\begin{equation}
\frac {d\sigma} {d\Omega} \propto 1+\lambda \cos^2\theta +\mu \sin2\theta
\cos \phi + \frac {\nu}{2} \sin^2\theta \cos 2\phi,
\label{eq:DY_ang}
\end{equation}
\noindent where $\theta$ and $\phi$ are the polar and azimuthal decay angle
of the $l^+$ in the dilepton rest frame. Boer showed that the $\cos 2\phi$
term is proportional to the convolution of the quark and antiquark
Boer-Mulders functions in the projectile and target~\cite{boer99}.
This can be understood by noting that the Drell-Yan cross
section depends on the transverse spins of the annihilating quark and
antiquark. Therefore, a correlation between the transverse spin and
the transverse momentum of the quark, as represented by the Boer-Mulders
function, would lead to a
preferred transverse momentum direction.

\begin{figure}[htb]
\begin{center}
\includegraphics[width=0.65\textwidth]{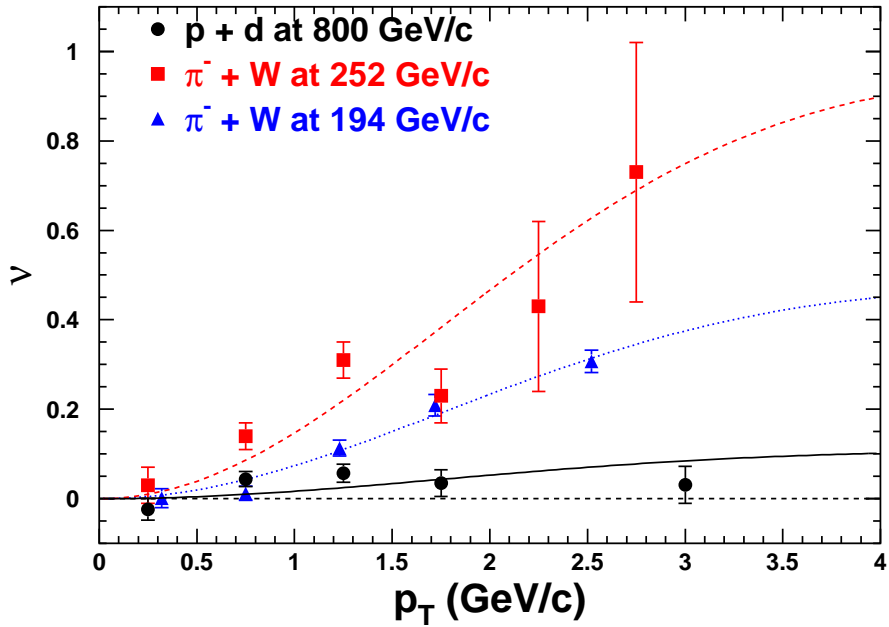}
\end{center}
\caption{The $\cos 2 \phi$ amplitude $\nu$ for pion- and
proton-induced Drell-Yan measurements. Curves are fits to
the data using an empirical parametrization discussed in 
Ref.~\cite{zhu07}.} 
\label{e866_bm}
\end{figure}

Pronounced $\cos 2 \phi$ dependence
were indeed observed in the NA10~\cite{falciano86} and E615~\cite{conway89}
pion-induced Drell-Yan experiments, and attributed to the
Boer-Mulders function.
The first measurement of the $\cos 2 \phi$
dependence of the proton-induced Drell-Yan process was reported for
$p+d$ and $p+p$ interactions
at 800 GeV/c~\cite{zhu07,zhu09}. In contrast to pion-induced Drell-Yan,
significantly smaller (but non-zero) cos$2\phi$ azimuthal angular dependence
was observed in the $p+d$ and $p+p$ reactions, as shown in
Fig.~\ref{e866_bm}. While the pion-induced 
Drell-Yan process
is dominated by annihilation between a valence antiquark in the pion
and a valence quark in the nucleon, the
proton-induced Drell-Yan process involves a valence quark in the proton
annihilating with a sea antiquark in the nucleon. Therefore, the
$p+d$ and $p+p$ results suggest~\cite{zhu07,zhu09} that the Boer-Mulders 
functions for
sea antiquarks are significantly smaller than those for valence quarks.

\begin{figure}[htb]
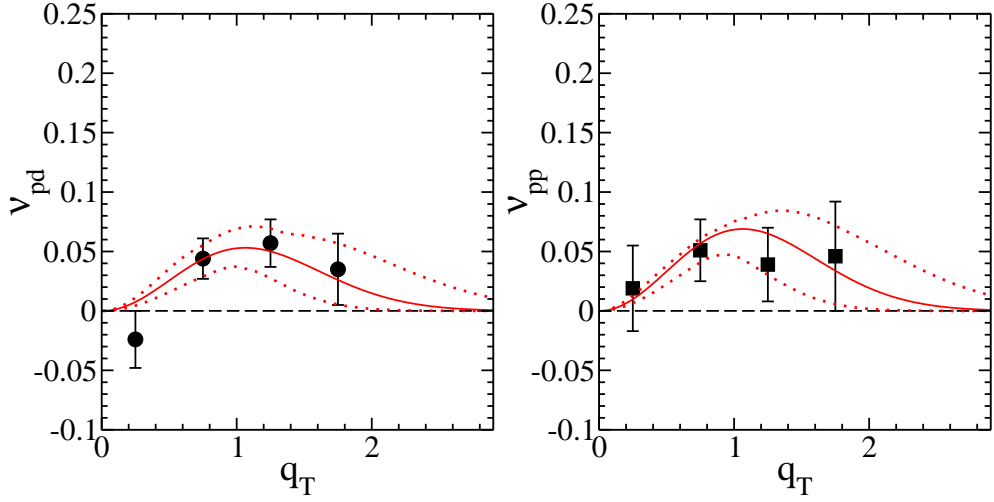

\centering
\begin{minipage}{0.35\textwidth}
\includegraphics[width=\textwidth,bb= 1 24 487 518]{fig/pd_cos2phi.eps}
\end{minipage}
\begin{minipage}{0.35\textwidth}
\includegraphics[width=\textwidth,bb= 1 24 487 518]{fig/pp_cos2phi.eps}
\end{minipage}
\caption{Fits to the $\cos 2 \phi$
amplitude $\nu$ for the
$p+d$ and $p+p$ Drell-Yan data~\cite{lu10}. The dotted lines
indicate the uncertainty of the fitted parameters. $q_T$
in the unit of GeV refers to the transverse momentum of the dimuon pair.}
\label{e866_bm_fit}
\end{figure}

Extractions of the Boer-Mulders functions from the $p+p$ and $p+d$ 
Drell-Yan data have been carried out by several 
groups~\cite{zhang08,lu10,barone10}. Figure~\ref{e866_bm_fit}
shows the 
fits obtained by Lu and Schmidt~\cite{lu10}. The statistical
accuracy of the data does not yet allow a precise extraction of 
Boer-Mulders functions. Nevertheless, the analysis suggests that
both the $\bar u$ and $\bar d$ Boer-Mulders functions can be
extracted from the Drell-Yan. Ongoing and future unpolarized
pion- and proton-induced Drell-Yan experiments are expected to provide 
new information on the Boer-Mulders functions.

\section{Experimental Perspectives}

In this section we outline the current experimental programs and the
future prospects which are related to the study of flavor structure of
nucleon sea. These activities are to be carried out by fixed-target
and collider experiments in various laboratories worldwide: Fermilab,
CERN, BNL, JLab, FAIR, NICA, and J-PARC.

\subsection{Fixed-Target Experiments}

\begin{figure}[H]
\begin{center}
\begin{minipage}{0.4\textwidth}
\includegraphics[width=\textwidth]{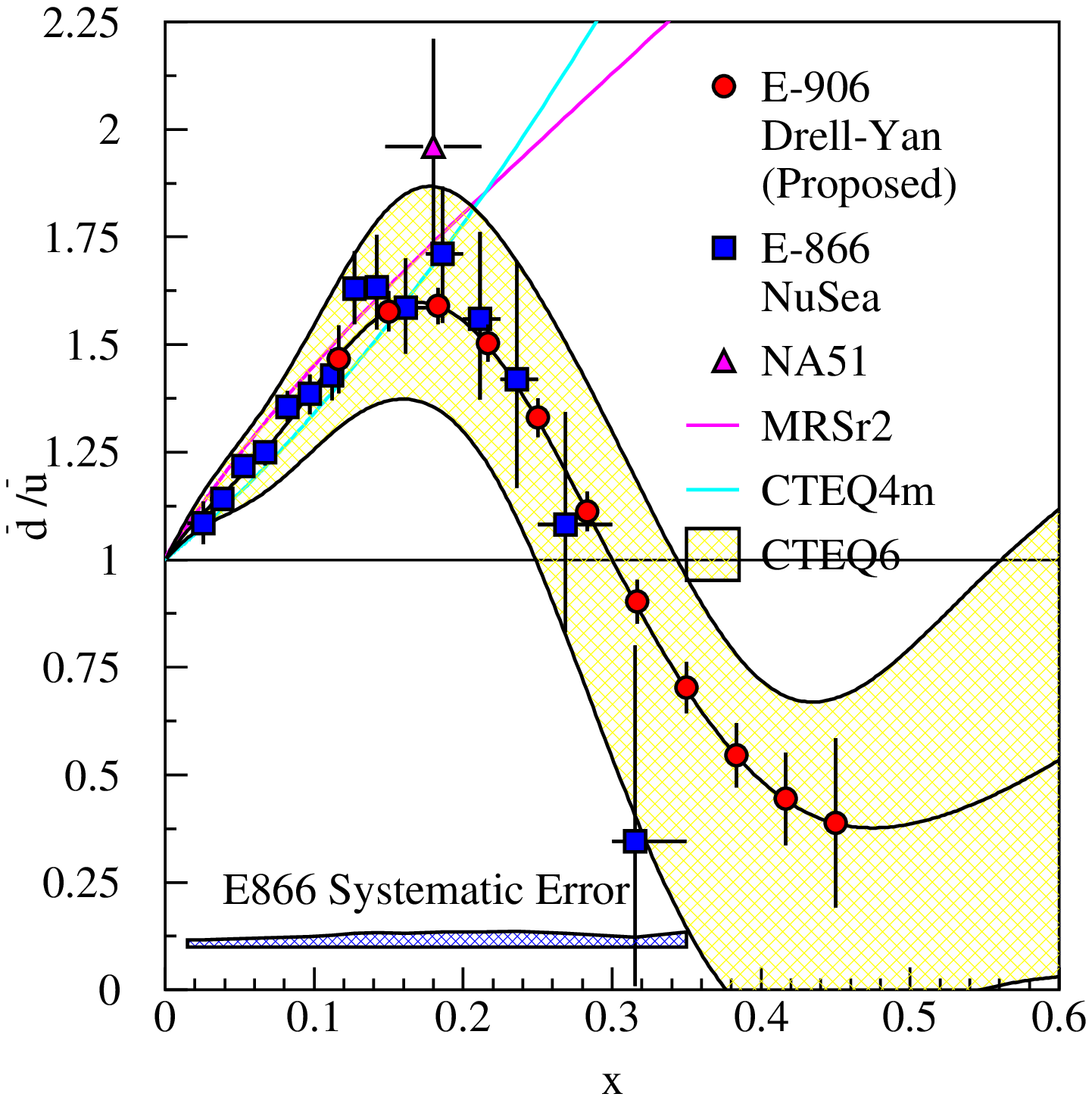}
\end{minipage}
\begin{minipage}{0.55\textwidth}
\includegraphics[width=\textwidth]{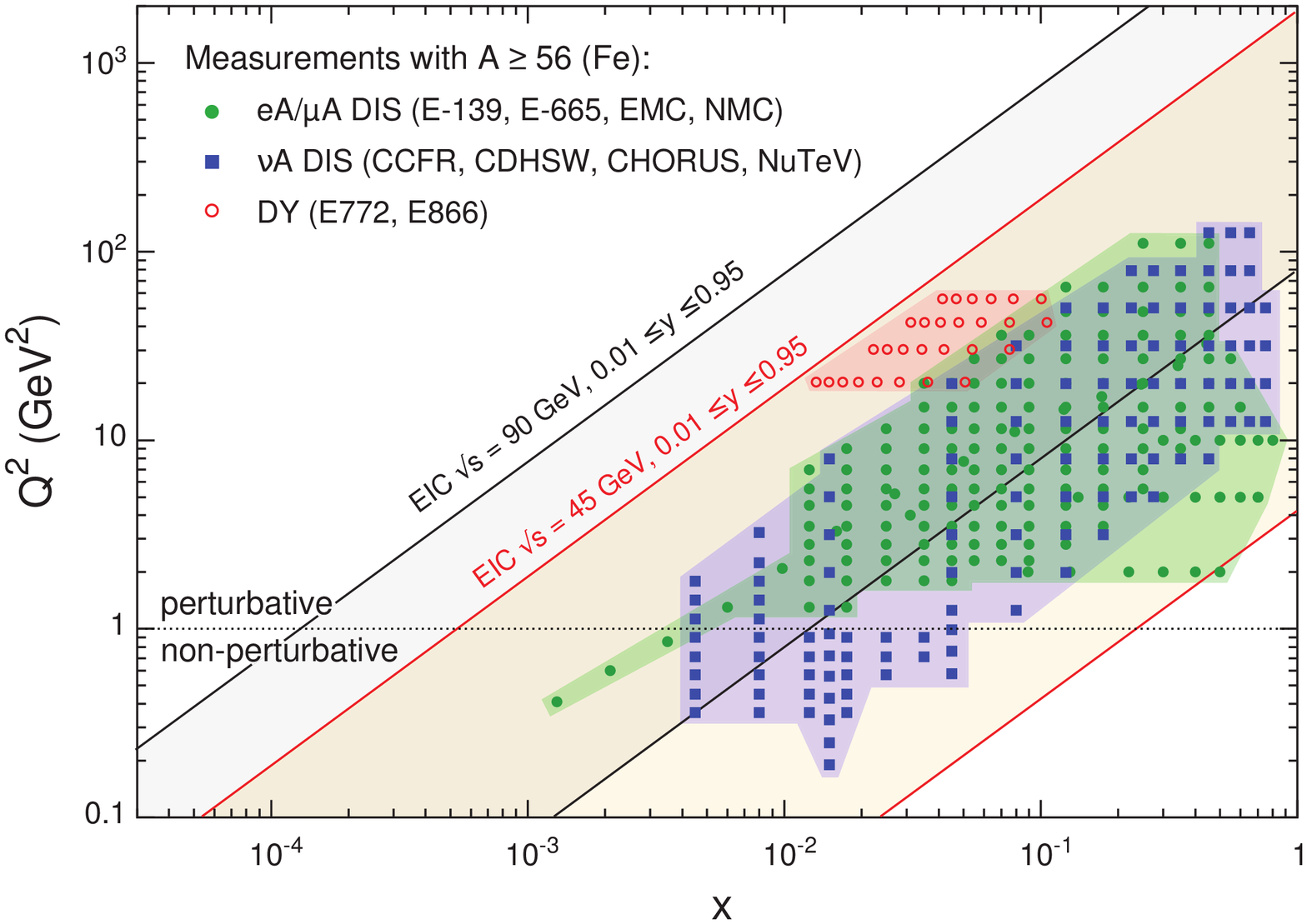}
\end{minipage}
\end{center}
\caption{Left (a): Expected sensitivity of $\bar d / \bar u$ as a
  function of $x$ for the Fermilab E906/SeaQuest
  experiment~\cite{reimer11}. Right (b): The kinematic acceptance in
  $x$ and $Q^2$ of completed lepton-nucleus DIS and fixed-target
  Drell-Yan experiments compared with EIC. Figure from~\cite{EIC2}.}
\label{fig:e906_EIC}
\end{figure}

While many theoretical models can describe the enhancement of $\bar d$
over $\bar u$, none of them predict that the $\bar d / \bar u$ ratio
falls below unity which is tentatively suggested by the E866 Drell-Yan data
at large $x$ ($x>0.2$) and the NMC data~\cite{peng14}. If confirmed by
more precise measurements, this intriguing $x$ dependence of the $\bar
d / \bar u$ asymmetry could shed important new light on the nature of
the nucleon sea. Thus, it would be very important to have new
measurements sensitive to the $\bar d / \bar u$ ratios at large $x$
($x>0.2$). The on-going Fermilab E906/SeaQuest Drell-Yan
experiment~\cite{e906,reimer11} utilizing 120 GeV proton beam will
hopefully extend the measurement of the $\bar d/ \bar u$ over the
region $0.25 < x < 0.5$ in the near future, as shown in
Fig.~\ref{fig:e906_EIC}(a). Furthermore there is an approved proposal
of measuring the Sivers functions of nucleon sea quarks using transversely
polarized $NH_3$ target and the existing E906 dimuon spectrometer at
Fermilab~\cite{P-1039}.

The new 50 GeV proton accelerator, J-PARC, presents opportunities for
extending the $\bar d/ \bar u$ measurement to even larger $x$ ($0.25 <
x < 0.7$)~\cite{peng00}. Since only 30 GeV proton beam is available at
the initial phase of J-PARC, the first measurements would focus on
$J/\Psi$ production at 30 GeV. An important feature of $J/\Psi$
production using 30 or 50 GeV proton beam is the dominance of the
quark-antiquark annihilation subprocess.  This is in striking contrast
to $J/\Psi$ production at 800 GeV (Fermilab E866) or at 120 GeV
(Fermilab E906), where the gluon-gluon fusion is the dominant
process. This suggests an exciting opportunity to use $J/\Psi$
production at J-PARC as an alternative method to probe antiquark
distribution.

With the capability of beam-particle identification, the usage of secondary
kaon beam in the Drell-Yan program of COMPASS-II experiment at
CERN~\cite{COMPASS:DY1,COMPASS:DY2} and high-momentum beam line in the
Hadron Hall at J-PARC~\cite{JPARC} will offer a clean and important
way of determining the strange sea.

There is a proposal of performing fixed-target experiments at LHC
using the 7 TeV proton beams~\cite{AFTER13}. With the measurement of
Drell-Yan production off nucleons at the very backward region, it will
measure the sea quark distributions of beam proton down to $x
\approx 10^{-3}$. 
The measurement of open-charm or
hidden-charm hadrons at large rapidities will be important for the
search of the intrinsic charm component.

\subsection{Collider Experiments}

The production of $W/Z$ bosons is known to provide strong constraints
on the light $u$ and $d$ quarks, both valence and sea. With an
increase of collision energy, the contributions of strange and charm
quark become much enhanced and the rapidity distribution of the $W/Z$
bosons provides a measure of the mix of valence and sea
quarks~\cite{kusina12}. Therefore, precision measurements of $W/Z$
production at the LHC may provide input to the global PDF fit to
constrain the strange and charm in the small $x$ region. In addition,
there are recent works showing the sensitivity to strange quark via
the charm production in association with an $W/Z$~\cite{stirling13}
and also to charm quark via the direct photon production in
association with a charm jet~\cite{stavreva09}.

\begin{figure}[htb]
\begin{center}
\begin{minipage}{.35\textwidth}
\includegraphics[width=\textwidth]{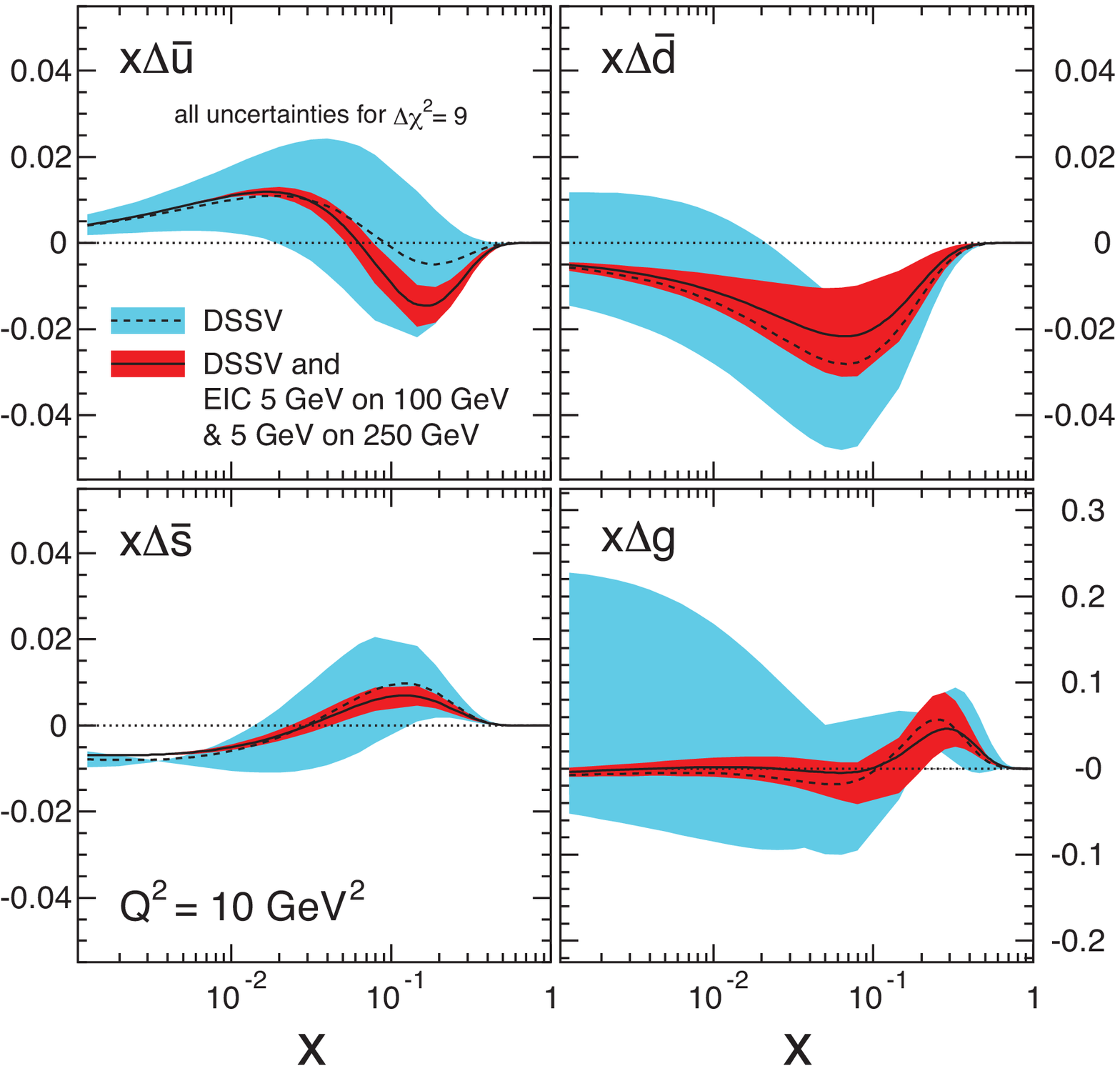}
\end{minipage}
\begin{minipage}{.40\textwidth}
\includegraphics[width=\textwidth]{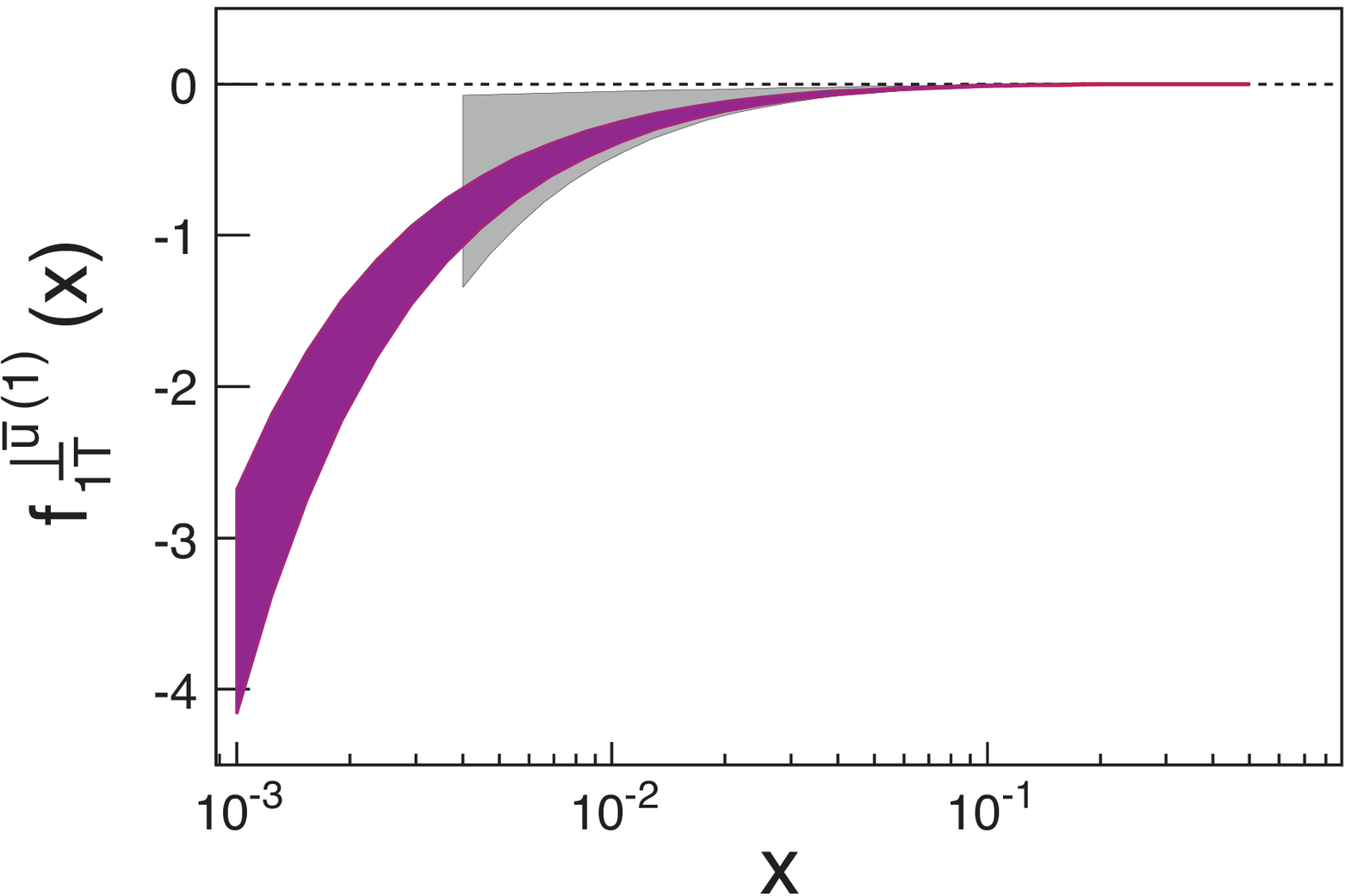}
\end{minipage}
\end{center}
\caption{Left (a): Uncertainty bands on helicity parton distributions,
  presently (light bands) and with EIC data (darker bands), using
  projected inclusive and semi-inclusive EIC data sets. Right (b):
  Comparison of the $2\sigma$ uncertainty of extracted Sivers function
  of $\bar u$ from currently available data (gray band) and from
  pseudo-data generated for the EIC with $\sqrt{s} = 45$ GeV and an
  integrated luminosity of 10 fb$^{-1}$ (purple band). Figures
  from~\cite{EIC2}.}
\label{fig:EIC23}
\end{figure}

There is a proposal of constructing an electron ion collider (EIC) with either
the CEBAF accelerator at JLab or the RHIC collider at
BNL~\cite{EIC1,EIC2}. This facility is aimed at addressing how the sea
quarks and gluons, and their spins, are distributed in spatial and
momentum space inside the nucleon with polarized ep and eA collisions
at $\sqrt{s}$=45-90 GeV. As shown in Figs.~\ref{fig:e906_EIC}(b) and
~\ref{fig:EIC23}, the availability of this facility will greatly
expand the explored kinematic regions of nucleon sea quarks in term of
distributions of partonic density, helicity density and TMDs, 
especially at small-$x$ regimes.

\section{Summary}
\label{sec:summary}

The nucleon sea is an indispensable part of the nucleon which is a
system bound by the strong interaction. As discussed in this review,
studies of the nucleon sea have revealed a rich flavor structure
beyond early simple pictures of gluon splitting and an SU(3)
symmetric sea. In conclusion, let us summarize what are learned so
far about the flavor structure of the nucleon sea.

\begin{itemize}

\item The observations of the Gottfried sum rule violation in a DIS
  experiment by NMC and the measurements of $\bar d(x)/\bar u(x)$ in
  Drell-Yan experiments by NA51 and E866 and SIDIS experiment by
  HERMES clearly establish the fact that the unpolarized distributions
  of $\bar u$ and $\bar d$ in the proton are strikingly different. At
  the region of $x>0.3$, the E866 data suggests that the $\bar d/\bar
  u$ falls below unity, albeit with large experimental uncertainty.

\item The non-perturbative models of a cloud of virtual mesons
  surrounding the nucleon provide quantitative descriptions of the
  observed $\bar d(x)/\bar u(x)$. None of them predicts
  that $\bar d/\bar u$ falls below unity at any value of $x$. New
  attempts of statistical and balance models based on the statistical
  properties of partons bound inside the nucleon could also reasonably
  describe the flavor structures of $\bar u$ and $\bar d$.  This
  flavor structure has been recently interpreted as the evidence of
  either intrinsic light sea abundant at valence region or sea quarks
  originated from the connected diagrams.

\item As for the quark sea with strange or charm flavors, fewer
  data are available. The momentum fraction carried by the strange sea is
  measured to be about half of that carried by $\bar u$ and $\bar d$
  at $Q^2$ = 1 GeV$^2$ in the neutrino DIS experiment by CCFR. This
  suggests the breaking of SU(3) flavor symmetric sea. Nevertheless
  the recent results of $s (x) + \bar s (x)$ from HERMES and ATLAS
  experiments suggest an SU(3) flavor symmetric quarks sea at $x \sim
  0.02$. The possible difference between $s(x)$ and $\bar s (x)$ was
  not excluded by the measurement of CCFR and NuTeV experiments. The
  charm production at the high $x$ and $Q^2$ regions in the DIS
  experiments by EMC shows possible evidences of a nonperturbative
  intrinsic charm component of the nucleon sea. However the most
  advanced global analysis shows that the currently available data
  could not effectively constrain the existence of this component and
  more precise experimental measurements are needed.

\item Experimental data from  polarized DIS, polarized SIDIS, and 
the single-spin asymmetry of $W$ boson production in $pp$ collision
show that sea quarks in the nucleons have small, but nonzero polarizations.
The SU(2) flavor asymmetry observed for  the unpolarized $\bar u$ and 
$\bar d$ sea is also found for the helicity distributions, as the data 
strongly favor $\Delta \bar u > \Delta \bar d$. For the strange quarks,
the polarized DIS and neutrino-nucleon elastic scattering data favor
a negative polarization, while the polarized SIDIS data suggest a 
positive strange quark polarization, at least in the measured $x$
region. Uncertainties in the kaon fragmentation functions remain
an obstacle in extracting the strange quark polarization in polarized
SIDIS reaction.

\item Impressive recent progress has been made in the study of the 
novel transverse-momentum and transverse-spin dependent parton distributions.
First information on the transversity, Sivers functions, and Boer-Mulders
functions has been successfully obtained in SIDIS and Drell-Yan experiments.
Currently, the focus of the TMD study is primarily on the valence quarks,
since the sea quarks are expected to have small contributions. Nevertheless,
indications for non-zero sea quark Sivers and Boer-Mulders
functions are already found in polarized and unpolarized SIDIS and
Drell-Yan experiments. A global effort to check the QCD prediction of the
sign-reversal for the T-odd Sivers and Boer-Mulders functions between
the SIDIS and Drell-Yan would very likely also lead to a better understanding
of the roles of sea quarks in the TMDs. The tantalizing connection between 
the sea-quark TMDs and the nucleon's orbital angular momentum remains to
be studied.

\end{itemize}

Major surprises were found in recent decades regarding the flavor and spin
structures of the nucleon sea. They have provided major challenges as well 
as important clues
for understanding the internal structure of the nucleons.
The flavor and spin structures of the nucleon sea will continue to offer
important new insights on how QCD works in the nonperturbative regime. 




\end{document}